\documentclass[aps,physrev,superscriptaddress,showkeys,floatfix,twocolumn]{revtex4-2}
\usepackage[colorlinks=true, allcolors=blue]{hyperref}
\usepackage{siunitx}	% for SI units
\usepackage{braket}
\usepackage{amsmath}
\usepackage{graphicx}
\usepackage{subcaption}
\usepackage{makecell}
\graphicspath{{images/}}

\begin{document}

%\preprint{}

\title{Modern approach to muonic x-ray spectroscopy demonstrated through the measurement of stable Cl radii}



\author{K.A.~Beyer}
\affiliation{Max-Planck-Institut für Kernphysik, Heidelberg, Germany}

\author{T.E.~Cocolios}
\affiliation{KU Leuven, Instituut voor Kern- en Stralingsfysica, Leuven, Belgium}

\author{C.~Costache}
\affiliation{Horia Hulubei National Institute for R\&D in Physics and Nuclear Engineering, Bucharest, Romania}

\author{M.~Deseyn}
\affiliation{KU Leuven, Instituut voor Kern- en Stralingsfysica, Leuven, Belgium}

\author{P.~Demol}
\affiliation{Université Libre de Bruxelles, Institut d’Astronomie et d’Astrophysique, Brussels, Belgium}
\affiliation{Brussels Laboratory of the Universe-BLU-ULB, Brussels, Belgium}

\author{A.~Doinaki}
\affiliation{PSI Center for Neutron and Muon Sciences, Villigen, Switzerland}
\affiliation{ETH Zürich, Institute for Particle Physics and Astrophysics, Zürich, Switzerland}

\author{O.~Eizenberg}
\affiliation{The Helen Diller Quantum Center, Department of Physics, Technion-Israel Institute of Technology, Haifa, Israel}

\author{M.~Gorshteyn}
\affiliation{Institut f\"ur Kernphysik, Johannes Gutenberg-Universit\"at Mainz, Mainz, Germany}
\affiliation{PRISMA$^+$ Cluster of Excellence, Johannes Gutenberg-Universit\"at Mainz, Mainz, Germany}

\author{M.~Heines}
\email[]{Corresponding author: michael.heines@kuleuven.be}
\affiliation{KU Leuven, Instituut voor Kern- en Stralingsfysica, Leuven, Belgium}

\author{A.~Herzáň}
\affiliation{Institute of Physics, Slovak Academy of Sciences, Bratislava, Slovakia}

\author{P.~Indelicato}
\affiliation{Laboratoire Kastler Brossel, Sorbonne Université, CNRS, ENS-PSL Research University, Collège de France, Paris, France}

\author{K.~Kirch}
\affiliation{PSI Center for Neutron and Muon Sciences, Villigen, Switzerland}
\affiliation{ETH Zürich, Institute for Particle Physics and Astrophysics, Zürich, Switzerland}

\author{A.~Knecht}
\affiliation{PSI Center for Neutron and Muon Sciences, Villigen, Switzerland}

\author{R.~Lica}
\affiliation{Horia Hulubei National Institute for R\&D in Physics and Nuclear Engineering, Bucharest, Romania}

\author{V.~Matousek}
\affiliation{Institute of Physics, Slovak Academy of Sciences, Bratislava, Slovakia}

\author{E.A.~Maugeri}
\affiliation{PSI Center for Nuclear Engineering and Sciences, Villigen, Switzerland}

\author{B.~Ohayon}
\affiliation{The Helen Diller Quantum Center, Department of Physics, Technion-Israel Institute of Technology, Haifa, Israel}

\author{N.S.~Oreshkina}
\affiliation{Max-Planck-Institut für Kernphysik, Heidelberg, Germany}

\author{W.W.M.M.~Phyo}
\affiliation{KU Leuven, Instituut voor Kern- en Stralingsfysica, Leuven, Belgium}

\author{R.~Pohl}
\affiliation{PRISMA$^+$ Cluster of Excellence, Johannes Gutenberg-Universit\"at Mainz, Mainz, Germany}
\affiliation{Institut f\"ur Physik, QUANTUM, Johannes Gutenberg-Universit\"at Mainz, Mainz, Germany}

\author{S.~Rathi}
\affiliation{The Helen Diller Quantum Center, Department of Physics, Technion-Israel Institute of Technology, Haifa, Israel}

\author{W.~Ryssens}
\affiliation{Université Libre de Bruxelles, Institut d’Astronomie et d’Astrophysique, Brussels, Belgium}
\affiliation{Brussels Laboratory of the Universe-BLU-ULB, Brussels, Belgium}

\author{A.~Turturica}
\affiliation{Horia Hulubei National Institute for R\&D in Physics and Nuclear Engineering, Bucharest, Romania}

\author{K.~von Schoeler}
\affiliation{ETH Zürich, Institute for Particle Physics and Astrophysics, Zürich, Switzerland}

\author{I.A.~Valuev}
\affiliation{Max-Planck-Institut für Kernphysik, Heidelberg, Germany}

\author{S.M.~Vogiatzi}
\affiliation{KU Leuven, Instituut voor Kern- en Stralingsfysica, Leuven, Belgium}

\author{F.~Wauters}
\affiliation{Institut f\"ur Kernphysik, Johannes Gutenberg-Universit\"at Mainz, Mainz, Germany}
\affiliation{PRISMA$^+$ Cluster of Excellence, Johannes Gutenberg-Universit\"at Mainz, Mainz, Germany}

\author{A.~Zendour}
\affiliation{PSI Center for Neutron and Muon Sciences, Villigen, Switzerland}
\affiliation{ETH Zürich, Institute for Particle Physics and Astrophysics, Zürich, Switzerland}

\date{\today}

\begin{abstract}
    Recent advances in muonic x-ray experiments have reinvigorated efforts in measurements of absolute nuclear charge radii. Here, a modern approach is presented, and demonstrated through determination of the charge radii of the two stable chlorine nuclides \textsuperscript{35}Cl and \textsuperscript{37}Cl. Knowledge of these radii has implications for fundamental studies in nuclear and atomic physics. For this purpose, a state-of-the-art experiment was performed at the $\pi$E1 beamline in the Paul Scherrer Institute (Switzerland), using a large-scale HPGe detector array in order to extract precise energies of the muonic \textsuperscript{35}Cl and \textsuperscript{37}Cl $np1s$ transitions. The nuclear charge radius extraction relies on modern calculations for QED effects and nuclear polarization with rigorous uncertainty quantification, including effects that were not accounted for in older studies. Additionally, we established a new method for applying the nuclear shape correction directly from energy density functionals, which are amenable to isotopes for which no high-quality electron scattering experiments are available. The resulting charge radii are $3.3334(23)~\si{\femto\meter}$ for \textsuperscript{35}Cl and $3.3444(23)~\si{\femto\meter}$ for \textsuperscript{37}Cl, thus improving the uncertainty of the available electron scattering values by a factor of seven. The correlation of several observables was evaluated between the different isotopes in order to produce a more precise value of the differential mean square charge radius $\delta \langle r^2 \rangle^{37, 35}=+0.0771(66)~\si{\femto\meter\squared}$. In this case, improvement of the uncertainty by more than one order of magnitude was achieved compared to the literature value. This precision is sufficient to use this differential as input for isotope shift factor determination.
\end{abstract}


\maketitle

\section{Introduction}
    The size of the atomic nucleus is one of its most fundamental properties. This quantity is most often expressed as the root-mean-square (RMS) nuclear charge radius, commonly referred to as simply \textit{the charge radius}. In itself, the charge radius is a sensitive probe of nuclear structure, revealing effects such as shell closures~\cite{koszorus2021charge, gorges2019laser} and nuclear deformations~\cite{verstraelen2019search}. In this scope, nuclear charge radii have been broadly studied throughout the nuclear landscape (see Ref.~\cite{angeli2013table}). While many studies probe changes in radii through isotope shifts using laser spectroscopy~\cite{yang2023laser}, they all require benchmark nuclear charge radii. Traditionally, such benchmarks were obtained for stable isotopes using muonic x-ray spectroscopy~\cite{fricke2004nuclear} and elastic electron scattering~\cite{de1987nuclear}. These techniques faded out around the turn of the millennium, as most stable isotopes had been probed and uncertainties were deemed sufficiently low for the majority of applications. 
    
    Since then, several physics cases have been evaluated that have the absolute charge radius as a leading systematic error. Among these are several fundamental nuclear and atomic physics experiments. Recent work showed that the absolute charge radius is critical for the determination of the V\textsubscript{ud} element of the Cabibbo-Kobayashi-Maskawa (CKM) matrix~\cite{plattner2023nuclear, seng2024data, ohayon2024reconciling}. Additionally, information about the absolute charge radius is used as input in studies on the equation of state for nuclear matter~\cite{ding2023investigation, pineda2021charge}, the mirror shift fit~\cite{ohayon2024critical}, the interpretation of atomic parity violation experiments~\cite{wansbeek2012charge}, and high-precision measurements in electronic atoms~\cite{sailer2022measurement}. Moreover, absolute charge radii are crucial ingredients in the parameter adjustment of several nuclear structure models: ab initio~\cite{hu2022ab, ekstrom2015accurate, arthuis2024neutron}, energy density functional-based~\cite{chabanat1998skyrme, grams2025skyrme}, and various types of phenomenological models~\cite{royer2009liquid, duflo1994phenomenological}.
    
    Finally, differential radius benchmarks are used in the determination of mass and field shift factors through King plots~\cite{king2013isotope}, which are critical for the radius extraction in laser spectroscopy studies~\cite{sahoo2024recent}. A lack of benchmark radii often leads to large systematic error bands which limit the precision of extracted differential mean square charge radii of isotopes far from stability.

    Currently, the charge radius compilation of Angeli and Marinova from 2013~\cite{angeli2013table} is the most commonly used for radius references. However, the accuracy of these tables has been a topic of debate in recent years~\cite{wiederhold2010equilibrium, schelfhout2022multiconfiguration, schelfhout2021isotope, sailer2022measurement}. Regarding muonic x-ray spectroscopy, the conventional compilation is that of Fricke and Heilig from 2004~\cite{fricke2004nuclear} (with the last measurements dated to 1993~\cite{fricke200410, fricke200450, fricke200466}), after which literature is sparse~\cite{saito2025muonic, antognini2020measurement, vogiatzi2023studies, vogiatzi2023studies, sun2025208}. From these, the only new measurement that has published radii is the work on Pd isotopes~\cite{saito2025muonic}. While these are the most recent muonic x-ray measurements, it should be noted that since 2010 an effort has been made to investigate light muonic atoms via laser spectroscopy, determining high-precision radii for H, D, and He~\cite{pohl2010size, pohl2016laser, krauth2021measuring, schuhmann2025helion}. New interest has also been sparked in the low-mass region, where high-resolution metallic magnetic microcalorimeters can be used to improve literature uncertainties substantially~\cite{ohayon2024towards}. Similar technology is being applied to probe the dynamics of the muonic cascade~\cite{yan2022absolute, okumura2021deexcitation}. Additionally, several ongoing projects utilize muonic atoms for material studies~\cite{gerchow2023germanium, tampo2024developments, aramini2020using} and the determination of nuclear matrix elements for neutrino-less double beta decay~\cite{araujo2024monument}. Similarly, the number of elastic electron scattering measurements reduced significantly around the turn of the millennium, primarily shifting its focus to the proton charge distribution. Apart from the identification of new physics cases that require high-precision absolute charge radii, renewed interest was sparked by experimental developments allowing for muonic x-ray spectroscopy and elastic electron scattering on long-lived radioactive isotopes~\cite{adamczak2023muonic, ohnishi2023scrit}.

    With growing interest in absolute radii with increased reliability and improved uncertainty treatment, it is timely to revisit the methods historically used for nuclear charge radius extraction~\cite{fricke2004nuclear, engfer1974charge}. 
    In this work, we provide a modern approach to muonic x-ray spectroscopy improving and addressing both experimental and theoretical aspects. 
    
    On the experimental side:

    \begin{itemize}
        \item Digital data acquisition systems for offline optimization
        \item Advanced data analysis techniques
        \item Extraction of multiple transition energies
        \item Improved muon beam infrastructure
    \end{itemize}

    On the theory side:

    \begin{itemize}
        \item Novel method for accounting for the nuclear shape
        \item More complete QED and NP than considered in earlier works (e.g., Ref.~\cite{fricke2004nuclear})
        \item Rigorous evaluation of associated uncertainties
        \item Treatment of correlation between theory uncertainties
    \end{itemize}
    
    The stable chlorine isotopes (\textsuperscript{35, 37}Cl) offer a compelling test case for this approach. Their radii serve as critical benchmarks for future laser spectroscopy experiments of exotic isotopes - particularly important given the complexity of atomic calculations for halogens. Charge radii of the chlorine isotopic chain would support nuclear structure investigations around the $N=20$ and $N=28$ neutron shell closures near the $Z=20$ proton shell closure. Another key motivation lies in the radioactive isotope \textsuperscript{34}Cl, which is of interest for precision studies of superallowed $0^+ \rightarrow 0^+$ Fermi $\beta$ decays used in the determination of the V\textsubscript{ud} matrix element of the CKM matrix~\cite{seng2024data}. At the present, calculations of the related $\mathcal{F}t$ values rely on an interpolated charge radius assuming negligible isospin symmetry breaking~\cite{seng2024data}. Such assumption is challenged by recent results in the $A=26$ isotriplet (see Ref.~\cite{ohayon2024critical}). In addition, \textsuperscript{35}Cl and \textsuperscript{37}Cl belong to the group of mirror nuclei that are highly sensitive to the slope of the symmetry energy, a key quantity in the equation of state of nuclear matter~\cite{ding2023investigation}. Despite this broad relevance, the charge radii of these isotopes were previously known only to a precision of approximately 0.5\%, nearly five times worse than the best-known reference isotope of any element in their region~\cite{ohayon2024critical}. This limited precision stems from the absence of muonic x-ray measurements, with earlier RMS values derived exclusively from electron scattering~\cite{briscoe1980elastic}. Furthermore, the accuracy of radii extracted from electron scattering has been a subject of debate, as the mostly unaccounted-for theoretical corrections such as two-photon-exchange lead to deviations in the order of 1\%~\cite{fricke2004nuclear}.

\section{Overview}
    Apart from its finite lifetime, the primary difference between muons and electrons is that the former has a mass ($m_\mu$) about 207 times larger than that of the latter ($m_e$). Consequently, the energy levels of the muon are scaled by $m_\mu/m_e$ compared to those of the electron, which leads to x-ray transition energies up to $10~\si{\mega\eV}$ for heavy systems. Additionally, the atomic orbitals are $m_\mu/m_e$ times smaller than for electronic atoms, which enhances the sensitivity to nuclear effects. Given that nuclear finite size effects are defined through the overlap of the nuclear and orbiting particle's wavefunctions, muonic atoms are a factor $(m_\mu/m_e)^3 \approx 10^7$ times more sensitive to such properties compared to electronic atoms. The atomic states with the highest sensitivity to finite size effects are generally the $s$ states, as their wavefunctions display the largest overlap with the nuclear wavefunction. In muonic x-ray spectroscopy, the $1s$ orbital is probed, typically through the $np1s$ transitions from the spontaneous x-ray cascade after a muon has been captured by an atom. The transition energies can then be compared to theoretical calculations in order to extract nuclear charge radii.

    \begin{figure}
        \centering
        \includegraphics[width=1.0\linewidth]{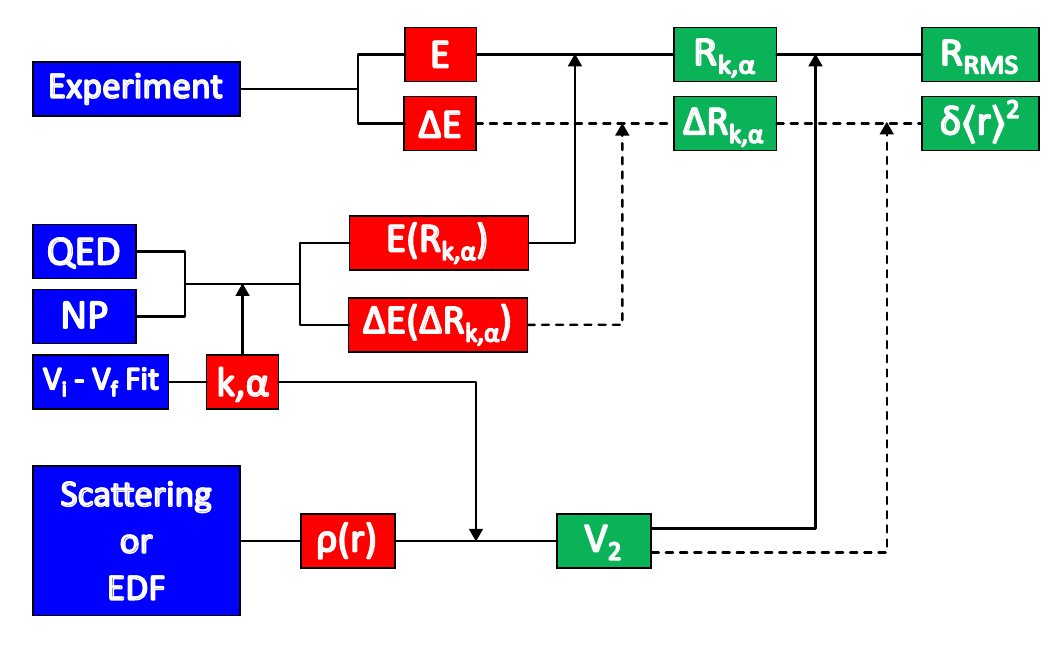}
        \caption{Flowchart of the radius extraction scheme. Blue boxes represent initial inputs, red boxes mark intermediate results, and green boxes mark the output. QED and nuclear polarization (NP) calculations are combined with Barrett parameters ($k$, $\alpha$) in order to provide energy dependencies of model independent Barrett radii ($R_{k, \alpha}$). After extracting the relevant Barrett radii, knowledge of the charge distribution from scattering or energy density functionals (EDF) are used to determine RMS radii.}
        \label{fig:theory_flowchart}
    \end{figure}

    For the radius extraction, several experimental and theoretical inputs are combined. A general overview of the radius extraction is shown in Fig.~\ref{fig:theory_flowchart}. From the experimental side, high-precision transition energies are obtained through a thorough gain correction, a careful line shape evaluation, and a high-precision local energy calibration of the spectra obtained by a large array of high-purity germanium (HPGe) detectors. The main differences compared to older literature lies in the use of offline optimization capabilities from digital data acquisition systems and more sophisticated analysis methods. Moreover, multiple $np1s$ transition energies were extracted, which allows to effectively average the effect of calibration uncertainty on the extracted radius. From the theoretical side, several inputs from nuclear and atomic theory must be combined. First, QED calculations are performed using different nuclear charge distributions. These calculations involve solving the Dirac equation including the following effects: 1) vacuum polarization (VP) of electronic, muonic, and hadronic origin, 2) a self-energy (SE) correction accounting for the muon exchanging virtual photons with itself, and 3) non-relativistic and relativistic nuclear recoil effects beyond the effective mass. On top of this, muonic atoms are subject to effects of the internal dynamic nuclear structure, commonly referred to as the nuclear polarization (NP). While NP gives the largest contribution to the final radius uncertainty, it's radius dependence is negligible within the current precision. 
    
    For the QED calculations, assumptions must be made on the nuclear charge distribution, typically considered to be a simple Fermi model. The assumptions of such a model introduce a charge distribution model dependence in extracted nuclear charge radii. Such issues can sometimes be solved by following the Barrett radius recipe~\cite{barrett1970model}, which introduces a moment that is directly related to the muonic transition energies (Barrett moment $\langle r^k e^{-\alpha r}\rangle$), and a corresponding equivalent radius (Barrett radius $R_{k \alpha}$, see Eq.~\eqref{eq:Bradius}). In order to calculate such a Barrett moment, the Barrett parameters $k$ and $\alpha$ are determined through a fit on the potential difference generated by the muon in the initial and final state. Incorporating this with the QED and NP calculations provides descriptions of the transition energy or muonic isotope shift (difference in transition energy between two isotopes) as a function of Barrett radius or Barrett radius difference. Subsequently, the transition energies can be combined with these dependencies to extract the Barrett radius of interest (or difference in Barrett radius). Finally, the Barrett radii are converted back into RMS quantities using a nuclear shape correction ($V_2$). This is rooted in knowledge of the shape of the charge distribution, which is conventionally taken from electron scattering experiments. In this work, we show that modern energy density functional (EDF) calculations, in particular Brussels-Skyrme-on-a-grid (BSkG) models~\cite{scamps2021skyrme, ryssens2022skyrme, grams2023skyrme, grams2025skyrme}, provide an excellent alternative for isotopes that do not have high-quality scattering measurements available in this region of the nuclear chart. For heavier systems, higher-lying transitions are sometimes used in order to extract more details of the charge distribution (e.g., also fitting for skin thickness). However, even such an approach can leave residual charge distribution model dependency as a specific distribution is still assumed~\cite{xie2023revisiting}.

\section{Experiment}
\subsection{Experimental setup}
    The muonic x-ray measurements presented in this work were performed at the $\pi$E1 beamline of the high-intensity proton accelerator (HIPA) facility~\cite{grillenberger2021high} at the Paul Scherrer Institute (PSI) in Switzerland. For these measurements, enriched targets containing \textsuperscript{35}Cl and \textsuperscript{37}Cl were provided by the Institute Laue-Langevin (ILL) and Argonne National Laboratory (ANL), respectively. The specifications of these targets are given in Table~\ref{tab:targets}. The momentum of the muon beam was optimized at  $32~\si{\mega\eV\per c}$ and $30~\si{\mega\eV\per c}$ for the \textsuperscript{35}Cl and \textsuperscript{37}Cl measurements, respectively. This was done by varying the momentum in steps of $2~\si{\mega\eV\per c}$ in order to find the highest $2p1s$ x-ray rate. Including the time needed for optimization, these targets were measured for $\sim5.5~\si{\hour}$ and $\sim 8.5~\si{\hour}$. The measurement on \textsuperscript{37}Cl was significantly more challenging due to 1) The small physical target size, leading to a smaller overlap with the muon beam, 2) The smaller target mass, leading to a lower x-ray rate, 3) Compton background from silver muonic x rays, and 4) The presence of Ag in the sample, which has a broader extended Coulomb potential, such that a smaller fraction of the muons are atomically captured by the chlorine nuclei. 

    \begin{table}
        \centering
        \caption{Information on the targets used for the muonic x-ray measurements.}
        \begin{ruledtabular}
        \begin{tabular}{c|c|ccc}
            Isotope & Chemical & Purity & Approx. Tot. & Approx. Cl \\
            & form & (\%) & mass ($\si{\milli\gram}$) & mass ($\si{\milli\gram}$)\\
            \hline
            \textsuperscript{35}Cl \rule{0pt}{2.5ex} & NaCl & 99.32(5) & 200 & 121 \\
            \textsuperscript{37}Cl \rule{0pt}{2.5ex} & AgCl & 99.28(5) & 70 & 18\\
        \end{tabular}
        \end{ruledtabular}
        \label{tab:targets}
    \end{table}

\begin{figure}
        \centering
        \includegraphics[width=0.7\linewidth]{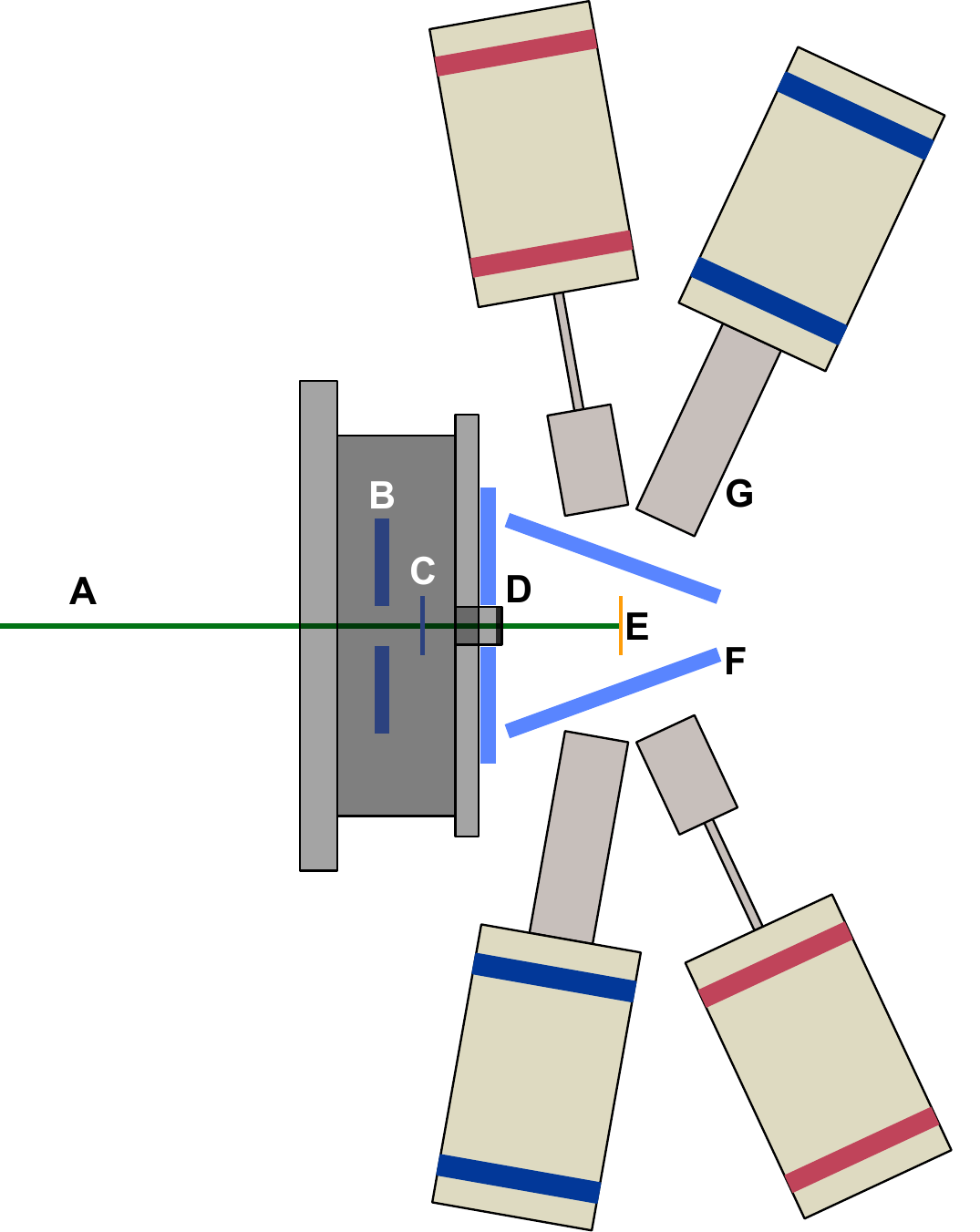}
        \caption{Simplified experimental setup used for the muonic x-ray measurements. The germanium detectors that are shown represent a total of 14 detectors (19 crystals). The components are as follows: A) Incoming muon beam, B) Muon veto detector, C) Muon entrance detector, D) Thin titanium window, E) Target mount, F) Electron veto detectors, and G) HPGe detectors.}
        \label{fig:setup}
    \end{figure}

    The experimental setup was based on the developments made in Refs.~\cite{gerchow2023germanium, adamczak2023muonic}. A graphic representation is shown in Fig.~\ref{fig:setup}. The data was collected using a 14-bit SIS3316 digitizer with a sampling rate of $250~\si{\mega\hertz}$ and digital signal processing through trapezoidal filters for the energy determination. For the detection of x rays, a variety of HPGe detectors were used and arranged in a detector array. The used detectors consisted of a Miniball cluster detector~\cite{warr2013miniball}; a TIGRESS-type clover detector~\cite{scraggs2005tigress}; reverse electrode coaxial germanium (REGe) detectors with relative efficiencies of 95\% (x1), and 70\% (x2); standard electrode coaxial germanium (SEGe) detectors with relative efficiencies of 50\% (x2), 58\% (x1), 75\% (x1), and 100\% (x2); and broad-energy germanium (BEGe) detectors (x3). In order to improve the timing resolution, the first $1.2~\si{\micro\second}$ of the raw traces was digitized for each germanium detector waveform for offline timing optimization. The HPGe-detector array's characteristics were as follows based on the detector-summed spectrum: 
    
    \begin{itemize}
        \item $\sim3\%$ total detection efficiency at $1332~\si{\kilo\eV}$
        \item Energy resolution (FWHM) $\sim2.6~\si{\kilo\eV}$ at $1332~\si{\kilo\eV}$
        \item  Time resolution (FWHM) $\sim14~\si{\nano\second}$ at approximately $700~\si{\kilo\eV}$ for scintillator-germanium timing.
    \end{itemize}

    This last value was determined by evaluating the time difference spectra for muonic x rays with respect to the entrance detector. As these occur effectively prompt after atomic muon capture, they provide an excellent probe for the time resolution.

    Apart from the HPGe detectors, a set of plastic scintillators was used for coincidence and veto logic. The first of these scintillators was a muon veto detector, which served to collimate the beam and veto muons that arrive with a lateral offset to the target. Downstream of this veto detector, a muon entrance detector was placed. This detector allows the photons detected in the HPGe detectors to be time correlated to incoming muons. Finally, a set of electron veto scintillator detectors was placed around the target position to veto photons originating from Bremsstrahlung induced by Michel electrons (electrons emitted during muon decay). The target material was first put in dedicated polystyrene target holders, which were in turn vacuum sealed to avoid spillage. This mounting structure was then placed in the center of the detector array. Given the x-ray energies of interest, self-absorption is not an issue. Furthermore, by choosing low-Z materials for the target holder and mounting structure, the absorption in these materials is minimal.

    During the measurements, several calibration sources ($\sim10~\si{\kilo\becquerel}$ \textsuperscript{133}Ba, $\sim1~\si{\kilo\becquerel}$ \textsuperscript{110m}Ag, and $\sim10~\si{\kilo\becquerel}$ \textsuperscript{60}Co) were placed near the target position. By doing so, a continuous calibration was available throughout the measurement.

\subsection{Data processing}
    In muonic x-ray experiments, three main time behaviors are present with respect to the muon arrival time: 
    
    \begin{itemize}
        \item Continuous in time from calibration sources and natural background
        \item Prompt in time from muonic x rays
        \item Michel electrons and muon capture-related events with decay times $\tau \in [70~\si{\nano\second}; 2.2~\si{\micro\second}]$ for muons captured in various materials.        
    \end{itemize}
    
    In order to make muonic x-ray spectra and calibration spectra without significant contamination peaks, a good time resolution is needed for the HPGe detectors. This time resolution was achieved via extrapolated leading edge timing followed by a time matching with respect to the muon entrance detector.

\begin{figure}
        \centering
        \includegraphics[width=1.0\linewidth]{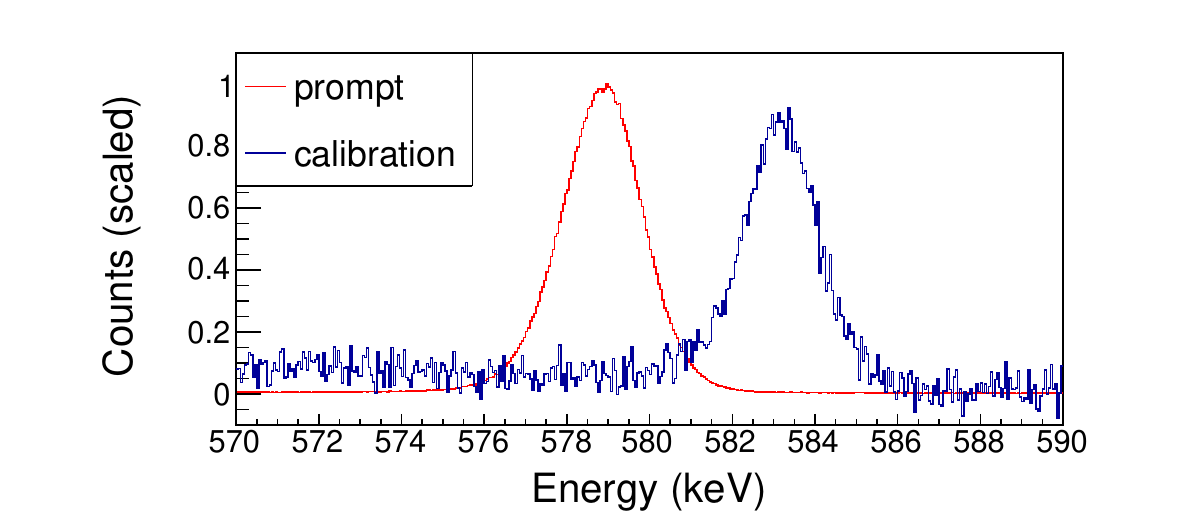}
        \caption{Overlay between the calibration spectrum and the muonic x-ray spectrum from the \textsuperscript{35}Cl measurement. For visualization purposes, the spectra were scaled and their offset aligned to the right of the peaks. The prompt spectrum shows the peak corresponding to the $2p1s$ transition in \textsuperscript{35}Cl, while the anticoincidence spectrum shows the $583~\si{\kilo\eV}$ peak from the decay of \textsuperscript{208}Tl (natural background in \textsuperscript{232}Th decay chain).}
        \label{fig:overlay_35Cl}
    \end{figure}

    For the fitting of the muonic x-ray peaks, a \textit{prompt time cut} was made by selecting events within the window $[-25; +25]~\si{\nano\second}$ with respect to a hit in the muon entrance detector alongside scintillator logic cuts. Similarly, an \textit{anticoincidence time cut} was taken for the calibration spectra by only selecting events that are not within the window $[-1, +3]~\si{\micro\second}$ with respect to any muon (muon entrance and muon veto detector). A comparison between the measured x-ray and calibration spectra from the \textsuperscript{35}Cl measurement are shown in Fig.~\ref{fig:timecuts}. A similar figure is given in the supplementary material for the \textsuperscript{37}Cl measurement. To visualize the absence of calibration peaks near the x-ray peaks and vice versa, an overlay comparison is given for the $2p1s$ region in Fig.~\ref{fig:overlay_35Cl}. Based on the respective time windows, the integral of the $583~\si{\kilo\eV}$ calibration line in the prompt spectrum was estimated to be below $5\times 10^{-5}$ for \textsuperscript{35}Cl and $5\times 10^{-4}$ for \textsuperscript{37}Cl compared to the muonic $2p1s$ line.    

    \begin{figure}
        \centering
        \includegraphics[width=1.0\linewidth]{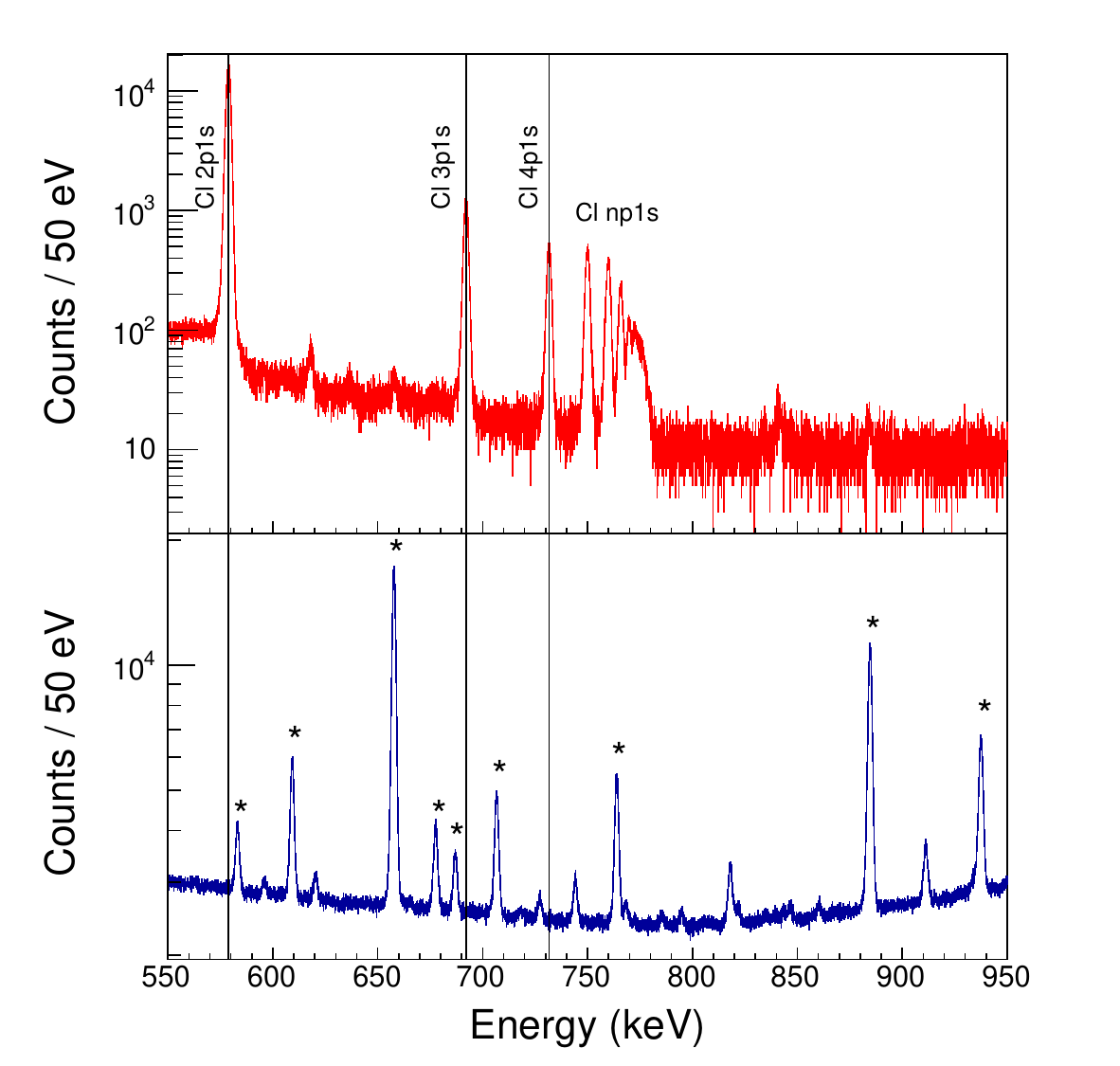}
        \caption{Prompt (top) and anticoincidence (bottom) spectra during the muonic x-ray measurements of \textsuperscript{35}Cl. The vertical lines mark the relevant peaks - the $np1s$ transition energies extracted in this work. The asterisk symbol labels the gamma-ray peaks used for the final energy calibration.}
        \label{fig:timecuts}
    \end{figure}

\subsection{Energy determination}
\begin{figure}
        \centering
        \includegraphics[width=1.0\linewidth]{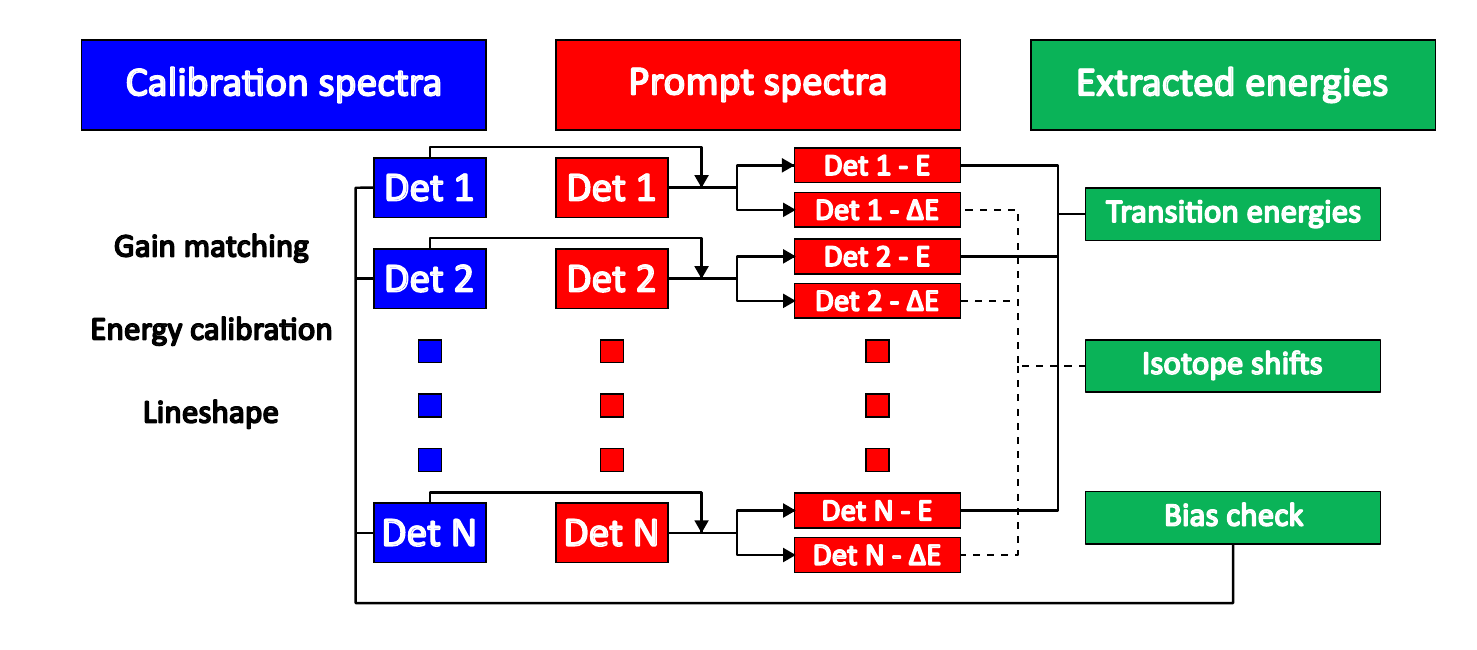}
        \caption{Graphical representation of the energy extraction process applied in this work.}
        \label{fig:exp_flowchart}
    \end{figure}

    With the calibration and muonic x-ray spectra separated, both can be fitted without contaminant lines in the spectrum. An overview of the process for the energy extraction is displayed in Fig.~\ref{fig:exp_flowchart}. 

    In order to maximize the accuracy of the extracted transition energies, the energies are determined in each detector separately before averaging the resulting values. This is important as different detectors have varying detection efficiencies, calibration uncertainty, and peak widths. Accordingly, the statistical and calibration error can be properly combined for each detector before averaging. First, the calibration spectra are used to perform a gain matching. Next, the corrected spectra can be used to constrain the line shape and perform a high-precision calibration. The former is used during the fitting of the x-ray lines, while the latter is combined with the extracted centroids to produce energies. In this process, the statistical uncertainty of the fit of the x-ray line and the systematic uncertainty from the calibration are added in quadrature. These values are then averaged, using the inverse variances as weights. Finally, the quality of the extraction process is tested by leaving calibration lines out of the calibration fit, determining their energy using the same fitting procedures as were applied for the x-ray lines, and evaluating the deviation between the extracted value and the literature value. \\

    For this gain correction, a linear recalibration was performed every 90 minutes worth of data using emission lines from \textsuperscript{60}Co, \textsuperscript{110m}Ag, and \textsuperscript{133}Ba. Here, a simple Gaussian + linear function was employed that was fitted once to extract the mean $\mu$ and spread $\sigma$ of the peak, and subsequently refitted starting from $E=\mu - \sigma$ to suppress the influence of the low energy tail of the peaks. Next, a high-precision calibration was performed in the range $[500; 1000]~\si{\kilo\eV}$ using emission lines from \textsuperscript{110m}Ag, complemented by natural background lines from the decay of \textsuperscript{208}Tl and \textsuperscript{214}Bi. An overview of the lines used in this calibration is given in Table~\ref{tab:calibration}.

    \begin{table}
        \centering
        \caption{The isotopes and their respective characteristic gamma-ray transition energies used for the high-precision energy calibration. Energies were taken from the table of radionuclides~\cite{be2010table}.}
        \begin{ruledtabular}
        \begin{tabular}{cl}
            Source & Energy $(\si{\kilo\eV})$ \\
            \hline
            \textsuperscript{208}Tl \rule{0pt}{2.5ex} & 583.187(2) \\
            \textsuperscript{214}Bi \rule{0pt}{2.5ex}  & 609.316(7) \\
            \textsuperscript{110m}Ag \rule{0pt}{2.5ex} & 657.7600(11) \\
            \textsuperscript{110m}Ag \rule{0pt}{2.5ex} & 677.6217(12) \\
            \textsuperscript{110m}Ag \rule{0pt}{2.5ex} & 687.0091(18) \\
            \textsuperscript{110m}Ag \rule{0pt}{2.5ex} & 706.6760(15) \\
            \textsuperscript{110m}Ag \rule{0pt}{2.5ex} & 763.9424(17) \\
            \textsuperscript{110m}Ag \rule{0pt}{2.5ex} & 884.6781(13) \\
            \textsuperscript{110m}Ag \rule{0pt}{2.5ex} & 937.485(3)
        \end{tabular}
        \end{ruledtabular}
        \label{tab:calibration}
    \end{table}

    For the high-precision energy calibration, the fitting of the calibration lines was performed with a linear background and a hypermet line shape, given by 

    \begin{equation}\label{eq:hypermet}
        f(E) = N_{sig} \left[ f_G \times g(E) + f_T \times t(E) + s(E) \right],
    \end{equation}

    where

    \begin{eqnarray*}
        f_T &&= 1 - f_G, \\
        g(E) &&= \frac{1}{\sqrt{2\pi}\sigma} \exp{\left(-\frac{1}{2}\left[\frac{E - \mu}{\sigma}\right]^2\right)}, \\
        t(E) &&= \frac{1}{2 \beta} \exp{\left(\frac{E - \mu}{\beta} + \frac{\sigma^2}{2\beta^2}\right)} \text{erfc}\left(\frac{E - \mu}{\sqrt{2} \sigma} + \frac{\sigma}{\sqrt{2}\beta}\right), \\
        s(E) &&= \frac{A}{2} \text{erfc}\left(\frac{E - \mu}{\sqrt{2}\sigma}\right).
    \end{eqnarray*}

\begin{figure}
        \centering
        \includegraphics[width=\linewidth]{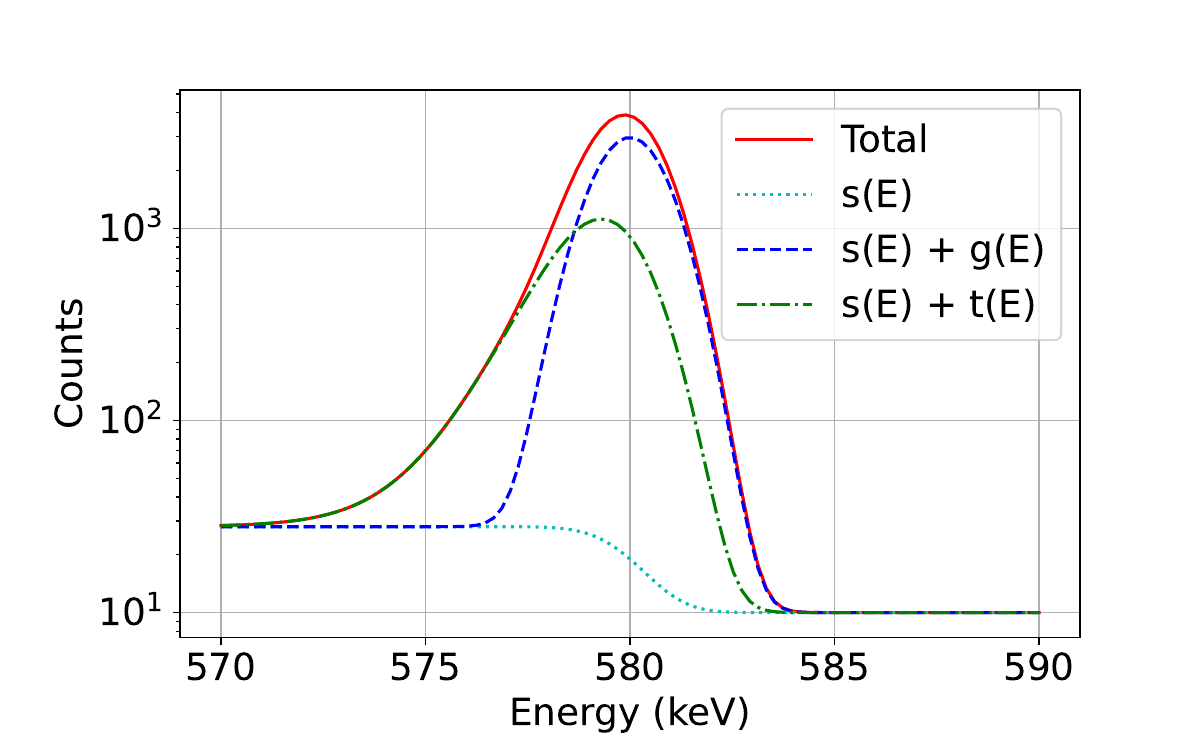}
        \caption{Visualization of the hypermet line shape using realistic parameters on a constant background.}
        \label{fig:hypermet}
    \end{figure}

    The three contributions are the ideal Gaussian detector response $g(E)$, a term accounting for incomplete charge collection $t(E)$ (primarily due to defects in the germanium crystal), and a step behavior $s(E)$ mainly induced by low-energy Compton scattering in the surrounding material~\cite{campbell1997cautionary, knoll2010radiation}. A graphical representation of this line shape it shown in Figure~\ref{fig:hypermet}.

\begin{figure}
        \centering
        \includegraphics[width=1.0\linewidth]{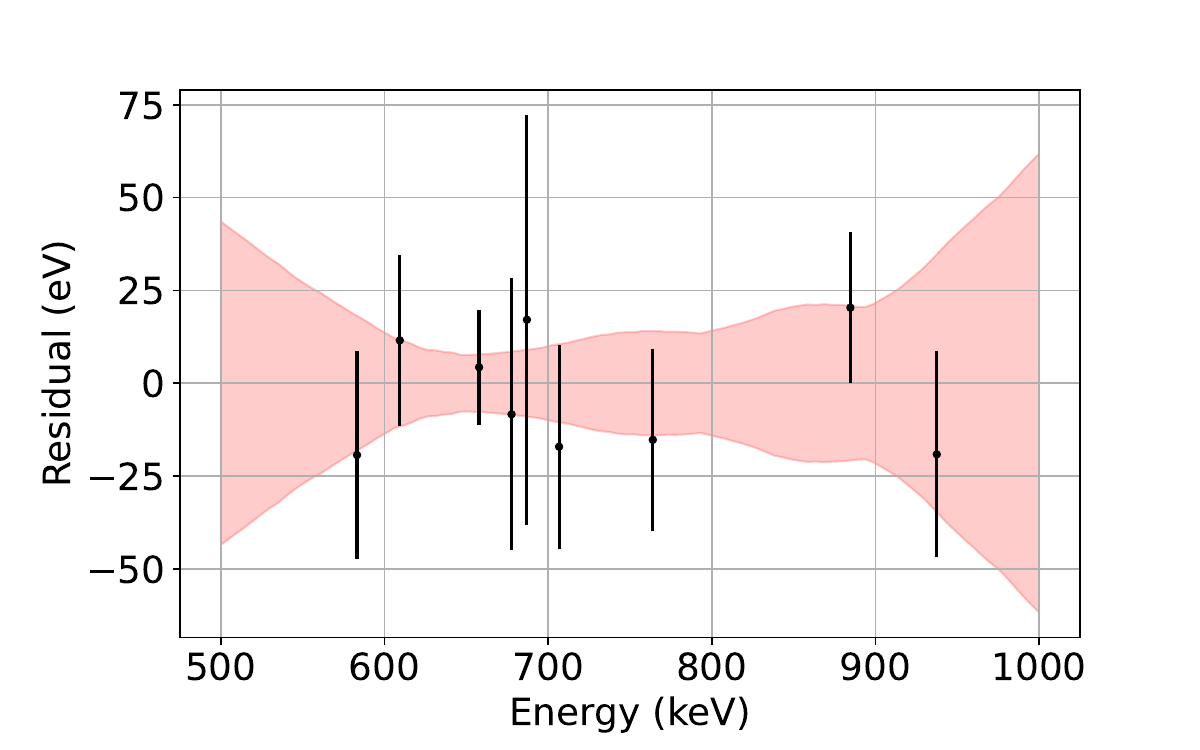}
        \caption{Example of calibration residuals for one of the detectors in the \textsuperscript{37}Cl measurement. The red shaded region corresponds to the $1\sigma$ confidence interval.}
        \label{fig:calib_residual}
    \end{figure}

    Each detector in the array was calibrated separately by performing a coupled likelihood fit on all lines, followed by a quadratic calibration fit. In this calibration fit, the uncertainty was determined by parametrically bootstrapping new calibration lines from the original dataset with repetition. By repeating this process a large number of times, the bootstrap uncertainty of the fit at a given energy is given by the distribution of mean fit results at this corresponding energy. As a fairly limited number of calibration lines are available, it may occur that the statistical uncertainty is not fully captured by the bootstrapping process. As such, the uncertainty of the calibration fit was taken to be the maximum of the bootstrap uncertainty and the statistical uncertainty. While the precision of the calibration depends heavily on the detector type, most detectors showed calibration uncertainties on the order of $10 \text{ - } 20~\si{\eV}$ in the region of interest. An example of the calibration residuals with the corresponding calibration confidence interval is given in Fig.~\ref{fig:calib_residual}. \\

    In this work, the $2p1s$, $3p1s$, and $4p1s$ emission energies were extracted. Higher $np1s$ lines are not quoted due to lower statistics and closer spacing between subsequent transitions. While the main radius sensitivity originates from the $1s$ state, quoting different $np1s$ transitions allows to effectively average experimental uncertainties on the radius. For the fitting, the same peak model was used as for the calibration. However, since the $np1s$ fine-structure splitting is not fully resolved, the signals were fitted as doublets. Furthermore, it was sufficient to replace the linear background by a constant term. The fitting of the $np1s$ lines was performed through Bayesian inference using a Poisson likelihood with the \texttt{PyMC} package in Python~\cite{patil2010pymc}. The use of priors is particularly beneficial in describing the fine structure, as it uses prior knowledge of the fine structure parameters (fine structure splitting and relative amplitude between the peaks) from QED calculations to guide the fit. The data may still overturn this prior if sufficient evidence is available, allowing for potential deviations from the theory predictions. Since the fine structure is not fully resolved, the uncertainty on the $np\textsubscript{3/2}1s$ energies is limited by the prior knowledge of the fine structure. To overcome this, we opted to quote the center-of-gravity energy of the $np1s$ transitions. These energies are extracted to a higher precision, as they are nearly uncorrelated to the fine structure parameters, while they still carry the nuclear charge radius sensitivity.

\begin{figure}
        \centering
        \subfloat[]{\includegraphics[width = 0.95 \linewidth]{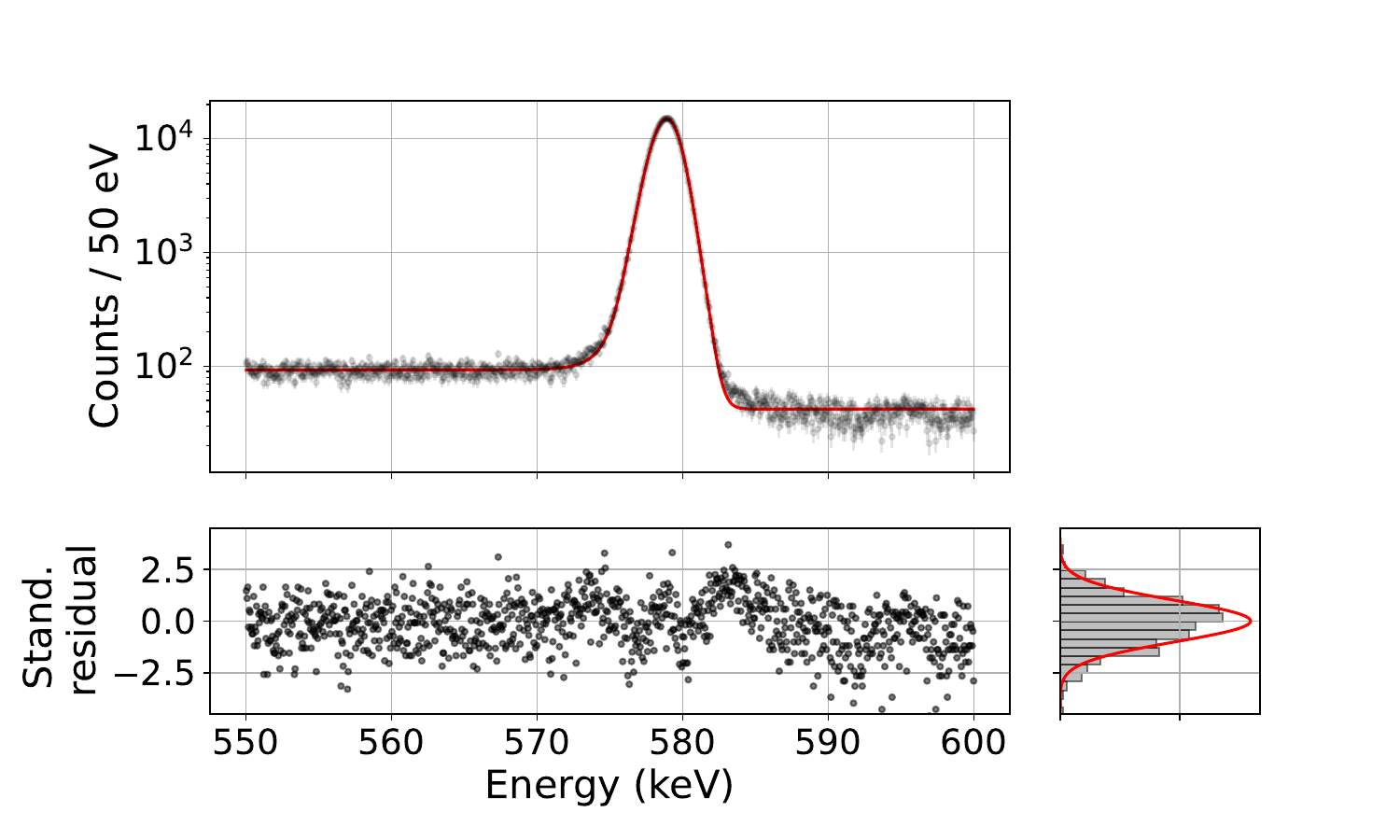}} \\
        \subfloat[]{\includegraphics[width = 0.95 \linewidth]{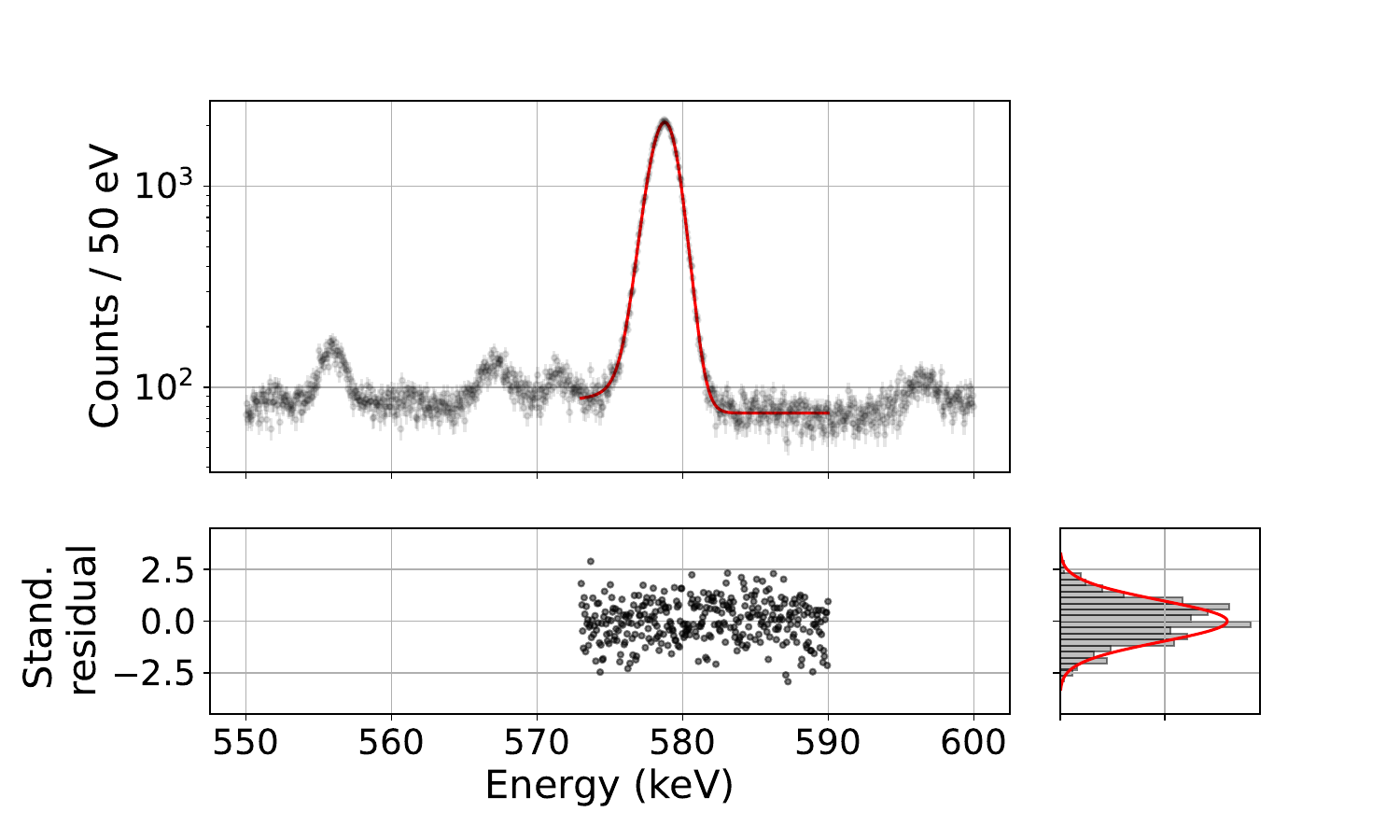}}
        \caption{Experimental spectra and corresponding fits of the $2p1s$ transition in (a) \textsuperscript{35}Cl and (b) \textsuperscript{37}Cl. These plots exclude one coaxial HPGe detector that displayed a high-energy tail. Standardized residuals are calculated using the total Poisson error on all detectors.}
        \label{fig:spectrum_2p1s}
    \end{figure}

    For visualization purposes, the sum of all detectors' data and the sum of the fits in each detector are shown for the $2p1s$ transition in Fig.~\ref{fig:spectrum_2p1s}. The $2p1s$ transition in \textsuperscript{37}Cl was fitted in a slightly narrower range ($[573; 590]~\si{\kilo\eV}$) due to the proximity of the $nf3d$ transitions in silver. The presence of these peaks does not introduce additional systematic errors, as the contaminant lines are sufficiently distant. A more detailed discussion on the analysis methods used for the $np1s$ energy extraction is given in the supplementary material.

\subsection{Uncertainty estimation}\label{sec:exp_unc}
    Due to detectors sharing certain digitizer components, their calibration uncertainties may show some level of correlation. To check the effect on the energy extraction, a bias investigation was performed. For every calibration peak that was used, the following procedure was performed: 1) Perform an energy calibration excluding one calibration line, 2) Extract the energy of the excluded line using the same fitting and averaging methods as was applied for the $np1s$ lines, 3) Compare to the literature energy of the excluded line, 4) Estimate how much more deviation is observed compared to the predicted error bars. This process revealed an additional systematic bias uncertainty $\sigma_{\text{bias}}$ of $\sim8.2~\si{\eV}$. Apart from this, a systematic uncertainty was added quadratically equal to the uncertainty of the best-known calibration line $\sigma_{\text{lit}} = 1.1~\si{\eV}$. These two systematic uncertainties are acting on broader energy regions, such that they can cancel out when determining muonic isotope shifts (difference in transition energies between isotopes).

    Additional systematic checks were performed that can not be accounted for by using calibration lines. The effect of the hyperfine splitting was investigated and deemed to contribute less than $0.1~\si{\eV}$, such that it can be neglected. This included both conventional hyperfine splitting and estimates on higher order hyperfine splitting, following the formalism from Ref.~\cite{pachucki2024second}. Since making a time cut preferentially selects certain rise times, the extracted energy can be shifted. This effect required a correction for each detector by comparing the utilized time cut and a broad time cut on high statistics data. Finally, the effect of the isotopic purity of the samples was determined by assuming the extracted experimental energies are the weighted average of the isotope of interest and the contaminant isotope. This introduced a minor shift of the order of a fraction of an $\si{\eV}$ on the $np1s$ energies. The uncertainty of the isotopic purity provides an additional uncertainty $\sigma_{\text{f}}$, which is in the order of a few $0.1~\si{\eV}$. A more in-depth explanation of these systematic checks is provided in the supplementary material.

    Table~\ref{tab:exp_uncertainties} gives a breakdown of the different sources of experimental uncertainty. Here, $\sigma_{\text{stat+cal}}$, $\sigma_{\text{bias}}$, $\sigma_{\text{lit}}$, $\sigma_{\text{f}}$, and $\sigma_{\text{tot}}$ respectively correspond to the summed statistical and calibration error after averaging, the systematic bias uncertainty induced by averaging over detectors, the uncertainty of the most precisely known calibration line, the uncertainty corresponding to the isotopic purity, and the total uncertainty obtained by adding the errors in quadrature. The resulting absolute $np1s$ energies and muonic isotope shifts are given in Table~\ref{tab:energies}. It should be noted that \textsuperscript{37}Cl was measured longer than \textsuperscript{35}Cl due to the limited target mass, which resulted in a better calibration uncertainty, while the statistical uncertainty on the $np1s$ energies are larger for each detector. Given that the $2p1s$ transition of Cl lies at the edge of the calibration interval, this region benefits more from increased calibration statistics. As such, $\sigma_{\text{stat+cal}}$ shows a non-standard behavior across the different transitions. Additionally, isotope shifts are obtained by averaging the value extracted in each detector separately, which results in a slightly different number than the difference in $np1s$ energy between different isotopes. Due to canceling systematic calibration errors, the uncertainty on the isotope shift is smaller than standard error propagation would imply. This effect is most prominent for the $2p1s$ transition, as it suffers least from statistical limitations.

    \begin{table}
        \centering
        \caption{A list of the experimental uncertainties on the muonic transition energies and isotope shifts.}
        \begin{ruledtabular}
        \begin{tabular}{cc|cccc|c}
            Isotope & Transition & $\sigma_{\text{stat+cal}}$ & $\sigma_{\text{bias}}$ & $\sigma_{\text{lit}}$ & $\sigma_{\text{f}}$ & $\sigma_{\text{tot}}$ \\
            & & $(\si{\eV})$ & $(\si{\eV})$ & $(\si{\eV})$ & $(\si{\eV})$ & $(\si{\eV})$ \\
            \hline
            \textsuperscript{35}Cl \rule{0pt}{2.5ex} & $2p1s$ & 13.0 & 8.2 & 1.1 & 0.11 & 15.5 \\
            & $3p1s$ & 06.6  & 8.2 & 1.1 & 0.13 & 10.6 \\
            & $4p1s$ & 09.4  & 8.2 & 1.1 & 0.20 & 12.5 \\
            \hline
            \textsuperscript{37}Cl \rule{0pt}{2.5ex} & $2p1s$ & 08.7  & 8.2 & 1.1 & 0.2 & 12.0 \\
            & $3p1s$ & 15.4 & 8.2 & 1.1 & 0.2 & 17.5 \\
            & $4p1s$ & 26.4 & 8.2 & 1.1 & 0.3 & 27.6 \\
            \hline
            \textsuperscript{37}Cl - \textsuperscript{35}Cl \rule{0pt}{2.5ex}  & $2p1s$ & 11.6 & / & / & 0.11 & 11.6 \\
            & $3p1s$ & 16.8 & / & / & 0.07 & 16.8 \\
            & $4p1s$ & 28.1 & / & / & 0.09 & 28.2
        \end{tabular}
        \end{ruledtabular}
        \label{tab:exp_uncertainties}
    \end{table}

    \begin{table}
        \centering
        \caption{Absolute $np1s$ energies and muonic isotope shifts of \textsuperscript{35, 37}Cl extracted in this work. The reduced $\chi^2$ is given for the fits ($\chi_{\nu, fit}^2$) and for the weighted average between the 19 detectors that were used ($\chi_{\nu, av}^2$).}
        \begin{ruledtabular}
        \begin{tabular}{cc|lcc}
            Isotope & Transition & Energy ($\si{\kilo \eV}$) & $\chi_{\nu, fit}^2$ & $\chi_{\nu, av}^2$ \\
            \hline
            \textsuperscript{35}Cl \rule{0pt}{2.5ex} & $2p1s$ & 578.869(16) & 1.13 & 2.53 \\
                                   & $3p1s$ & 692.096(11) & 0.68 & 1.36 \\
                                   & $4p1s$ & 731.665(13) & 0.73 & 1.23 \\
            \hline
            \textsuperscript{37}Cl \rule{0pt}{2.5ex} & $2p1s$ & 578.734(12) & 1.13 & 1.70 \\
                                   & $3p1s$ & 691.997(18) & 1.10 & 0.60 \\
                                   & $4p1s$ & 731.548(28) & 0.96 & 0.85 \\
            \hline
            \hline
            \textsuperscript{37}Cl $-$ \textsuperscript{35}Cl \rule{0pt}{2.5ex} & $2p1s$ & $-$0.144(12) & / & 0.38 \\
                                                     & $3p1s$ & $-$0.102(17) & / & 0.58 \\
                                                     & $4p1s$ & $-$0.116(29) & / & 0.78
        \end{tabular}
        \end{ruledtabular}
        \label{tab:energies}
    \end{table}

\section{Theory}
\subsection{QED}\label{sec:QED}
    %Corrections

    The older QED calculations used for radius extraction in muonic atoms is described in Engfer et al.~\cite{engfer1974charge}. Compared to those calculations, several improvements were made: 
    
    \begin{itemize}
        \item Numerically improved Wichmann-Kroll (WK) correction
        \item Relativistic finite size self-energy (SE$_{\text{FNS}}$) instead of non-relativistic
        \item Inclusion of previously neglected hadronic vacuum polarization (HVP)
        \item Improved recoil correction with relativistic and finite size contributions
    \end{itemize}
    
    The theoretical estimation of the $np1s$ transition energies with  $n = 2,3,4 $ was performed using the state-of-the-art multiconfiguration Dirac Fock and general matrix elements (MDFGME) code~\cite{indelicato2024mcdfgme}. This code treats the bound muon as a relativistic Dirac particle interacting with the nucleus, governed by the Dirac equation:

    \begin{equation}\label{eq:theo_Dirac}
        \left[ \boldsymbol{\alpha} \cdot \mathbf{p} + \beta m_{\mu} + V(r) \right] | \psi_{n\kappa m} \rangle = E_{n\kappa} | \psi_{n\kappa m} \rangle,
    \end{equation}

    \noindent where $| \psi_{n\kappa m} \rangle$ and $E_{n\kappa}$ are the muon Dirac wavefunction and energy, respectively, for a level with quantum numbers $n$ (principal), $\kappa$ (relativistic angular momentum), and $m$ (magnetic), $\boldsymbol{\alpha}$ and $\beta$ are the Dirac matrices, $\mathbf{p}$ is the muon momentum operator, and $m_{\mu}$ is the mass of the muon. The total potential $V(r)$ used in the Dirac equation includes several contributions. The first contribution comes from the Coulomb interaction (V$_\text{Coulomb}$) of the muon with the extended nuclear charge distribution. For this, we used the most commonly used potential in the literature for spherical nuclei~\cite{fricke2004nuclear}, which is to use a 2-parameter Fermi distribution

    \begin{equation}\label{eq:theo_Fermi2}
        \rho(r) = \frac{\rho_0}{1 + \exp\left[4\ln{3} \left(\frac{r-c}{t}\right)\right]}.
    \end{equation}

    \noindent Here, $\rho_0$ is the normalization factor, $c$ is the nuclear radius at half density, and $t$ is the skin thickness (distance over which $\rho$ drops from 90\% to 10\% of its maximal value).
    
    In order to estimate the effect of the nuclear size on the transition energies, we performed calculations for a range of $c$ values corresponding to a change in RMS radii of 2\% centered around the literature value from Ref.~\cite{briscoe1980elastic}. Given the low sensitivity to $t$ in medium-mass systems, the skin thickness was kept fixed at $t =2.3~\si{\femto\meter}$ in these calculations. The model dependency coming from the choice of this simple distribution is dealt with separately by the Barrett-moment recipe described in Section~\ref{sec:theo_Barrett}.

    The next dominant contribution to the potential originates from vacuum polarization (VP), with the leading term given by the Uehling potential of order $\alpha(Z\alpha)$, as described in Ref.~\cite{uehling1935polarization, klarsfeld1977analytical, indelicato2013nonperturbative}. We include both electronic Uehling (eUe) and muonic Uehling ($\mu$Ue) potentials in the Dirac equation, which also capture higher-order loop-after-loop effects. Hence, our total potential is then given by
    
    \begin{equation}\label{eq:theory_potential}
        V(r) = V_\text{Coulomb} + \text{eUe} + \mu \text{Ue}.
    \end{equation}

    \noindent Substituting Eq.~\eqref{eq:theory_potential} in Eq.~\eqref{eq:theo_Dirac}, the Dirac equation was solved for the range of input charge distributions.

    The higher order corrections, e.g., the Wichmann-Kroll (WK) correction~\cite{wichmann1956vacuum, huang1976calculation}, along with other higher-order terms were evaluated perturbatively using the previously obtained muonic wavefunction. Similarly, the K\`{a}ll\"{e}n and Sabry (KS) potential~\cite{fullerton1976accurate} for electron and muon vacuum pairs, the HVP~\cite{breidenbach2022hadronic} and virtual-Delbrück scattering~\cite{korzinin2018light} were accounted for. The correction from the one-loop muon self-energy (SE) along with the finite-size correction were evaluated following the methods described in Refs.~\cite{indelicato2013nonperturbative, karshenboim2018alpha}.

    For the nuclear recoil correction two methods have been used. The first is the rigorous QED treatment~\cite{2023-uRec} in the first order in the mass ratio $m_\mu / m_N~\approx~0.3\%$ (with $m_N$ the nuclear mass). The second relies on the inclusion of the  reduced mass in the Dirac equation. Here, additional corrections beyond the reduced mass ($\Delta$Rec) are evaluated based on the methodology outlined in Refs.~\cite{indelicato2013nonperturbative, sapirstein1990qed, mohr2025codata}. Both methods are in the good agreement within the required accuracy of less than $0.5~\si{\eV}$.
    
    Finally, we evaluate the electron screening effect by considering fully populated $1s$ and $2s$ electron orbitals. The total screening on the individual levels is substantial (order $\si{\kilo\eV}$). However, most of this cancels out when looking at transitions. The effect on the transitions of interest was calculated to be $-0.41~\si{\eV}$, $-1.12~\si{\eV}$, and $-1.53~\si{\eV}$ on $2p1s$, $3p1s$, and $4p1s$, respectively. These values were consistent between the two chlorine isotopes within the current precision.

    \begin{table}
        \centering
        \caption{Overview of QED Contribution (in $\si{\eV}$) for the $1s$ state of \textsuperscript{35, 37}Cl. Calculations are shown for the resulting Barrett radii, given in Section~\ref{sec:results}. These correspond to 2pF distributions with parameters ($c=3.4933~\si{\femto\meter}$, $t=2.3~\si{\femto\meter}$) and ($c=3.5127~\si{\femto\meter}$, $t=2.3~\si{\femto\meter}$) for \textsuperscript{35}Cl and \textsuperscript{37}Cl, respectively. The screening was calculated separately at the literature RMS radius from Ref.~\cite{angeli2013table}.}
        \begin{ruledtabular}
        \begin{tabular}{r|rr}
            Contribution & \textsuperscript{35}Cl & \textsuperscript{37}Cl \\
            \hline
            Coulomb + Uehling & $-$782,471.8 & $-$782,383.4 \\
            SE\textsubscript{Point} & 204.8 & 204.9 \\
            SE\textsubscript{FNS} & $-$62.7 & $-$63.0 \\
            HVP & $-$6.3 & $-$6.3 \\
            WK & 3.2 & 3.2 \\
            KS & $-$38.0 & $-$38.0 \\
            $\Delta$Rec & $-$0.1 & $-$0.1\\
            Screening & 1,296.1 & 1,295.4 \\
            \hline
            Total & $-$781,074.7 & $-$780,987.2
        \end{tabular}
        \end{ruledtabular}
        \label{tab:QED_overview}
    \end{table}

    Although extensive QED corrections have been considered in the present work, there are still some higher-order effects that are not accounted for but can have a sizable (order $1~\si{\eV}$) effect on the muonic transition energies~\cite{2011-RadRec, 2011-DF, 2011-RadDef}. We estimated the uncertainty of QED effects for the $1s$ state to be about $3~\si{\eV}$ from neglected effects and higher order contributions. The uncertainty on the $np$ states is negligible compared to that of the $1s$ state. More information about these calculations and their uncertainty estimation can be found in the supplementary material.

\subsection{Nuclear polarization}\label{sec:NP}
    While the reduced distance to the nucleus is the prime benefit of muonic atoms, it also amplifies the effects of the internal dynamic nuclear structure beyond the finite size. These effects are commonly referred to as the nuclear polarization (NP). The NP represents the most challenging and uncertain correction to muonic energy levels, as its calculation requires knowledge of the entire nuclear spectrum of the isotope of interest. The summation over nuclear excitations can be divided into three contributions: 1)~Low-lying states (LLS), 2)~High-frequency collective excitations known as giant resonances (GR), and 3)~Excitations in the hadronic range corresponding to the so-called nucleon polarization (nP). For the nuclear part (LLS + GR), the same methods were used as those presented in Refs.~\cite{valuev2022evidence, 2024_Valuev, 2007_Haga}. The nP has only recently been considered for systems beyond \textsuperscript{4}He~\cite{gorchtein2025hitchhiker}, and it provides a shift equivalent to half of the NP uncertainty for the isotopes in this work. The distinction between the LLS and GR is made due to the different ways of taking them into account when fully microscopic calculations of a nuclear spectrum are not available, which is the case for most odd-mass nuclei like \textsuperscript{35}Cl and \textsuperscript{37}Cl. The nuclear parameters for low-lying nuclear excitations are taken from nuclear data sheets~\cite{Nucl_data_35, Nucl_data_37}, while in the case of the giant resonances one has to resort to phenomenological energy-weighted sum rules~\cite{rinker1978nuclear}. In such cases, one also needs to choose an appropriate model for nuclear transition charge densities, which is of major importance for muonic atoms. More details on the approach adopted in this work are provided in the supplementary material.

    The uncertainty treatment for nuclear polarization in existing literature is rather ad hoc. The most commonly quoted uncertainties are based on those in the book of Fricke and Heilig~\cite{fricke2004nuclear}. There, an uncertainty of 30\% is assigned to the NP correction. In this work, we attempt to root the uncertainty estimates more firmly in statistical evaluations. A differentiation is made between the three contributions of the NP. The relative uncertainty of the nuclear part (LLS + GR) is based on the calculations for the nearby doubly-magic \textsuperscript{40}Ca, for which fully microscopic calculations are possible~\cite{skyrme_rpa, 2024_Valuev} and can thus be used as a reliability check. We benchmarked our adopted approach by comparing it to fully microscopic calculations with different Skyrme parametrizations. Here, the selection of Skyrme models was made covering the same criteria as taken in Ref.~\cite{valuev2022evidence}: They should cover significant portions of the constraints on the saturation properties, and represent different groups and fitting protocols. Such checks showed a maximum deviation of about 23\% (equivalent to $39.3~\si{\eV}$), which was taken as the fractional uncertainty on the nuclear part of the NP.  It should be noted that this relative uncertainty estimate is for the sum of the two nuclear parts, while for the contribution from the LLS likely has a larger relative uncertainty than that from the GR due to incompleteness of the information in Nuclear Data Sheets~\cite{Nucl_data_35, Nucl_data_37}. For the nucleon part, values were taken from Ref.~\cite{gorchtein2025hitchhiker} ($\sim$10\% uncertainty). When adding the different contributions, their uncertainties were linearly added as a conservative estimate. The resulting NP corrections are given in Table~\ref{tab:NP}. Given that NP is dominated by the $1s$ state, it is nearly identical for all $np1s$ transitions.

    \begin{table}
        \centering
        \caption{Calculations for the nuclear polarization (in $\si{\eV}$) for the relevant transitions in \textsuperscript{35}Cl and \textsuperscript{37}Cl. $f_{\text{Corr}}$ (\%) denotes the fraction of the nuclear polarization that behaves in a correlated way between the two isotopes.}
        \begin{ruledtabular}
        \begin{tabular}{cc|cc|cc}
            Isotope & Transition & Nuclear & Nucleon & Total & $f_{\text{Corr}}$ (\%) \\
            \hline
            \textsuperscript{35}Cl \rule{0pt}{2.5ex} & $2p1s$ & 103(24) & 11.9(1.2) & 115(25) & 90.4 \\
                                   & $3p1s$ & 104(24) & 11.9(1.2) & 116(26) & 90.4 \\
                                   & $4p1s$ & 104(24) & 11.9(1.2) & 116(26) & 90.4 \\
            \hline
            \textsuperscript{37}Cl \rule{0pt}{2.5ex} & $2p1s$ & 99(23)  & 12.6(1.3) & 112(25) & 97.6 \\
                                   & $3p1s$ & 100(23) & 12.6(1.3) & 112(25) & 97.6 \\
                                   & $4p1s$ & 100(23) & 12.6(1.3) & 113(25) & 97.6 \\
        \end{tabular}
        \end{ruledtabular}
        \label{tab:NP}
    \end{table}

    While studying muonic isotope shifts, part of the nuclear polarization uncertainty should cancel out due to correlation (primarily from the GR and nP). The book of Fricke and Heilig~\cite{fricke2004nuclear} treats this by assuming that the uncertainty on the difference in NP is equal to 10\% of the largest NP value. Instead of directly quoting an uncertainty on the difference in NP, we opted for a determination of the correlation between the NP of the two isotopes. The GR and nP are strongly correlated between different isotopes, while the contribution from the LLS is rather uncorrelated. Accordingly, it was assumed that the former two are maximally correlated, while the latter is completely uncorrelated. This leads to a correlated fraction $f_{\text{Corr}}$ of 90.4\% and 97.6\% for \textsuperscript{35}Cl and \textsuperscript{37}Cl, respectively. The correlation between the nuclear polarization of the two isotopes was then estimated by the multiplication of these correlated fractions, resulting in 88.2\% correlation. This is equivalent to a nuclear polarization difference of \textsuperscript{37}Cl with respect to \textsuperscript{35}Cl of $-3(12)~\si{\eV}$. The uncertainties predicted by our method are nearly identical (not by construction) to those obtained from the assumptions in Fricke and Heilig~\cite{fricke2004nuclear}.

\subsection{Barrett radii}\label{sec:theo_Barrett}
    Up to now, the extracted quantities are charge distribution parameters belonging to a specific simple charge distribution model. The next step is to take this charge distribution and extract mean-square charge radii or root mean square (RMS) charge radii. By definition, one can achieve this by solving 

    \begin{equation}
        \langle r^2 \rangle = \frac{\int \rho(r) r^4 dr}{\int \rho(r) r^2 dr}.
    \end{equation}

    This however, leads to a large error from the charge distribution model dependency. Changing $t$ by 10\%, which is a commonly assumed uncertainty~\cite{fricke2004nuclear}, may drastically alter the RMS radius. Applying such an approach for the case of Cl leads to an uncertainty of 0.15\%, larger than any other contribution to the total uncertainty budget. A solution to this problem was proposed by Barrett based on perturbation theory~\cite{barrett1970model}. The general concept relies on the fact that the difference in potential generated by the muon in the initial and final states, $V_\mu^i(r)$ and $V_\mu^f(r)$, can be approximated within the region where $r^2 \rho(r)$ is large by the relation

    \begin{equation}\label{eq:V_diff}
        V_\mu^i(r) - V_\mu^f(r) = B r^k e^{-\alpha r},
    \end{equation}

    \noindent where $B$, $k$, and $\alpha$ are parameters used to fit this potential difference. Using the Barrett parameters $k$ and $\alpha$, the Barrett moment is defined as

    \begin{equation}\label{eq:Bmoment}
        \langle r^k e^{-\alpha r} \rangle = \frac{\int \rho(r) r^k e^{-\alpha r} r^2 dr}{\int \rho(r) r^2 dr}.
    \end{equation}
    
    According to perturbation theory, the Barrett moment is in first order independent of the \textit{shape} of the charge distribution, while retaining the size sensitivity. In good approximation (in this region), any two charge distributions with the same Barrett moment predict the same transition energy. Alongside this Barrett moment, the Barrett equivalent radius $R_{k \alpha}$ is defined as the radius of a homogeneously charged sphere with the same Barrett moment as the nucleus. This can be determined by iteratively solving 

    \begin{equation}\label{eq:Bradius}
        \frac{3}{R_{k \alpha}^3} \int_0^{R_{k \alpha}} r^k e^{-\alpha r} r^2 dr = \langle r^k e^{-\alpha r} \rangle.
    \end{equation}

    \noindent The resulting Barrett radii are considered charge distribution model independent~\cite{fricke2004nuclear}. 
    
    For the determination of the parameters, the potential differences were fitted with Eq.~\eqref{eq:V_diff} in the range $r \in [0; 30]~\si{\femto\meter}$. The fits were made using least squares minimization with associated weights $r^2 \rho(r)$. Here, $\rho(r)$ is taken to be a two-parameter Fermi distribution (Eq.~\eqref{eq:theo_Fermi2}) with $t=2.3~\si{\femto\meter}$ and $c$ chosen such that the RMS radius corresponds to the literature value~\cite{briscoe1980elastic, angeli2013table}. Small variations in the charge distribution used for such weights do not substantially impact the extracted values for $k$ and $\alpha$. Varying the input RMS radius by 1\% changes $k$ by 0.02\% and $\alpha$ by 0.13\%. The choice of weight was made because it corresponds to the multiplicative factor required in the relevant integrand in Eq.~\eqref{eq:Bmoment}. In the past, authors claimed that weights of $r^4 \rho(r)$ could give less shape sensitivity~\cite{engfer1974charge}, but we could not reproduce this argument and the extracted RMS radii remained identical (though, with different values for $k$ and $\alpha$). We expect that such arguments may not be relevant in the medium mass region considered in this study. The fits described the potential difference within 1\% accuracy within the range $r \in [1.5; 15]~\si{\femto\meter}$, where the fit weights are the largest. The resulting values for $k$ and $\alpha$ are given in Table~\ref{tab:k_alpha}.
    
    \begin{table}[h]
        \centering
        \caption{Calculated values for the Barrett parameters. The values for the $np1s$ transitions are obtained by performing a weighted average over the fine structure levels. Uncertainty on the average was taken to be the maximal deviation from an individual transition.}
        \begin{ruledtabular}
        \begin{tabular}{cc|ccc|c}
            Isotope & Parameter & $2p1s$ & $3p1s$ & $4p1s$ & Average \\
            \hline
            \textsuperscript{35}Cl \rule{0pt}{2.5ex} 
                & $k$      & 2.0941 & 2.0936 & 2.0934 & 2.0937(4) \\
                & $\alpha$ & 0.0561 & 0.0559 & 0.0558 & 0.0559(2) \\
            \hline
            \textsuperscript{37}Cl \rule{0pt}{2.5ex} 
                & $k$        & 2.0944 & 2.0939 & 2.0937 & 2.0940(4) \\
                & $\alpha$   & 0.0560 & 0.0557 & 0.0557 & 0.0558(2) \\            
        \end{tabular}
        \end{ruledtabular}
        \label{tab:k_alpha}
    \end{table}

    One should note that the Barrett radii and parameters carry no physical meaning. The only aspect of interest is that a set of parameters is obtained for which the shape sensitivity is minimized. Different sets of $k$ and $\alpha$ may exist that achieve this (e.g., when taking different weighting factors or including including different QED corrections), which lead to different Barrett radii. Additionally, $k$ and $\alpha$ may, in general, vary between different transitions or isotopes. Accordingly, $R_{k, \alpha}$ does not necessarily represent the same physical observable across different transitions or isotopes. In our particular case of \textsuperscript{35, 37}Cl, the values of $k$ and $\alpha$ are very similar across the transitions, such that their average value could be used. Further tests showed that this did not introduce additional model dependencies.

\begin{figure}
        \centering
        \includegraphics[width=0.9\linewidth]{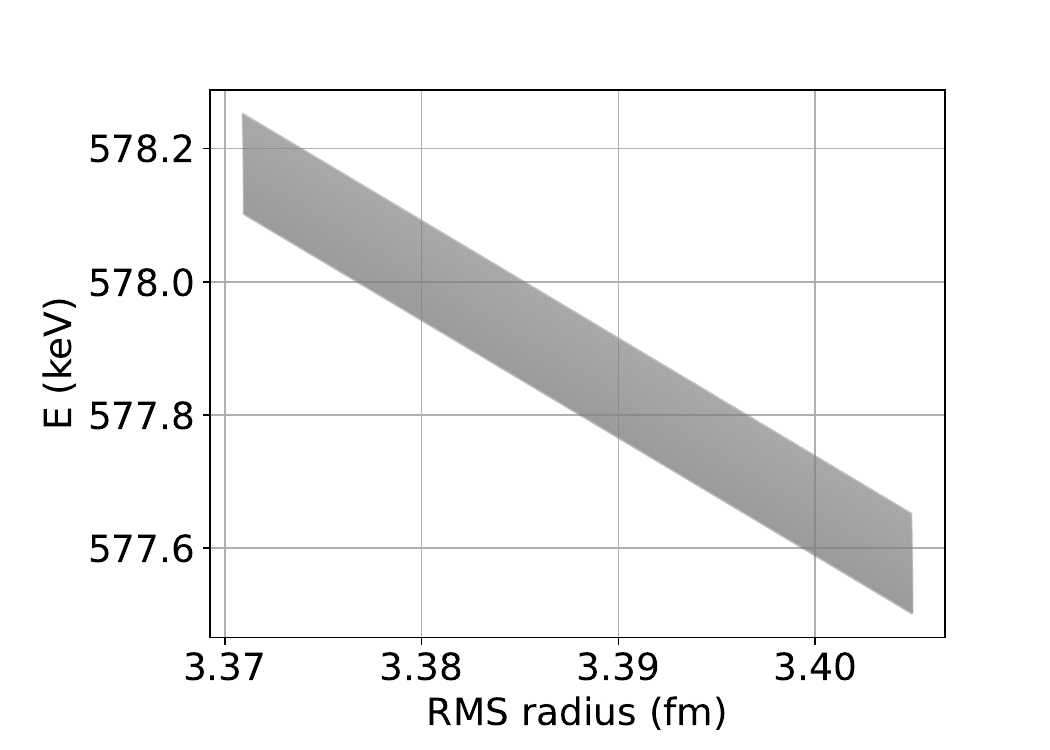}
        \includegraphics[width=0.9\linewidth]{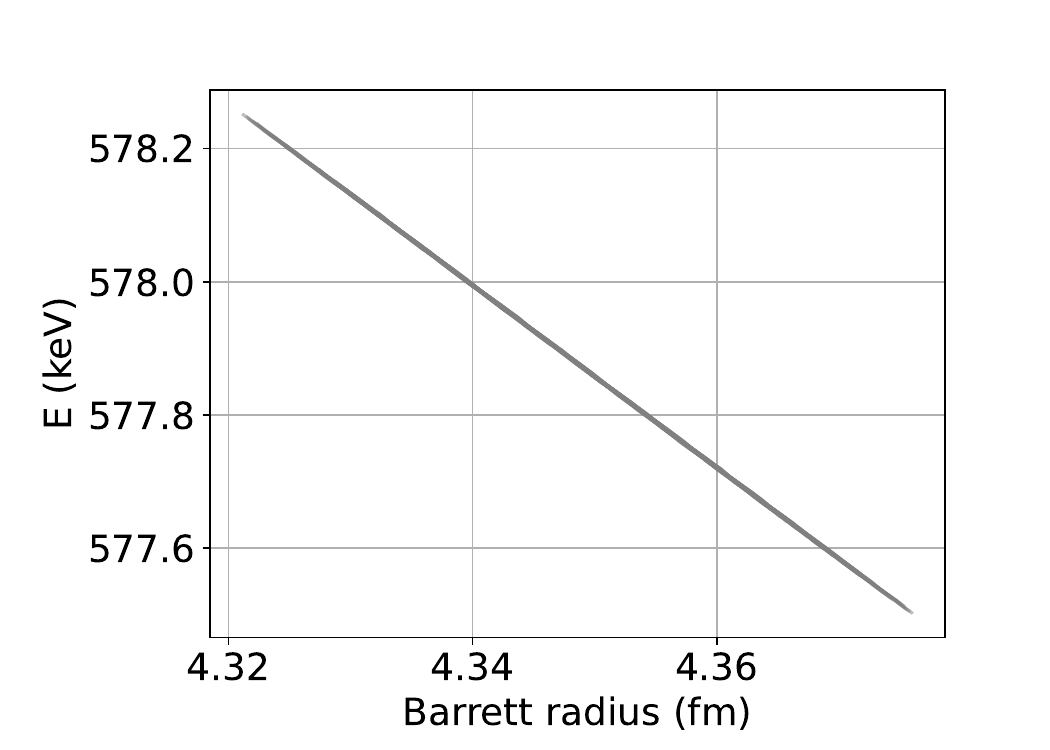}
        \caption{Muonic $2p1s$ energies as a function of RMS radius (top) and Barrett radius (bottom) for $t=2.3~\si{\femto\meter} \pm 10\%$.}
        \label{fig:Barrett_1D}
    \end{figure}

    In order to visualize the shape sensitivity and probe residual charge distribution model dependence, mudirac~\cite{sturniolo2021mudirac} calculations for the $np1s$ energies of \textsuperscript{35}Cl were performed using different 2-parameter Fermi distributions. These distributions were chosen to have a variation of 1\% on the mean square radius and 10\% on the skin thickness. For each of these distributions, the previously obtained Barrett parameters were used to calculate the corresponding Barrett radius. In Fig.~\ref{fig:Barrett_1D}, the transition energies are shown as a function of radius (RMS and Barrett) for different values of $t$. For the RMS radius, a relatively broad range of radii can reproduce the same transition energy by varying $t$. In contrast, the Barrett radius does not show such a trend, suppressing the model dependence substantially. The residual model dependency was determined to be at most $2~\si{\eV}$ for the $np1s$ lines of interest, deemed negligible compared to other experimental and theoretical uncertainties.

    Using the determined optimal values for $k$ and $\alpha$, the half-height radii ($c$ in Eq.~\eqref{eq:theo_Fermi2}) used in the QED calculations can be transformed into Barrett radii using Eq.~\eqref{eq:Bmoment} and Eq.~\eqref{eq:Bradius}. The integration required for the Barrett radius determination was performed using Riemann summing with a step size of $10^{-6}~\si{\femto\meter}$ and a cutoff range at large $r$ of $25~\si{\femto\meter}$ was used.

\subsection{Combining theory inputs}\label{sec:combining}
    In order to extract experimental Barrett radii, the different theoretical inputs must be combined. After adding the QED and NP values, one arrives at the transition energies. This, however, differs from the experimentally measured emission energies due to photon recoil, where the nucleus carries a small fraction of the total transition energy. This effect is energy and mass dependent, such that it varies between different transitions and isotopes. The calculated correction corresponding to the measured emission energy is given in Table~\ref{tab:photon_recoil}. The shifts due to the photon recoil are in the same order of magnitude as the experimental error, such that they can not be ignored.

    \begin{table}
        \centering
        \caption{Difference between transition and emission energies (in $\si{\eV}$) due to photon recoil.}
        \begin{ruledtabular}
        \begin{tabular}{c|ccc}
            Isotope & $2p1s$ & $3p1s$ & $4p1s$ \\
            \hline
            \textsuperscript{35}Cl \rule{0pt}{2.5ex}& 5.14 & 7.35 & 8.22 \\
            \textsuperscript{37}Cl \rule{0pt}{2.5ex}& 4.86 & 6.95 & 7.77
        \end{tabular}
        \end{ruledtabular}
        \label{tab:photon_recoil}
    \end{table}

    Next, the emission energy was fitted as a function of the Barrett radius using a second degree polynomial. Given that the calculations are made far from $R_{k \alpha} = 0$, the fit parameters are expected to be highly correlated due to multicollinearity, which can result in larger rounding errors. While this does not affect our extracted radius, it may cause complexities for future works trying to reproduce our results. To reduce this effect, a recentering around $R_{\text{cen}} = 4.3~\si{\femto\meter}$ was employed, transforming the model into
    
    \begin{equation}\label{eq:E_Rka}
        E = a_0 + a_1 (R_{k\alpha} - R_{\text{cen}}) + a_2 (R_{k \alpha} - R_{\text{cen}})^2.
    \end{equation}
    
    The resulting fit parameters are given in Table \ref{tab:QED_Rka}. As the fit residuals were smaller than $0.1~\si{\eV}$, the uncertainty on this fit is negligible compared to other uncertainties in this work.

    \begin{table}
        \centering
        \caption{Fit parameters for the center-of-gravity energy of the $2p1s$, $3p1s$, and $4p1s$ lines in \textsuperscript{35}Cl and \textsuperscript{37}Cl, see Eq.~\eqref{eq:E_Rka}.}
        \begin{ruledtabular}
        \begin{tabular}{c|l|ccc}
            Isotope & Parameter & $2p1s$ & $3p1s$ & $4p1s$\\
            \hline
            \textsuperscript{35}Cl \rule{0pt}{2.5ex} & $a_0~(\si{\kilo\eV})$  & 578.56192 & 691.78601 & 731.37157 \\
                & $a_1~(\si{\kilo\eV\per\femto\meter})$  & $-$13.629  & $-$13.636  & $-$13.638  \\
                & $a_2~(\si{\kilo\eV\per\femto\meter\squared})$ & $-$0.555 & $-$0.557 & $-$0.557 \\
            \hline
            \textsuperscript{37}Cl \rule{0pt}{2.5ex} & $a_0~(\si{\kilo\eV})$  & 578.65141 & 691.89543 & 731.48796 \\
                & $a_1~(\si{\kilo\eV\per\femto\meter})$  & $-$13.636  & $-$13.642  & $-$13.644  \\
                & $a_2~(\si{\kilo\eV\per\femto\meter\squared})$ & $-$0.548 & $-$0.549 & $-$0.550 \\
        \end{tabular}
        \end{ruledtabular}
        \label{tab:QED_Rka}
    \end{table}

\subsection{\texorpdfstring{$V_2$}{V2} correction}\label{sec:V2}
    In order to translate Barrett radii into more meaningful RMS radii, more information on the nuclear shape is needed. This information is typically provided by electron scattering measurements. Within such studies, the $V_2$ factor is defined as the ratio between the RMS radius ($R_{\text{RMS}}$) and the Barrett radius ($R_{k \alpha}$), such that

    \begin{equation}
        V_2^{(e)} \equiv \frac{R_{k \alpha}^{\text{(e)}}}{R_{\text{RMS}}^{\text{(e)}}} = \frac{\int \rho(r) r^{k} e^{-\alpha r} r^2 dr}{\int \rho(r) r^4 dr}.
    \end{equation}

    It is believed that by taking this ratio, most of the systematic uncertainties from the electron scattering measurements cancel out~\cite{fricke2004nuclear}. As a result, the combined analysis of muonic atoms and electron scattering provides the most precise RMS charge radius using

    \begin{equation}\label{eq:combined_rms}
        R_{\text{RMS}}^{(\mu, \text{e})} \equiv \frac{R_{k \alpha}^{(\mu)}}{V_2^{\text{(e)}}}.
    \end{equation}

    In the current literature, the $V_2$ correction is most often made assuming no uncertainty (e.g., Ref\cite{fricke2004nuclear}). However, recent work has shown that the uncertainty on $V_2$ is the dominant uncertainty for many radii~\cite{ohayon2024critical}. In many cases, no electron scattering is available, or the available data is not of sufficient quality. This is the case in Cl isotopes, where a limited momentum range was studied~\cite{briscoe1980elastic}, leading to a relative uncertainty of 0.12\% on the $V_2$ correction using determination from Ref.~\cite{ohayon2024critical}. At this precision, they provide a similar uncertainty as taking a simple 2pF with $t=2.3~\si{\femto\meter} \pm 10\%$.
    
    Given the success of energy density functionals (EDF) in describing nuclear observables (e.g., Ref.~\cite{colo2020nuclear}), we opted to test the reliability of the $V_2$ prediction from monopole averaged charge distributions extracted from BSkG models~\cite{scamps2021skyrme, ryssens2022skyrme, grams2023skyrme, grams2025skyrme}. In general, EDF models have an easy access to the one-body charge density needed to calculate $V_2$. Additionally, they carry the benefit of being universally applicable across the nuclear chart. This is particularly important for future works involving odd nuclei and heavier systems. The BSkG models were chosen since they are based on a fit that successfully and simultaneously describes many observables (including the charge radius) across the nuclear landscape. Given that no significant differences were found between the different models, the most recent version was used (BSkG4). 

    The reliability of these theoretical models was determined by comparing the extracted $V_2$ to experimental values for isotopes in the region that have high $q_{\text{max}}$ measurements available (\textsuperscript{31}P, \textsuperscript{32, 34, 36}S, \textsuperscript{39}K, and \textsuperscript{40, 48}Ca). For the calculation of these $V_2$ factors, a step size of $10^{-6}~\si{\femto\meter}$ was used. This revealed that the values from BSkG4 agreed within experimental errors (estimated using the empirical formula in Ref.~\cite{ohayon2024critical}), showing an average deviation in $V_2$ of 0.05\%. This average deviation was then taken as the uncertainty for the predictions of \textsuperscript{35}Cl and \textsuperscript{37}Cl, which reduces the uncertainty by approximately a factor 2.5 compared to the estimated uncertainty from literature electron scattering~\cite{briscoe1980elastic}.
    
    The extracted $V_2$ factors, using $k$ and $\alpha$ determined in Section~\ref{sec:theo_Barrett}, are given in Table \ref{tab:V2}. Apart from the values determined by BSkG4, this table includes those calculated from the existing electron scattering data~\cite{briscoe1980elastic} and those determined by a \textit{basic} charge distribution model. This last model assumed a 2-parameter Fermi distribution (Eq.~\eqref{eq:theo_Fermi2}) with $t=2.3~\si{\femto\meter}$ and $c$ chosen such that the RMS radius matches that of the literature value~\cite{briscoe1980elastic, angeli2013table}. The uncertainty on this value was estimated as the deviation in $V_2$ from varying $t$ by 10\%. For the radius extraction, the $V_2$ from the EDF calculations was used, as they are considered more precise and no less accurate than the existing electron scattering~\cite{briscoe1980elastic}.

    \begin{table}
        \centering
        \caption{Calculated values for $V_2$ based on a basic charge distribution model, electron scattering~\cite{briscoe1980elastic}, and BSkG models.}
        \begin{ruledtabular}
        \begin{tabular}{c|ccc}
            Isotope & $V_2$ (Basic) & $V_2$ (\cite{briscoe1980elastic}) & $V_2$ (BSkG4) \\
            \hline
            \textsuperscript{35}Cl \rule{0pt}{2.5ex} & 1.28365(172) & 1.28237(150) & 1.28328(65) \\
            \textsuperscript{37}Cl \rule{0pt}{2.5ex} & 1.28371(170) & 1.28306(150) & 1.28382(65)
        \end{tabular}
        \end{ruledtabular}
        \label{tab:V2}
    \end{table}

    It should be noted that $V_2$ correction factors that are extracted under the same experimental conditions (setup and maximal momentum transfer) typically show some level of correlation. As such, the difference $\Delta V$ (or equivalently, the ratio) between the $V_2$ factors of two isotopes is determined to a higher precision than uncorrelated error propagation would indicate. Similarly, the theory predictions are expected to predict differences and ratios to a higher precision than absolute values.

    By making comparisons between electron scattering measurements and model predictions, an estimate was made for the correlation between $V_2$ of different isotopes extracted from BSkG4. One can approximate the error on $\Delta V$ by taking the difference between electron scattering data and the theoretical model of choice. By combining this with the assumed uncertainty of $V_2$ from BSkG4 (0.05\%), the correlation between the $V_2$ extracted for different isotopes can be inferred from correlated error propagation. Applying this method using electron scattering data~\cite{briscoe1980elastic} showed a correlation of 97.4\%. This estimated correlation is similar between all BSkG models, giving a minimal value of $\sim 95\%$ on the slightly less-advanced BSkG2. A similar check was performed on \textsuperscript{34, 36}S, which have the same neutron numbers as \textsuperscript{35, 37}Cl and one fewer proton. Comparing to literature electron scattering~\cite{rychel1983charge, de1987nuclear} resulted in an approximate correlation of 98.0\%. As a conservative estimate, the correlation between $V_2$ factors predicted from EDF models for different isotopes was assumed to be 97.0\%. While the $V_2$ factor is typically similar for neighboring isotopes, estimates on the correlation are not straightforward due to nuclear structure effects. As such, the $V_2$ factors from the basic charge distribution model were assumed to be uncorrelated.

\subsection{Differential mean square radius}\label{sec:theo_diff}
    As highlighted in Section~\ref{sec:exp_unc}, muonic isotope shifts can be determined more precisely than absolute energies. Similarly, Section~\ref{sec:NP} and Section~\ref{sec:V2} indicated the strong correlation between the NP and $V_2$ uncertainties of different isotopes. As such, the differential mean square charge radii can be extracted to a substantially higher precision than uncorrelated error propagation would indicate. One can rewrite the differential mean square radius of an isotope $A$ and a reference isotope $A'$ as

    \begin{eqnarray}
        \delta \langle r^2 \rangle^{A, A'} && \equiv R_{A}^2 - R_{A'}^2 \nonumber \\
        && = (R_{A} + R_{A'})(R_{A} - R_{A'}) \nonumber\\
        && = (2 R_{A'} + \Delta R) \Delta R \nonumber \\
        %& = 2 R_{A} \Delta R + \Delta R ^2 \\
        && = \frac{2 R_{k \alpha, A'} \Delta R}{V_{2, A'}} + \Delta R ^2 \label{eq:combined_dR2},
    \end{eqnarray}

    where

    \begin{eqnarray}
        \Delta R && \equiv R_{A} - R_{A'} \nonumber \\
        && = \frac{R_{k \alpha, A}}{V_{2, A}} - \frac{R_{k \alpha, A'}}{V_{2, A'}} \nonumber \\
        && = \frac{1}{V_{2, A'}} \left[ R_{k \alpha, A} \frac{V_{2, A'}}{V_{2, A}} - R_{k \alpha, A'}\right] \nonumber \\
        && = \frac{1}{V_{2, A'}} \left[ \Delta R_{k \alpha} + \left(\frac{V_{2, A'}}{V_{2, A}} - 1\right) R_{k \alpha, A} \right] \label{eq:combined_dR}
    \end{eqnarray}

    and 

    \begin{equation}
        \Delta R_{k \alpha} \equiv R_{k \alpha, A} - R_{k \alpha, A'}.
    \end{equation}

    Due to the more accurate muonic isotope shift and differential nuclear polarization compared to their absolute counterparts, an increased sensitivity can be reached for the difference in Barrett radius $\Delta R_{k \alpha}$. Additionally, the correlation between $V_2$ factors results in an enhanced precision in $\left(\frac{V_{2, A'}}{V_{2, A}} - 1\right)$. Using the same QED calculations described in Section~\ref{sec:QED}, the muonic isotope shift was fitted as a function of the Barrett radius difference and the Barrett radius of the reference isotope $A'$. Omitting the latter component results in a minor residual trend in the order of $5~\si{\eV}$. The best fitting model was determined to be

    \begin{eqnarray}
        \Delta E = && b_0 + b_1 \Delta R_{k\alpha} + b_2 (R_{k \alpha, 1} - R_{\text{cen}})\label{eq:E_dRka} \\
        && + b_3 (\Delta R_{k\alpha})^2 + b_4 (R_{k \alpha, 1} - R_{\text{cen}}) \Delta R_{k \alpha}, \nonumber
    \end{eqnarray}

    where the Barrett radius of the reference isotope is again recentered around $R_{\text{cen}} = 4.3~\si{\femto\meter}$ to reduce multicollinearity. This resulted in residuals smaller than $0.1~\si{\eV}$, negligible compared to the other uncertainties in this work. The resulting fit parameters are given in Table~\ref{tab:fit_dRka}.
    
    \begin{table}
        \centering
        \caption{Fit parameters for the center-of-gravity energy difference of the $2p1s$, $3p1s$, and $4p1s$ lines between \textsuperscript{35}Cl and \textsuperscript{37}Cl $(37 - 35)$, see Eq.~\eqref{eq:E_dRka}.}
        \begin{ruledtabular}
        \begin{tabular}{l|ccc}
            Parameter & $2p1s$ & $3p1s$ & $4p1s$ \\
            \hline
            $b_0~(\si{\kilo\eV})$ 
            & 0.08907 & 0.10889 & 0.11580 \\
            $b_1~(\si{\kilo\eV\per\femto\meter})$ 
            & $-$13.6557 & $-$13.6625 & $-$13.6644 \\
            $b_2~(\si{\eV\per\femto\meter})$ 
            & $-$6.04 & $-$6.04 & $-$6.04 \\
            $b_3~(\si{\kilo\eV\per\femto\meter\squared})$ 
            & 0.552 & 0.554 & 0.555 \\
            $b_4~(\si{\kilo\eV\per\femto\meter\squared})$ 
            & $-$1.109 & $-$1.112 & $-$1.114
        \end{tabular}
        \end{ruledtabular}
        \label{tab:fit_dRka}
    \end{table}

\section{Results}\label{sec:results}
    Combining the experimental energies with the parametrized theoretical input from Section~\ref{sec:theo_Barrett} and Section~\ref{sec:theo_diff} provides the Barrett radii and differential Barrett radii, which are listed in Table~\ref{tab:result_barrett}. The experimental energy uncertainty and NP uncertainty are translated to an error in radius by dividing them by the slope of the fit functions in Eq.~\eqref{eq:E_Rka} and Eq.~\eqref{eq:E_dRka} for the Barrett radius and difference in Barrett radius, respectively. Since the same values for $k$ and $\alpha$ were used in the different transitions, the Barrett radii across these transitions represent the same physical observable. Accordingly, they were averaged using the inverse of the experimental variance as weights. The reduced $\chi^2$ of the average over different transitions (shown in Table~\ref{tab:result_barrett}) provides an indication that the uncertainty on the experimental energies is estimated well. One should note that the uncertainties from QED and NP are primarily from the $1s$ state, such that its uncertainty cannot be reduced by averaging across different $np1s$ transitions.

    \begin{table*}
        \centering
        \caption{Extracted Barrett radii (in $\si{\femto\meter}$) from this work. The round brackets (), square brackets [], and curly brackets \{\} represent the uncertainty originating from the experimental energy determination, the NP correction, and the QED. $\chi_{\nu, av}^2$ denotes the reduced chisquare of the average over transitions.}
        \begin{ruledtabular}
        \begin{tabular}{l|lll|ll}
            Radius & $2p1s$ & $3p1s$ & $4p1s$ & Average & $\chi_{\nu, av}^2$ \\
            \hline
            $R_{k \alpha}^{35}$ \rule{0pt}{2.5ex} 
            & 4.2774(12)[18]\{3\} & 4.2773(8)[19]\{3\} & 4.2785(10)[19]\{3\} & 4.2777(6)[19]\{3\} & 0.53 \\
            $R_{k \alpha}^{37}$ 
            & 4.2940(9)[18]\{3\} & 4.2926(13)[18]\{3\} & 4.2956(21)[18]\{3\} & 4.2938(7)[18]\{3\} & 0.83 \\
            $R_{k \alpha}^{37} - R_{k \alpha}^{35}$ 
            & 0.0171(9)[9] & 0.0155(13)[9] & 0.0170(21)[9] & 0.0166(7)[9] & 0.62
        \end{tabular}
        \end{ruledtabular}
        \label{tab:result_barrett}
    \end{table*}

    The Barrett radii were subsequently transformed into RMS quantities using Eq.~\eqref{eq:combined_rms}, Eq.~\eqref{eq:combined_dR}, and Eq.~\eqref{eq:combined_dR2}. A breakdown of the involved uncertainties is given in Table~\ref{tab:rms_err}. This table shows values for the $V_2$ estimates from the basic charge distribution (described in Section~\ref{sec:V2}) and BSkG4. The former results in a \textit{pure muonic} charge radius, which has a significantly larger uncertainty. As mentioned before, almost all published studies ignore the uncertainties on the nuclear shape correction, while it is the largest source of uncertainty for the absolute radius and of similar magnitude as the other uncertainties for the difference in radius. The resulting RMS radii are given in Table~\ref{tab:result_rms} comparing our results to the existing literature values and radius estimates based on mirror pairs.

    \begin{table}
        \centering
        \caption{Uncertainty breakdown for the extracted RMS radii. All errors are in $\si{\atto\meter}$ (attometer) besides those on the differential mean square radius, which are in $10^{-3}~\si{\femto\meter\squared}$.}
        \begin{ruledtabular}
        \begin{tabular}{c|c|cccc|c}
            Radius & $V_2$ model & $\sigma_{\text{Exp}}$ & $\sigma_{\text{NP}}$ & $\sigma_{\text{QED}}$ & $\sigma_{V_2}$ & $\sigma_{\text{tot}}$ \\
            \hline
            $R^{35}$ \rule{0pt}{2.5ex}
            & Basic & 0.41 & 1.43 & 0.23 & 4.47 & 4.72 \\
            & BSkG4 & 0.41 & 1.43 & 0.23 & 1.69 & 2.27 \\
            \hline
            $R^{37}$ \rule{0pt}{2.5ex}
            & Basic & 0.53 & 1.38 & 0.23 & 4.43 & 4.68 \\
            & BSkG4 & 0.53 & 1.38 & 0.23 & 1.69 & 2.26 \\
            \hline
            $R^{37} - R^{35}$ \rule{0pt}{2.5ex}
            & Basic & 0.52 & 0.68 & / & 6.29 & 6.35 \\
            & BSkG4 & 0.52 & 0.68 & / & 0.48 & 0.98 \\
            \hline
            $\delta \langle r^2\rangle^{37, 35}$ \rule{0pt}{2.5ex}
            & Basic & 3.5 & 4.5 & / & 42.0 & 42.4 \\
            & BSkG4 & 3.5 & 4.5 & / & 3.2 & 6.6
        \end{tabular}
        \end{ruledtabular}
        \label{tab:rms_err}
    \end{table}

    \begin{table*}
        \centering
        \caption{Resulting RMS radii of \textsuperscript{35}Cl and \textsuperscript{37}Cl (in $\si{\femto\meter}$ and $\si{\femto\meter\squared}$). For the pure muonic RMS radii, the $V_2$ is taken from the basic charge distribution model introduced in Section~\ref{sec:V2}.}
        \begin{ruledtabular}
        \begin{tabular}{l|llll}
            Radius & Pure muonic & Using BSkG4 & Literature~\cite{briscoe1980elastic} & Mirror estimates~\cite{ohayon2024critical} \\
            \hline
            $R^{35}$ \rule{0pt}{2.5ex} 
            & 3.3325(48) & 3.3334(23) & 3.388(17) & 3.323(11) \\
            $R^{37}$ & 
            3.3448(47) & 3.3444(23) & 3.384(17) & 3.338(7) \\
            $R^{37} - R^{35}$ 
            & 0.0128(64) & 0.01154(98) & $-$0.004(24) & 0.015(11) \\
            $\delta \langle r^2\rangle^{37, 35}$ 
            & 0.085(43) & 0.0771(66) & $-$0.03(16) & 0.103(70)
        \end{tabular}
        \end{ruledtabular}
        \label{tab:result_rms}
    \end{table*}

    The results provide a major improvement on the knowledge of chlorine radii. Compared to literature electron scattering results, the radii are shifted by respectively $3.2\sigma$ and $2.3\sigma$ for \textsuperscript{35}Cl and \textsuperscript{37}Cl, while reducing the uncertainties by a factor of seven. We suspect this discrepancy may originate from underestimated systematics in the literature electron scattering measurements. In contrast, our values are in agreement with estimates from the phenomenological mirror shift fit~\cite{ohayon2024critical}, showing the predictive power of such estimates. Such an approach can be used to determine radii of isotopes for which the corresponding mirror pair (opposing proton and neutron number) has a known radius. Moreover, the uncertainty on the difference in radius is improved by a factor 25. At this level of precision, this result is valuable for extracting charge radii of chlorine isotopes from future isotope shifts measurements by calibrating the ratio of isotope shift factors.

    Given that the uncertainties of the stable chlorine radii have now been reduced substantially, they can be used as an additional data point for the mirror shift fit~\cite{ohayon2024critical}. This model describes the behavior of the radius difference in a mirror pair $\Delta_I$ as a function of the isospin asymmetry $I = \frac{Z - N}{A}$. The behavior is expected to be approximately linear due to neutron/proton skin effects~\cite{novario2023trends}. In principle, one would expect that the intercept would be zero, as a nucleus with no isospin asymmetry is its own mirror isotope, and as such gives no difference in radius. It was recently suggested that a proportional fit could be used to predict radii of isotopes that have no experimental data, if their corresponding mirror counterpart has been measured~\cite{ohayon2024critical}. The author does not include mirror pairs involving \textsuperscript{35}Cl and \textsuperscript{37}Cl due to doubts about the reliability of their RMS radii in the literature. If the reduced $\chi^2$ of this fit is statistically distributed when many mirror pairs are included, it provides an indication that predictions using this fit are reliable. With new measurements of the chlorine isotopes, we can assess their impact on the mirror shift fit. The radii of the relevant mirror isotopes (\textsuperscript{35}Ar and \textsuperscript{37}Ca) have been measured in laser spectroscopy studies~\cite{blaum2008nuclear, miller2019proton}, and were recently reevaluated~\cite{ohayon2024critical}. Combining these with our chlorine radii results in radius differences of $0.030(11)~\si{\femto\meter}$ and $0.107(7)~\si{\femto\meter}$, respectively for the (17, 18) and (17, 20) mirror pairs. These uncertainties are dominated by the radii of \textsuperscript{35}Ar and \textsuperscript{37}Ca. Here, both a proportional model ($\Delta_I = c_1 I$) and a linear model ($\Delta_I = d_0 + d_1 I$) were considered. For these fits, data for the other mirror pairs were taken from Ref.~\cite{ohayon2024critical}. Furthermore, they were performed under three conditions: 1) Omitting chlorine-related mirror pairs, 2) using literature chlorine radii~\cite{briscoe1980elastic}, and 3) using our updated chlorine radii.

\begin{figure}
        \centering
        \includegraphics[width=0.9\linewidth]{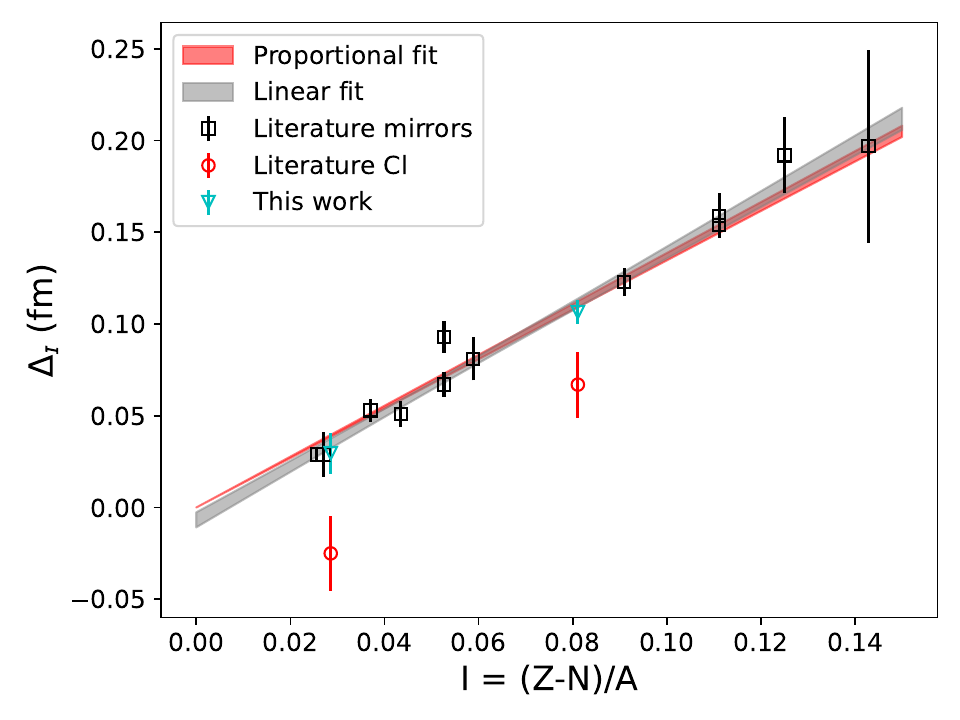}
        \caption{Mirror shift fit updated with chlorine radii from this work. Our measurement provides a substantial shift of the two mirror pairs involving stable Cl isotopes. The new values are in excellent agreement with trend of the other mirror pairs.}
        \label{fig:mirrorfit}
    \end{figure}

    The fit is plotted in Fig.~\ref{fig:mirrorfit} and the corresponding parameters are listed in Table~\ref{tab:mirrorfit}. First, the results show a clear preference for the radii extracted in this work over those from literature~\cite{briscoe1980elastic}. The reduced $\chi^2$ is well within the statistical distribution for the fit excluding Cl and the fit with the Cl radii from this work. However, the reduced $\chi^2$ of the fits including literature Cl radii are in the upper 1\% (proportional fit) and 2\% (linear fit) quantile of the statistical $\chi^2$ distribution. Additionally, the intercept of the fit is not significantly different from zero. The Akaike information criterion~\cite{akaike1998information} shows a very slight preference to the linear model over the proportional model. Compared to the case where chlorine-related mirror pairs are excluded, the value of the proportionality factor $c_1$ shifts by about $0.4\sigma$, while the uncertainty is reduced by 10\%.

    \begin{table}
        \centering
        \begin{ruledtabular}
        \begin{tabular}{c|c|c|c}
            Parameter        & Excluding Cl & Literature Cl & This work \\
            \hline
            $\nu$            & 11          & 13          & 13          \\
            $\chi_\nu^2$     & 1.08        & 2.15        & 1.01        \\
            $c_1$            & 1.380(36)   & 1.362(50)   & 1.367(32)   \\
            \hline
            $\nu$            & 10          & 12          & 12          \\
            $\chi_\nu^2$     & 0.98        & 2.06        & 0.90        \\
            $d_0 (10^{-3})$  & $-$6.4(4.4) & $-$8.0(6.3) & $-$6.7(4.2) \\
            $d_1$            & 1.465(68)   & 1.468(98)   & 1.457(63)   \\
            Corr($d_0, d_1$) & -86.8\%     & -86.9\%     & -87.9\%
        \end{tabular}
        \end{ruledtabular}
        \caption{Results from the mirror shift fit performed under different conditions, see Fig.~\ref{fig:mirrorfit}. The horizontal line marks separates the proportional and linear model.}
        \label{tab:mirrorfit}
    \end{table}

    Additional evaluations of mirror pairs with high accuracy and reliability would contribute to this fit. The most interesting mirror pairs would be those with a very low isospin asymmetry (for evaluating the potential intercept) and at high isospin asymmetry (to better constrain the slope). Currently, the only significantly deviating point is that of the \textsuperscript{19}F-\textsuperscript{19}Ne mirror pair. For both of these isotopes, muonic x-ray measurements are planned using microcalorimeters~\cite{ohayon2024towards}.

\section{Conclusion}
    Given renewed interest in absolute charge radius inputs and some outstanding questions on the systematics in muonic atoms, it is critical to reassess the radius extraction methods. Using chlorine, we demonstrated a modern approach to RMS nuclear radius extraction through muonic x rays, improving both experimental and theoretical methods compared to older literature. This included a more rigorous uncertainty evaluation and considerations of correlated uncertainties. A high-precision measurement was made of the muonic $2p1s$, $3p1s$, and $4p1s$ energies of isotopically pure \textsuperscript{35}Cl and \textsuperscript{37}Cl, which resulted in a substantial improvement compared to literature radii. Our values show an improvement in the precision of the RMS radii by a factor seven, and showed a disagreement of \textbf{$3.2 \sigma$} and $2.3 \sigma$ compared to the literature values~\cite{briscoe1980elastic} of \textsuperscript{35}Cl and \textsuperscript{37}Cl. These updated radii agree much better with the mirror shift fit, adding confidence such a fit formalism in this region of the nuclear chart. Furthermore, the muonic isotope shifts were used to extract a more accurate value for the difference in RMS radii and the difference in mean square radius, revealing that \textsuperscript{37}Cl has a significantly larger charge radius than \textsuperscript{35}Cl. We demonstrated how novel methods are beneficial to obtaining better results on nuclear structure and hope that the present investigation will trigger a revival of highly precise muonic x-ray experiments.

\section*{Acknowledgments}
    The experiments were performed at the $\pi$E1 beam line of PSI. We would like to thank the accelerator and support groups for the excellent conditions. The germanium detector setup is shared with the MIXE project at PSI (\url{https://www.psi.ch/en/smus/muon-induced-x-ray-emission-mixe-project}), which has greatly contributed to its construction, providing a fantastic platform for several muonic atom experiments taking place at PSI. This research used targets provided by the Center for Accelerator Target Science at Argonne National Laboratory, which is a DOE Office of Science User Facility and supported by the U.S. Department of Energy, Office of Nuclear Physics, under Award No. DE-AC02-06CH11357.

    The authors acknowledge the following funding institutions. The Swiss National Science Foundation, Sinergia project “Deep$\mu$”, Grant: 193691 (MIXE); KU Leuven BOF under contract number C14/22/104; the European Research Council (ERC) through proposal number 101088504 (NSHAPE)); Fonds de la Recherche Scientifique - FNRS, under project No~F.4553.25; The romanian ministry of education and scientific research under project number PN 23 21 01 02; FWO Vlaanderen, through proposal numbers G0G3121N (NSHAPE), and 11P6V24N (M.D.); the ETH Research Grant 22-2 ETH-023 (K.v.S.).; The Technion postdoctoral fellowship (S.R.). A.H. would like to thank the Slovak Research and Development Agency under contract No. APVV-20-0532, and Slovak grant agency VEGA (contract No. 2/0175/24). M.~G. acknowledges support  by the Deutsche Forschungsgemeinschaft (DFG) under grant agreement GO 2604/3-1.
    
    %Center of Excellence in High energy Physics and Astrophysics, Suranaree University of Technology (N.R.); 

    W.R. is a Research Associate of the F.R.S.-FNRS (Belgium). Nuclear calculations were performed using computational resources from the Tier-1 supercomputer Lucia of the Fédération Wallonie-Bruxelles, infrastructure funded by the Walloon Region under the grant agreement nr 1117545, and the clusters Consortium des Équipements de Calcul Intensif (CÉCI), funded by F.R.S.-FNRS under Grant No.~2.5020.11 and by the Walloon Region.

\section*{Contribution statement}
    E.A.M. arranged the production of the \textsuperscript{110m}Ag calibration source; The experiment was performed by T.E.C., C.C., M.D., A.D., M.H., A.H., A.K., R.L., V.M., A.T., S.M.V., and K.v.S.; The analysis was performed by M.H., with input provided in discussions with T.E.C., C.C., M.D., A.D., O.E., A.K., R.L., B.O., W.W.M.M.P., R.P., S.M.V., K.v.S., F.W., and A.Z.; QED calculations and interpretation involved M.H., B.O., N.S.O., P.I., and S.R.; NP calculation and interpretation were performed by M.G., M.H., N.O., and I.V.; Barrett parameters were determined by K.A.B. and M.H.; The correction for the nuclear shape was evaluated by P.D., M.H., B.O., and W.R.; The manuscript was written by M.H.; All co-authors reviewed the manuscript and were involved in practical discussions.

\section*{Competing interests}
    The authors declare no competing interests.

\section*{Data availability}
    Processed data is made available on Zenodo~\cite{dataset_Cl}.

\bibliography{biblio.bib}

\end{document}

%\twocolumngrid

%\preprint{}

\title{Modern approach to muonic x-ray spectroscopy demonstrated through the measurement of stable Cl radii \texorpdfstring{\\}{linebreak} Supplementary material}


\author{K.A.~Beyer}
\affiliation{Max-Planck-Institut für Kernphysik, Heidelberg, Germany}

\author{T.E.~Cocolios}
\affiliation{KU Leuven, Instituut voor Kern- en Stralingsfysica, Leuven, Belgium}

\author{C.~Costache}
\affiliation{Horia Hulubei National Institute for R\&D in Physics and Nuclear Engineering, Bucharest, Romania}

\author{M.~Deseyn}
\affiliation{KU Leuven, Instituut voor Kern- en Stralingsfysica, Leuven, Belgium}

\author{P.~Demol}
\affiliation{Université Libre de Bruxelles, Institut d’Astronomie et d’Astrophysique, Brussels, Belgium}
\affiliation{Brussels Laboratory of the Universe-BLU-ULB, Brussels, Belgium}

\author{A.~Doinaki}
\affiliation{PSI Center for Neutron and Muon Sciences, Villigen, Switzerland}
\affiliation{ETH Zürich, Institute for Particle Physics and Astrophysics, Zürich, Switzerland}

\author{O.~Eizenberg}
\affiliation{The Helen Diller Quantum Center, Department of Physics, Technion-Israel Institute of Technology, Haifa, Israel}

\author{M.~Gorshteyn}
\affiliation{Institut f\"ur Kernphysik, Johannes Gutenberg-Universit\"at Mainz, Mainz, Germany}
\affiliation{PRISMA$^+$ Cluster of Excellence, Johannes Gutenberg-Universit\"at Mainz, Mainz, Germany}

\author{M.~Heines}
\email[]{Corresponding author: michael.heines@kuleuven.be}
\affiliation{KU Leuven, Instituut voor Kern- en Stralingsfysica, Leuven, Belgium}

\author{A.~Herzáň}
\affiliation{Institute of Physics, Slovak Academy of Sciences, Bratislava, Slovakia}

\author{P.~Indelicato}
\affiliation{Laboratoire Kastler Brossel, Sorbonne Université, CNRS, ENS-PSL Research University, Collège de France, Paris, France}

\author{K.~Kirch}
\affiliation{PSI Center for Neutron and Muon Sciences, Villigen, Switzerland}
\affiliation{ETH Zürich, Institute for Particle Physics and Astrophysics, Zürich, Switzerland}

\author{A.~Knecht}
\affiliation{PSI Center for Neutron and Muon Sciences, Villigen, Switzerland}

\author{R.~Lica}
\affiliation{Horia Hulubei National Institute for R\&D in Physics and Nuclear Engineering, Bucharest, Romania}

\author{V.~Matousek}
\affiliation{Institute of Physics, Slovak Academy of Sciences, Bratislava, Slovakia}

\author{E.A.~Maugeri}
\affiliation{PSI Center for Nuclear Engineering and Sciences, Villigen, Switzerland}

\author{B.~Ohayon}
\affiliation{The Helen Diller Quantum Center, Department of Physics, Technion-Israel Institute of Technology, Haifa, Israel}

\author{N.S.~Oreshkina}
\affiliation{Max-Planck-Institut für Kernphysik, Heidelberg, Germany}

\author{W.W.M.M.~Phyo}
\affiliation{KU Leuven, Instituut voor Kern- en Stralingsfysica, Leuven, Belgium}

\author{R.~Pohl}
\affiliation{PRISMA$^+$ Cluster of Excellence, Johannes Gutenberg-Universit\"at Mainz, Mainz, Germany}
\affiliation{Institut f\"ur Physik, QUANTUM, Johannes Gutenberg-Universit\"at Mainz, Mainz, Germany}

\author{S.~Rathi}
\affiliation{The Helen Diller Quantum Center, Department of Physics, Technion-Israel Institute of Technology, Haifa, Israel}

\author{W.~Ryssens}
\affiliation{Université Libre de Bruxelles, Institut d’Astronomie et d’Astrophysique, Brussels, Belgium}
\affiliation{Brussels Laboratory of the Universe-BLU-ULB, Brussels, Belgium}

\author{A.~Turturica}
\affiliation{Horia Hulubei National Institute for R\&D in Physics and Nuclear Engineering, Bucharest, Romania}

\author{K.~von Schoeler}
\affiliation{ETH Zürich, Institute for Particle Physics and Astrophysics, Zürich, Switzerland}

\author{I.A.~Valuev}
\affiliation{Max-Planck-Institut für Kernphysik, Heidelberg, Germany}

\author{S.M.~Vogiatzi}
\affiliation{KU Leuven, Instituut voor Kern- en Stralingsfysica, Leuven, Belgium}

\author{F.~Wauters}
\affiliation{Institut f\"ur Kernphysik, Johannes Gutenberg-Universit\"at Mainz, Mainz, Germany}
\affiliation{PRISMA$^+$ Cluster of Excellence, Johannes Gutenberg-Universit\"at Mainz, Mainz, Germany}

\author{A.~Zendour}
\affiliation{PSI Center for Neutron and Muon Sciences, Villigen, Switzerland}
\affiliation{ETH Zürich, Institute for Particle Physics and Astrophysics, Zürich, Switzerland}


\date{\today}

\begin{abstract}
    This document provides more detailed information about the radius extraction of \textsuperscript{35}Cl and \textsuperscript{37}Cl. First, more information is given on the determination of the experimental muonic $np1s$ transition energies. Next, a more in-depth explanation is provided for the specific theoretical calculations used for the radius extraction.
\end{abstract}

\maketitle
\newpage
\tableofcontents
\newpage

%%%%%%%%%%%%%%%%%%%%%%%
%%%%%%%%%%%%%%%%%%%%%%%
%%%%% E N E R G Y %%%%%
%%%%%%%%%%%%%%%%%%%%%%%
%%%%%%%%%%%%%%%%%%%%%%%
\section{Energy extraction}
    A main challenge for this work was the accurate determination of $np1s$ energies. This section describes the analysis steps taken to extract the energies quoted in the manuscript. The following steps are used to determine the transition energies: 1) timing optimization process for the ELET algorithm, 2) choice of anticoincidence spectrum, 3) gain correction, 4) data filtering with coincidence and veto cuts, 5) high-precision calibration and reliability checks, 6) fitting of $np1s$ peaks. Additional systematic investigations were performed considering the hyperfine splitting, the isotopic purity of the used targets, and the choice of time cut. Finally, comparisons were made between experimental values and QED predictions for non-radius sensitive energy differences as an additional self-consistency check.

    The experimental data were taken in a broader experiment, also focused on the \textsuperscript{39, 40, 41}K triplet. Given that these were measured in near-identical conditions, certain optimizations were performed on higher calibration statistics in \textsuperscript{40}K (longer measurements) or higher signal statistics in \textsuperscript{39}K (larger target). The higher calibration statistics is particularly interesting, as it allows for systematic checks of the calibration to a higher precision.

\subsection{Timing optimization}
\begin{figure}[h]
        \centering
        \includegraphics[width=0.5\linewidth]{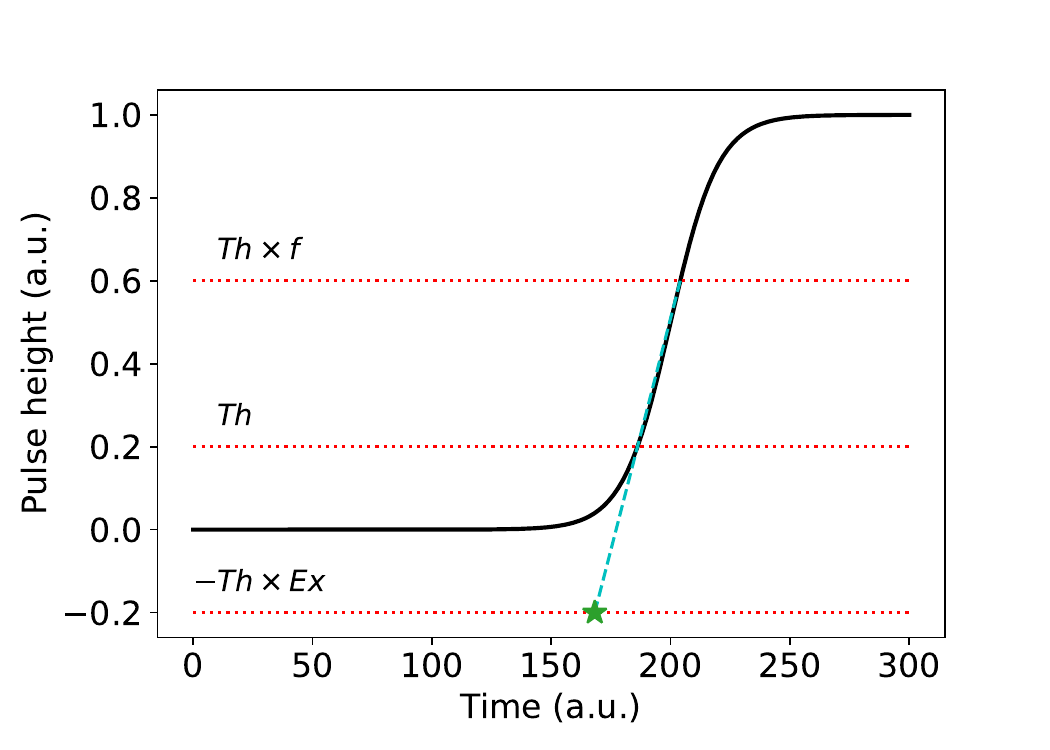}
        \caption{Graphical representation of the used ELET algorithm. The dashed red lines represent the two thresholds at $t_L$ and $t_U$. The green star represents the calculated ELET time $t_{ELET}$ for this pulse using the different methods. Parameters for this example are set to $Th = 0.2$, $f=3.0$, $Ex = 1.0$.}
        \label{fig:exp_ELET_principle}
    \end{figure}

    As a first step, the timing of the germanium detectors was improved using extrapolated leading edge timing (ELET). In such an approach the waveform between two thresholds is used to linearly extrapolate the value of the waveform in order to determine its timing. First, the baseline $b$ is determined for each event by 
    averaging over the start of the waveform. Next, the threshold ($Th$) and factor ($f$) are defined, such that they describe two thresholds above the baseline ($Th$ and $Th \times f$). These parameters can be optimized for the best time resolution in a certain energy range. The germanium waveform crosses the lower and upper thresholds at times $t_L$ and $t_U$, respectively. For convenience, the time difference between these crossing points is defined as $\Delta t = t_U - t_L$. Using these two crossings, one can estimate time of the pulse $t_{\text{ELET}}$ as the linear extrapolation to the baseline, given by $t_{ELET} = t_L - \Delta t$. It was, however, observed that the time resolution could be slightly improved by extrapolating to $-Ex \times Th$, where $Ex$ is a third parameter to be optimized. Accordingly, the time of a given pulse is determined using Eq.~\ref{eq:exp_tElet}. A graphical representation of the ELET timing is shown in Figure~\ref{fig:exp_ELET_principle}.

    \begin{equation}\label{eq:exp_tElet}
        t_{ELET} = t_L - \frac{1 + Ex}{f - 1}\,\Delta t
    \end{equation}

    Using a small sample of the data, the timing of each event is calculated for a given parameter space of $Th$, $f$, and $Ex$. Next, the time difference is taken with respect to the trigger muon for each set of parameters. To evaluate the time resolution, a prompt peak is selected within the energy range of interest. For this dataset, the timing was optimized using data from a measurement on $500~\si{\milli\gram}$ \textsuperscript{39}KCl with the $2p1s$ peak of \textsuperscript{39}K around $713~\si{\kilo\eV}$. The resulting time distributions were then fitted with a Gaussian model, such that the width of the time distribution could be plotted as a function of $Th$, $f$, and $Ex$.

\begin{figure}[h]
        \centering
        \begin{subfigure}[b]{0.45\textwidth}
             \centering
             \includegraphics[height=0.7\textwidth]{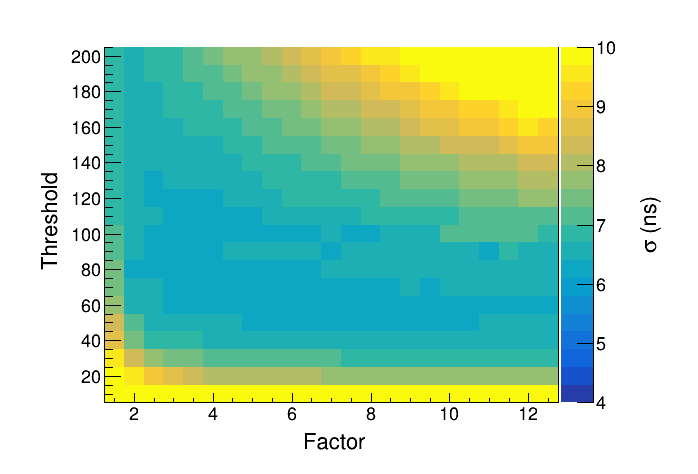}
             \caption{$Ex=0$, $Th=60$ selected for next step}
         \end{subfigure}
         \quad
         \begin{subfigure}[b]{0.45\textwidth}
             \centering
             \includegraphics[height=0.7\textwidth]{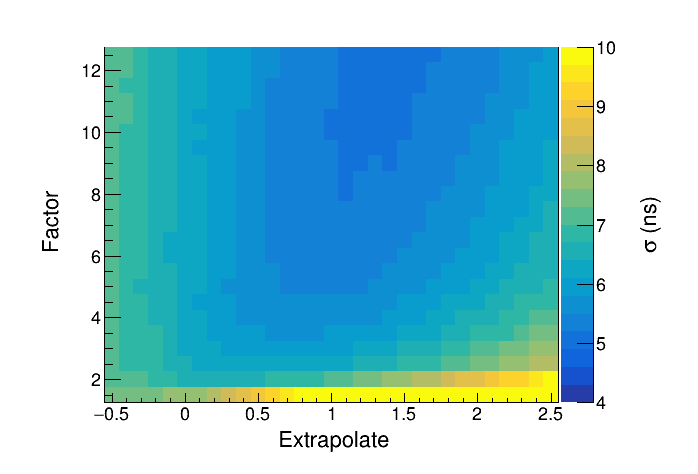}
             \caption{$Th=60$, $Ex=1.1$ selected as optimal value}
         \end{subfigure}
         \\
         \begin{subfigure}[b]{0.45\textwidth}
             \centering
             \includegraphics[height=0.7\textwidth]{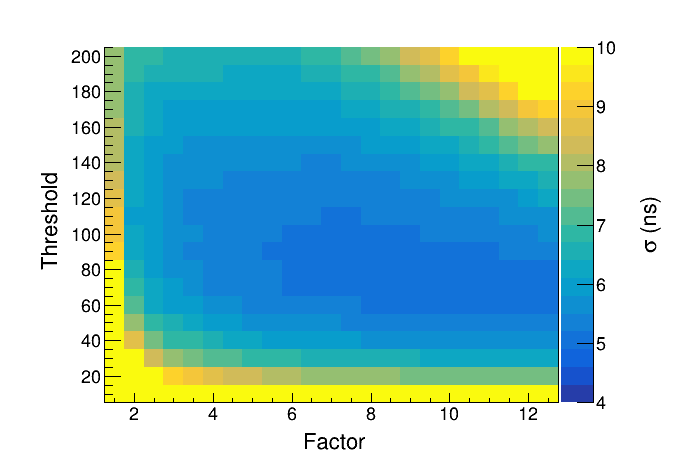}
             \caption{$Ex=1.1$, final parameters $Th=80$, $f=8.0$ selected}
         \end{subfigure}
        
        \caption{Minimization procedure used for the ELET parameters. The final chosen parameters are $Th=80$, $f=8.0$, $Ex=1.1$.}
        \label{fig:exp_optimize_ELET}
    \end{figure}

    For the optimization, the following steps are taken: 1) Using $Ex=0$, find a good initial value for $Th$ by plotting the time resolution as a function of $Th$ and $f$. 2) Using the initial value for $Th$, find the optimal value for $Ex$ by plotting the time resolution as a function of $Ex$ and $f$. 3) Using the optimal value of $Ex$ find the best value for $Th$ and $f$. The optimization sequence is shown for one the detectors in Figure~\ref{fig:exp_optimize_ELET}.

    For BEGe (Broad Energy Germanium) detectors, a more manual approach had to be taken due to their peculiar waveforms. In such detectors, multiple groups of rise times occur due to the slow charge collection in combination with photons converting at different places in the detector. Consequently, the fitted time resolutions are not always reliable. In order to find a good set of parameters, $Ex$ was set to zero, while only small values of $Th$ and $f$ were considered. From experience, a low value of $Th$ and $f$ will be less likely to produce multiple time structures (from the different groups of rise times). Finally, the fits are checked to ensure no such time structures are present in the time difference spectrum.
    
    After time aligning all detectors, the summed germanium spectrum shows a time distribution with $\text{FWHM} \approx 13.7~\si{\nano\second}$ at $713~\si{\kilo\eV}$. It should be noted that a few prompt events \textit{leak} to earlier times due to noise triggering the crossing of the first threshold. Furthermore, some slow rise-time pulses accumulate in a slight tailing at positive times.

\subsection{Anticoincidence window}
\begin{figure}[h]
        \centering
        \begin{subfigure}[b]{0.7\textwidth}
             \centering
             \includegraphics[width=\textwidth]{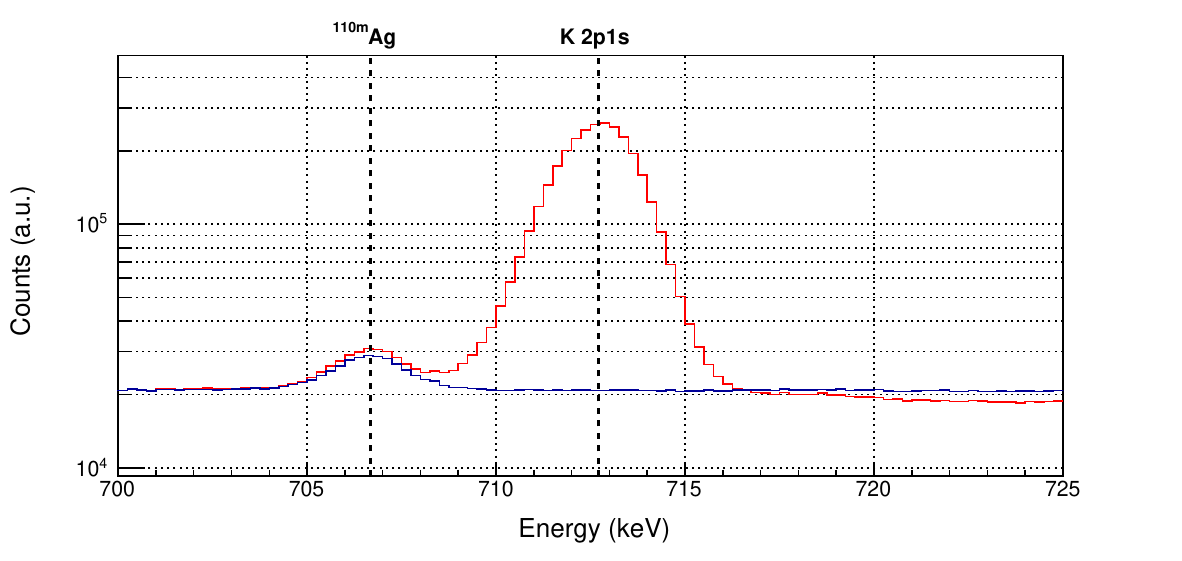}
         \end{subfigure}
         \\
         \begin{subfigure}[b]{0.7\textwidth}
             \centering
             \includegraphics[width=\textwidth]{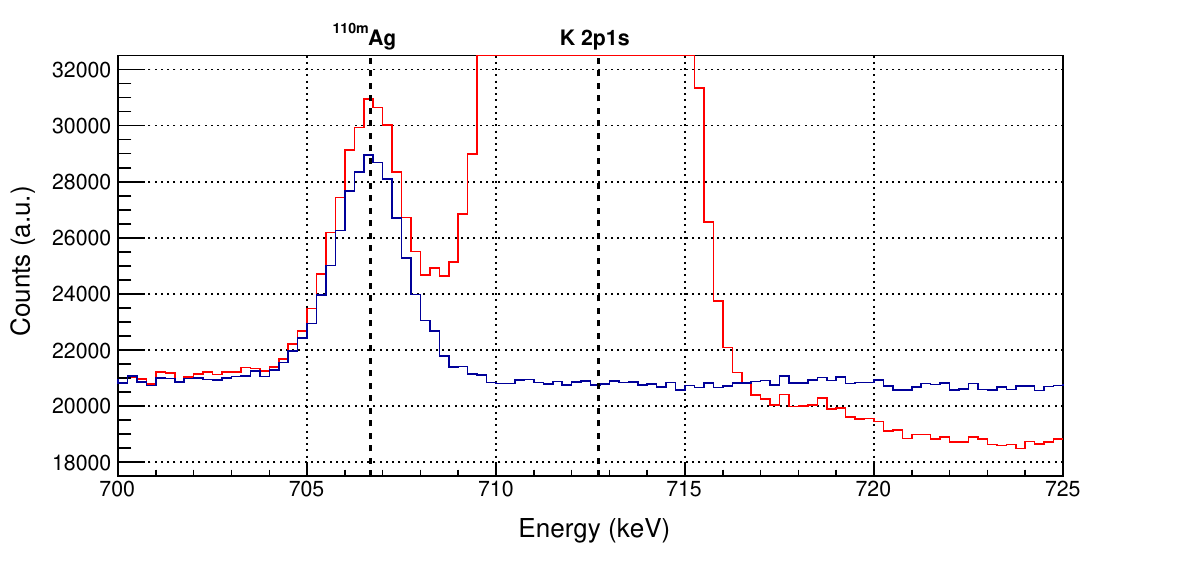}
         \end{subfigure}
        \caption{Comparison of uncorrelated (red) and anticoincidence (blue) spectrum. For the linear-scale plot, the anticoincidence spectrum was shifted upwards with a constant term to match to the uncorrelated spectrum.}
        \label{fig:exp_anticoinc_effect}
    \end{figure}

    The total energy spectra in the germanium detectors show strong peaks originating from events correlated to the muon. This makes it rather difficult to reliably fit certain calibration peaks for the gain drift correction or the final calibration due to larger background and/or underlying peaks. By constructing a spectrum of events that are anticoincident with incoming muons, the spectrum is cleaned of muon-related events, essentially resulting in a continuously running calibration measurement. An example of the effect of the anticoincidence window is shown in Figure~\ref{fig:exp_anticoinc_effect}. Here, the total spectrum in a germanium detector is compared to a spectrum constructed with events that have no muons in the window $[-3, 1]~\si{\micro\second}$ with respect to the detected photon.

    Here, the spectrum evolves from an intense $2p1s$ peak from \textsuperscript{39}K with a minute shoulder of the $708~\si{\kilo\eV}$ calibration peak to a clean spectrum of the calibration peak. It should be noted that the background for the correlated spectrum displays a significant step behavior, which is originating from the intense $2p1s$ peak. A broader anticoincidence window corresponds to fewer muon-related attributes remaining in the spectrum, primarily from non-prompt events. However, the statistics of calibration peaks are also cut. Assuming that the muons are uniformly distributed in time, one can show that the fraction of calibration events kept in the anticoincidence spectrum is given by

    \begin{equation}
        f = e^{-N_{\mu} \Delta t}.
    \end{equation}

    Where $N_{\mu}$ is the muon rate and $\Delta t$ is the width of the anticoincidence window. Since the measurements of these macroscopic targets is partially limited by calibration statistics, it remains of interest to minimize losses. By scanning the window before and after an event in a germanium detector where no muon can be present, the statistics were maximized while retaining very few muon related events.

\begin{figure}[h]
        \centering
        \begin{subfigure}[b]{0.7\textwidth}
             \centering
             \includegraphics[width=\textwidth]{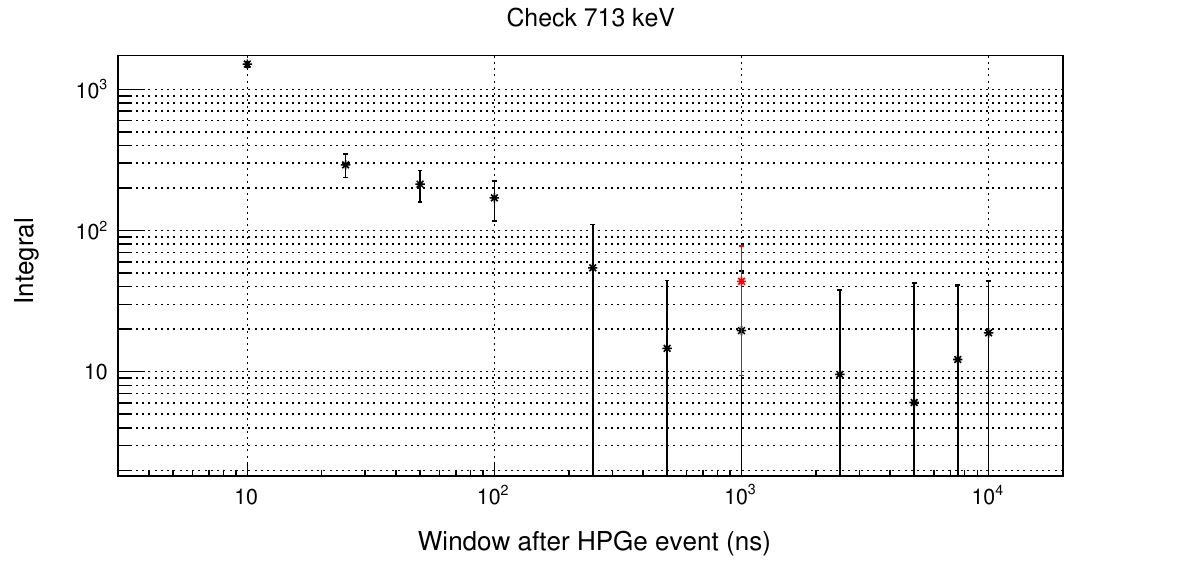}
             \caption{Muon hit after germanium hit}
         \end{subfigure}
         \\
         \begin{subfigure}[b]{0.7\textwidth}
             \centering
             \includegraphics[width=\textwidth]{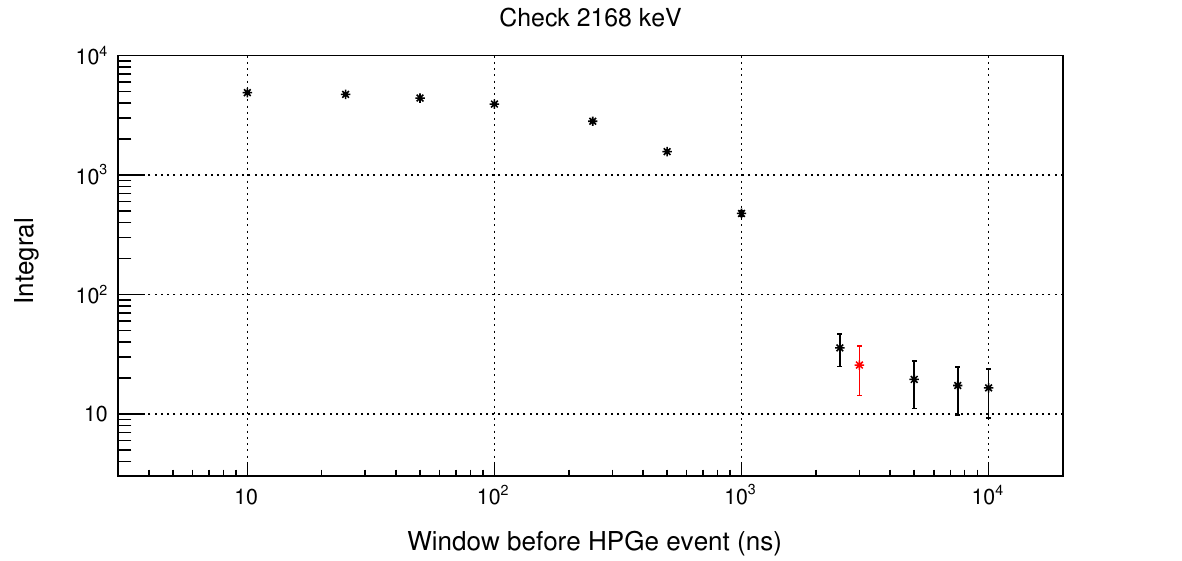}
             \caption{Muon hit before germanium hit}
         \end{subfigure}
        \caption{Window optimization for the anticoincidence spectrum. The black points show the rates when setting the opposite window to $10~\si{\micro\second}$. The red points represent the chosen anticoincidence window of $[-3, 1]~\si{\micro\second}$.}
        \label{fig:exp_anticoinc_window}
    \end{figure}

    If the event in the germanium detector occurs before the muon, no muon-related events should be present. Hence, the width of the window is only limited by the time resolution of the detector. Accordingly, the intense muonic x-ray signal at $713~\si{\kilo\eV}$ was probed from the \textsuperscript{39}KCl measurement for the time window after the photon detection. After the muon has hit the target, some capture peaks decay exponentially with the lifetime of the muonic atom in which they were caught. Such a muon capture peak can be used to optimize the time window before the photon detection. One of these capture peaks at $\sim2170~\si{\kilo\eV}$ (originating from \textsuperscript{39}K($\mu$, $\nu$ + n)\textsuperscript{38}Ar) was used for the optimization due to the low background in this region and the relatively high intensity. The number of counts is plotted as a function of the window edges in Figure~\ref{fig:exp_anticoinc_window}.

    From these plots, an anticoincidence window was set for muons of $[-3, 1]~\si{\micro\second}$ with respect to the germanium events. This window is rather conservative on the positive side in order to avoid triggers on noise in the ELET.

\subsection{Gain drift correction}
    The calibration of the used HPGe detectors varies over time due to temperature fluctuations, humidity changes, and vibrations. In order to account for these variations, a gain drift correction or gain matching is applied. For this dataset, the data were separated in two-hour blocks. For each block, the data within this time window and the neighboring time windows were taken to provide a higher statistics moving average. Next, the anticoincidence spectrum was used to fit several calibration peaks from \textsuperscript{60}Co, \textsuperscript{110m}Ag, \textsuperscript{133}Ba, and natural background. For this purpose, a Gaussian + linear fitting model was used. First, a preliminary fit was made to extract the centroid $\mu$ and width $\sigma$, before refitting starting from $\mu - \sigma$ to avoid effects induced by the low-energy tail. The resulting centroid positions were subsequently used for a linear energy calibration. This calibration does not use the full statistics, nor does it account for the full line shape. As a result, the calibration is aligned for different time slices, but not sufficiently accurate for the centroid positions.
    
\begin{figure}[h]
        \centering
        \includegraphics[width=0.7\linewidth]{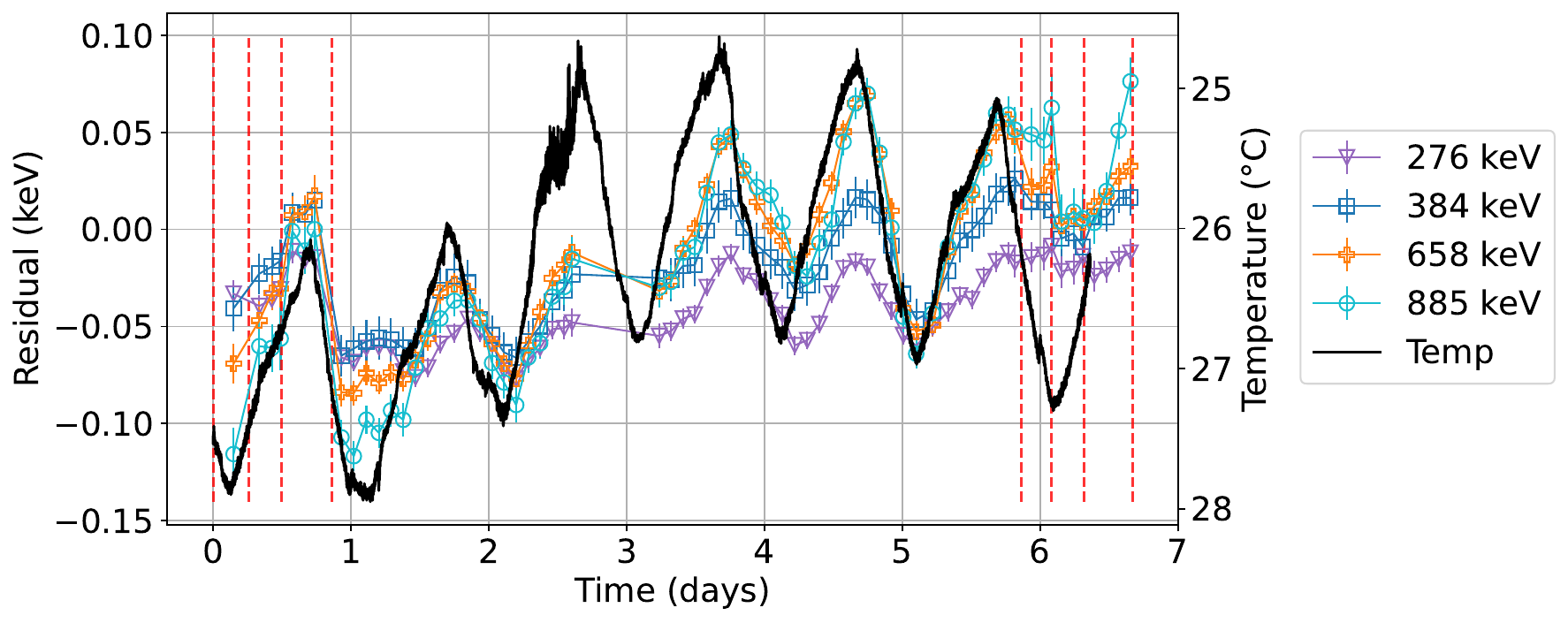}
        \caption{Residuals over time before gain drift correction in Ge02. The dashed red lines represent sample changes. (Several hours of no beam around the 3-day mark.)}
        \label{fig:exp_gaindrift_before}
    \end{figure}

    To assess the gain drift, a single calibration was first made for each detector. Next, the difference between fitted centroids with respect to their corresponding literature energies were plotted versus time, as shown in Figure~\ref{fig:exp_gaindrift_before}. The resulting graphs indicate some level of periodicity, which likely relates to the day-night cycles. As an additional diagnostic tool, the temperature at the array was plotted on a secondary axis, showing substantial correlation with the centroid position.

\begin{figure}[h]
        \centering
        \includegraphics[width=0.7\linewidth]{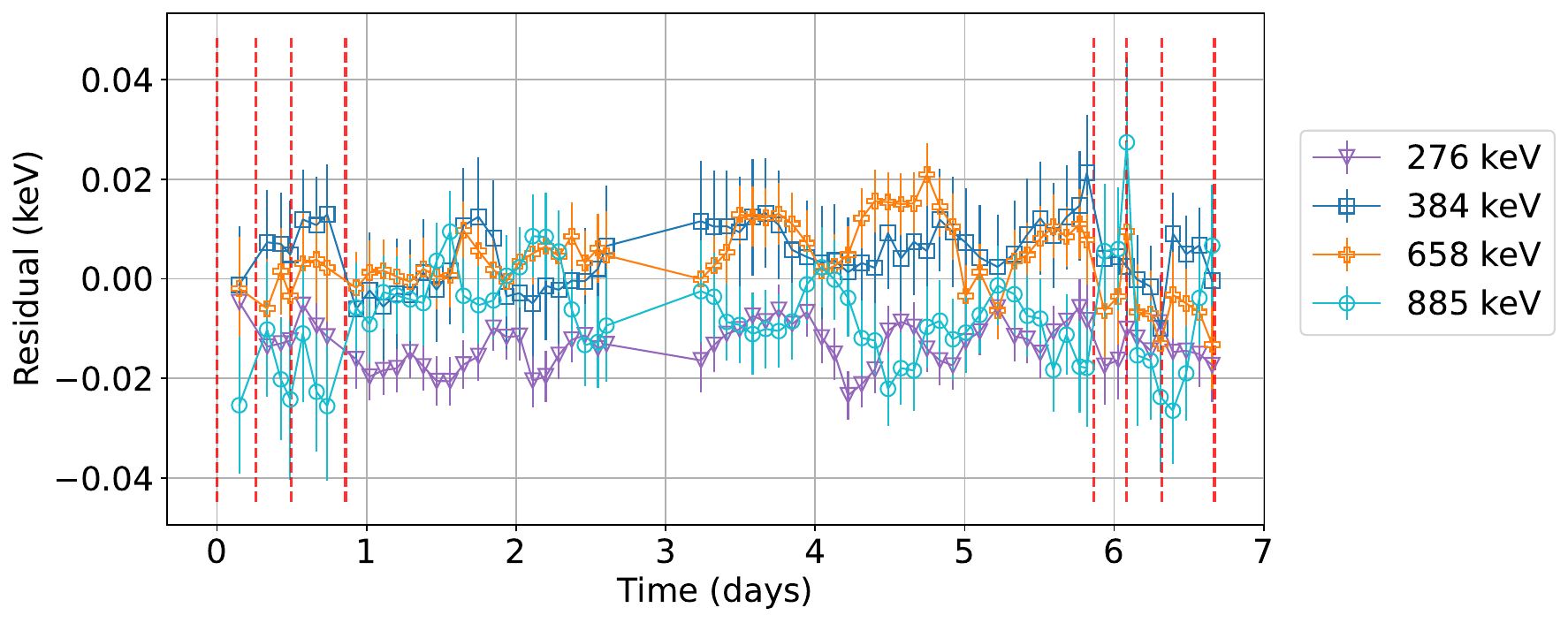}
        \caption{Residuals over time after gain drift correction in Ge02. The dashed red lines represent sample changes. There were several hours of no beam around the 3-day mark.}
        \label{fig:exp_gaindrift_after}
    \end{figure}

    For this particular detector, the displayed centroids deviate by up to $200~\si{\eV}$ at different times in the measurement. Since this is an order of magnitude larger than the precision aim of the study, a gain drift correction is critical for this analysis. The residual energies after gain drift correction are shown in Figure~\ref{fig:exp_gaindrift_after}.
    
    After the gain drift correction, the trends in time were removed almost entirely. Note that since a moving average was used to increase statistics, some autocorrelation remains between subsequent points. This leads to an apparent local trend apart from statistical fluctuation. The scatter of a particular centroid is also not necessarily centered around zero, as the calibration is only performed with a linear fit and the peaks are not fitted with the proper line shape.

\subsection{Calibration}\label{sec:exp_calibration}
    After a linear calibration is applied for the gain drift, a final calibration was performed to account for the residual differences. The procedure applied here relies on coupled fitting, making certain line shape parameters a function of energy or shared between different peaks. This is then followed by a quadratic calibration using bootstrapping. The quality of this calibration is checked by omitting one line from the energy calibration, evaluating the centroid position with similar fitting methods as used for the $np1s$ lines, and evaluating the deviation compared to the literature value. The deviation of these centroids may be used as an estimation of potential bias in the energy calibration. The calibration peaks used for these purposes are given in Table~\ref{tab:exp_calibration_lines}. Some calibration peaks are used in Section \ref{sec:exp_hypermet} and not in Section \ref{sec:exp_boostrapping}. These peaks can benefit as high-leverage points to help constrain the line shape, but they lay outside of the energy window of interest ($\sim [550, 950]~\si{\kilo\eV}$). 

    \begin{table}[h]
        \centering
        \caption{Used calibration peaks for the calibration of the germanium detectors. Energies were taken from Ref.~\cite{be2010table}.}
        \begin{tabular}{c|c|c|c|c}
            Energy ($\si{\kilo\eV}$) & Source & Resolution fit & Coupled fit & Calibration fit \\
            \hline
            356.0129(7)  & \textsuperscript{133}Ba \rule{0pt}{2.5ex}  & \checkmark & \checkmark & $\times$ \\
            583.187(2)   & \textsuperscript{208}Tl \rule{0pt}{2.5ex}  & $\times$   & \checkmark & \checkmark \\
            609.316(7)   & \textsuperscript{214}Bi \rule{0pt}{2.5ex}  & \checkmark & \checkmark & \checkmark \\
            657.7600(11) & \textsuperscript{110m}Ag \rule{0pt}{2.5ex} & \checkmark & \checkmark & \checkmark \\
            677.6217(12) & \textsuperscript{110m}Ag \rule{0pt}{2.5ex} & $\times$   & \checkmark & \checkmark \\
            687.0091(18) & \textsuperscript{110m}Ag \rule{0pt}{2.5ex} & $\times$   & \checkmark & \checkmark \\
            706.6760(15) & \textsuperscript{110m}Ag \rule{0pt}{2.5ex} & $\times$   & \checkmark & \checkmark \\
            763.9424(17) & \textsuperscript{110m}Ag \rule{0pt}{2.5ex} & $\times$   & \checkmark & \checkmark \\
            884.6781(13) & \textsuperscript{110m}Ag \rule{0pt}{2.5ex} & \checkmark & \checkmark & \checkmark \\
            937.485(3)   & \textsuperscript{110m}Ag \rule{0pt}{2.5ex} & \checkmark & \checkmark & \checkmark \\
            1173.228(3)  & \textsuperscript{60}Co \rule{0pt}{2.5ex}   & \checkmark & $\times$ & $\times$ \\
        \end{tabular}
        \label{tab:exp_calibration_lines}
    \end{table}

    \subsubsection{Coupled hypermet fitting}\label{sec:exp_hypermet}
        The individual peaks are fitted with a \texttt{hypermet} function (Eq. \eqref{eq:exp_hypermet}). 
        
        \begin{equation}\label{eq:exp_hypermet}
            f(E) = N_{sig} \left[ f_G \cdot g(E) + f_T \cdot t(E) + s(E) \right] + b(E),
        \end{equation}
    
        where
    
        \begin{align*}
            f_T &= 1 - f_G, \\
            g(E) &= \frac{1}{\sqrt{2\pi}\sigma} \exp{\left(-\frac{1}{2}\left[\frac{E - \mu}{\sigma}\right]^2\right)}, \\
            t(E) &= \frac{1}{2 \beta} \exp{\left(\frac{E - \mu}{\beta} + \frac{\sigma^2}{2\beta^2}\right)} \text{erfc}\left(\frac{E - \mu}{\sqrt{2} \sigma} + \frac{\sigma}{\sqrt{2}\beta}\right), \\
            s(E) &= \frac{A}{2} \text{erfc}\left(\frac{E - \mu}{\sqrt{2}\sigma}\right), \\
            b(E) &= \text{Linear Chebychev with integral } N_{Bg} = N_{data} - N_{sig}.
        \end{align*}
    
\begin{figure}
            \centering
            \includegraphics[width=0.6\linewidth]{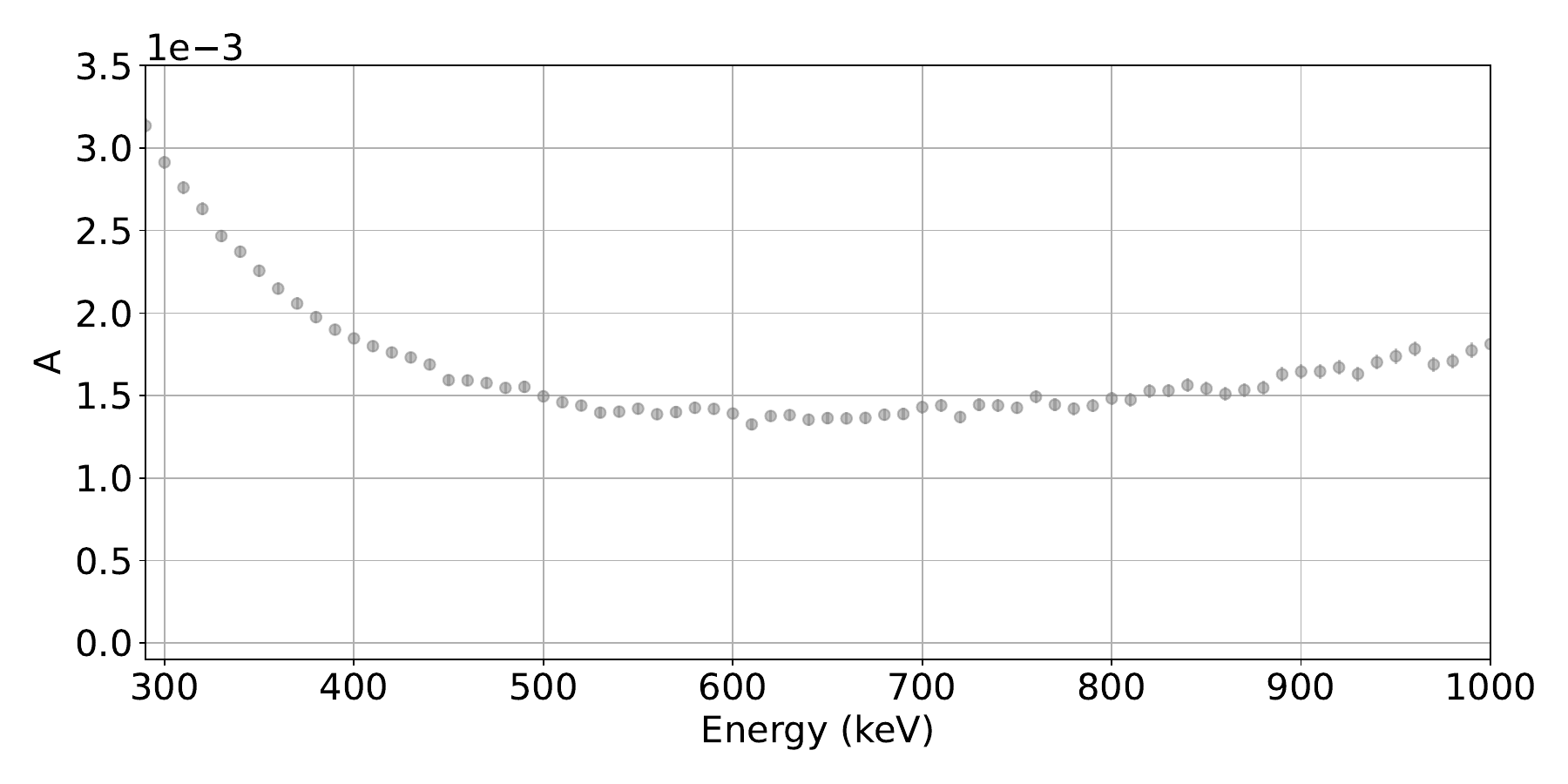}
            \caption{Geant4 simulation of the step height as a function of energy for a 75\% relative efficiency coaxial HPGe detector (Ge12).}
            \label{fig:exp_step_energy}
        \end{figure}

        While the line shape depends on energy (such that it is different depending on which peak to fit), it is possible to share some parameters between peaks. According to Ref.~\cite{campbell1997cautionary}, $\beta/\sigma$ and $f_T/f_G$ vary smoothly as a function of energy. Since a relatively narrow energy region is fitted, we can thus approximate these as linear relationships. However, the variation of $f_T/f_G$ seemed to be negligible in the fitting process, such that it could be reliably omitted. The size of the step $A$ relative to the peak integral was evaluated by keeping it at the same value for all peaks. The results showed no benefit leaving this parameter free, while complicating the fit substantially. It should be noted that this is primarily due to statistics in the calibration and that the step height is not generally constant as a function of energy. In later steps of the analysis, the step can still be varied. Figure~\ref{fig:exp_step_energy} shows Geant4 results for the step height as a function of energy, which shows a relatively constant value in the energy region of interest.
    
        For the width of the Gaussian, a first set of regular Gaussian fits was performed on \textit{clean} peaks without side structures. Similarly to the gain correction, these peaks were fitted and then refitted from $\mu - \sigma$ to avoid effects induced by the low energy tail. The resulting $\sigma (E)$ was subsequently fitted using the model given in Ref.~\cite{knoll2010radiation}:
    
        \begin{equation}
            \sigma = a + b \cdot E + c \cdot \sqrt{E}.
        \end{equation}
    
\begin{figure}
            \centering
    
            \begin{subfigure}[b]{0.65\textwidth}
                 \centering
                 \includegraphics[width=\textwidth]{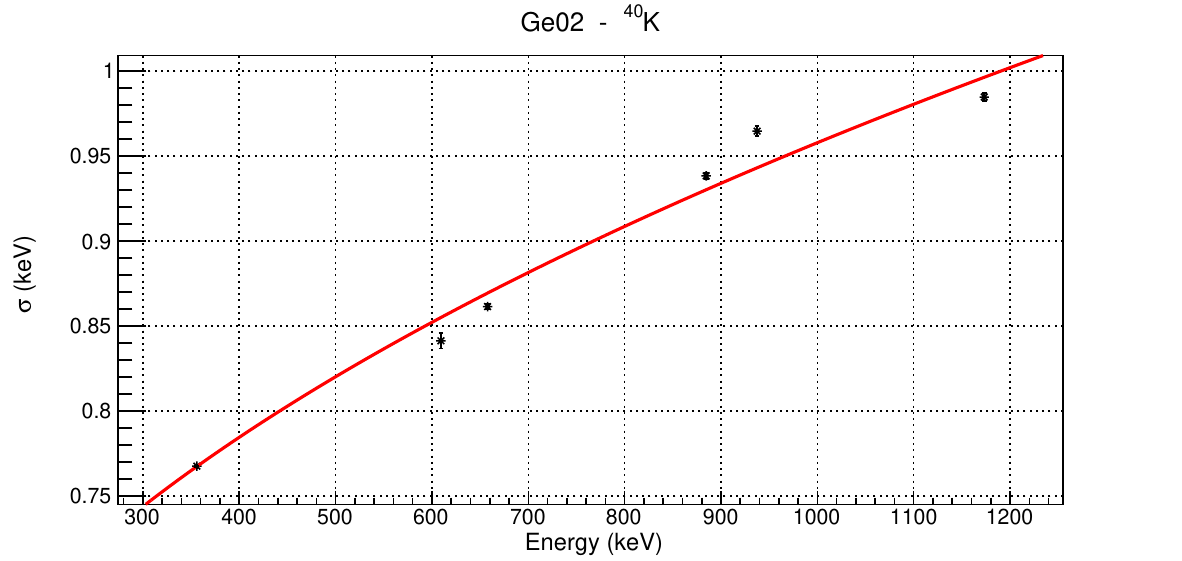}
             \end{subfigure}
             \\
             \begin{subfigure}[b]{0.65\textwidth}
                 \centering
                 \includegraphics[width=\textwidth]{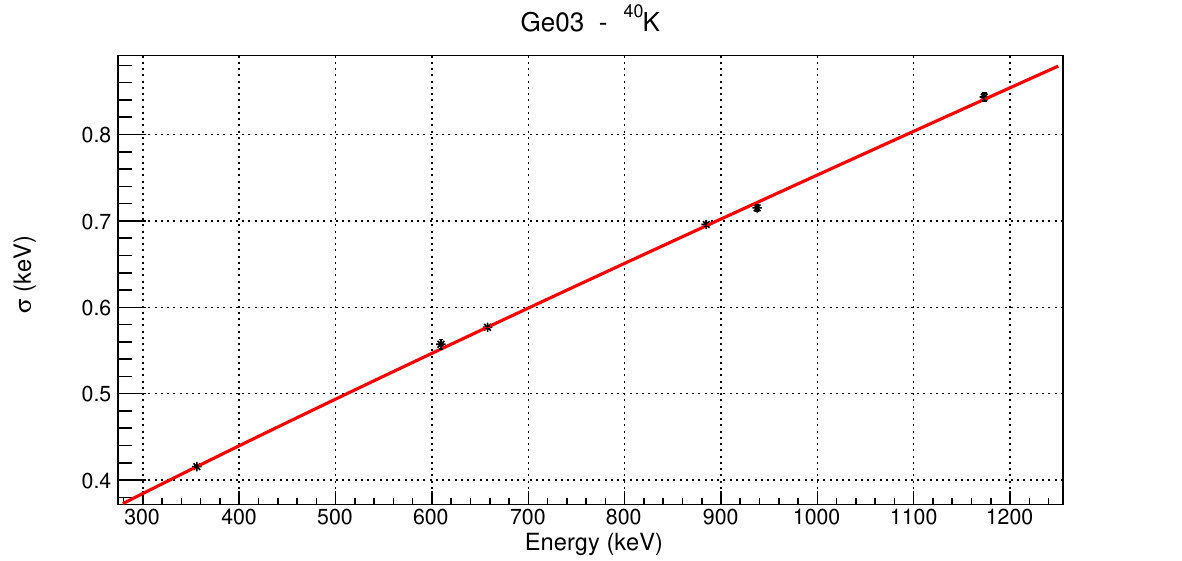}
             \end{subfigure}
    
            \caption{Fit for $\sigma$ as a function of energy for Ge02 (REGe) and Ge03 (BEGe) during the \textsuperscript{40}K run.}
            \label{fig:exp_sigma_E}
        \end{figure}

        Two examples of such fits are shown in Figure \ref{fig:exp_sigma_E} (for a BEGe and a REGe type detector). In this fit, detectors with a mostly Gaussian response provide near statistical spread in the residuals, while detectors with a more pronounced low-energy tail show a larger scatter ($< 5\%$). While the $356~\si{\kilo\eV}$ peak from \textsuperscript{133}Ba and the $1173~\si{\kilo\eV}$ peak from \textsuperscript{60}Co were not used in the final calibration, they ensure realistic values of $\sigma$ at the edges of the energy range of interest.
        
        For the fitting with the hypermet line shape, the value of $\sigma$ of the individual peaks was allowed to vary up to 5\% with respect to the calculated value at its literature energy to account for imperfections in this resolution fit. With some tuning of the initial parameters, the fit converges and provides line shape parameters and centroid positions for all of the calibration peaks. While the intense \textsuperscript{133}Ba peak at $356~\si{\kilo\eV}$ was not used in the calibration, it was included in the coupled fit to aid the convergence of the step height parameter. None of the fits converged near the extremities of the parameter windows, such that the uncertainty estimates should be appropriate.

    \subsubsection{Bootstrap fitting}\label{sec:exp_boostrapping}
        As a next step, a calibration fit was performed. If we proceed with standard least-squares fitting methods, it is assumed that the estimated centroid uncertainties are accurate. When the calibration is statistics limited, this is a valid assumption. In many cases, however, the limiting aspect of the calibration is an underlying non-linearity. As a result, the estimated confidence intervals for the fit will be poor. In order to deal with this, a non-parametric bootstrap approach was taken in combination with orthogonal distance regression (ODR) fits. In this approach, one can sample new datasets of calibration centroids from the original dataset with repetition. By doing so, the standard deviation of an observable $\sigma_{Boot}$ can be probed by varying the importance of the different points. When a sufficiently large number of data points are available, $\sigma_{Boot}$ should also fully encapsulate the statistical uncertainty of a single fit $\sigma_{Stat}$. Since this is not the case here, it may occur that $\sigma_{Stat} > \sigma_{Boot}$. Accordingly, the uncertainty of the calibration is determined at every energy using
    
        \begin{equation}
            \sigma_{tot} = Max \{\sigma_{Boot}, \sigma_{Stat}\}.
        \end{equation}
    
\begin{figure}
            \centering
            \begin{subfigure}[b]{0.45\textwidth}
                 \centering
                 \includegraphics[width=\textwidth]{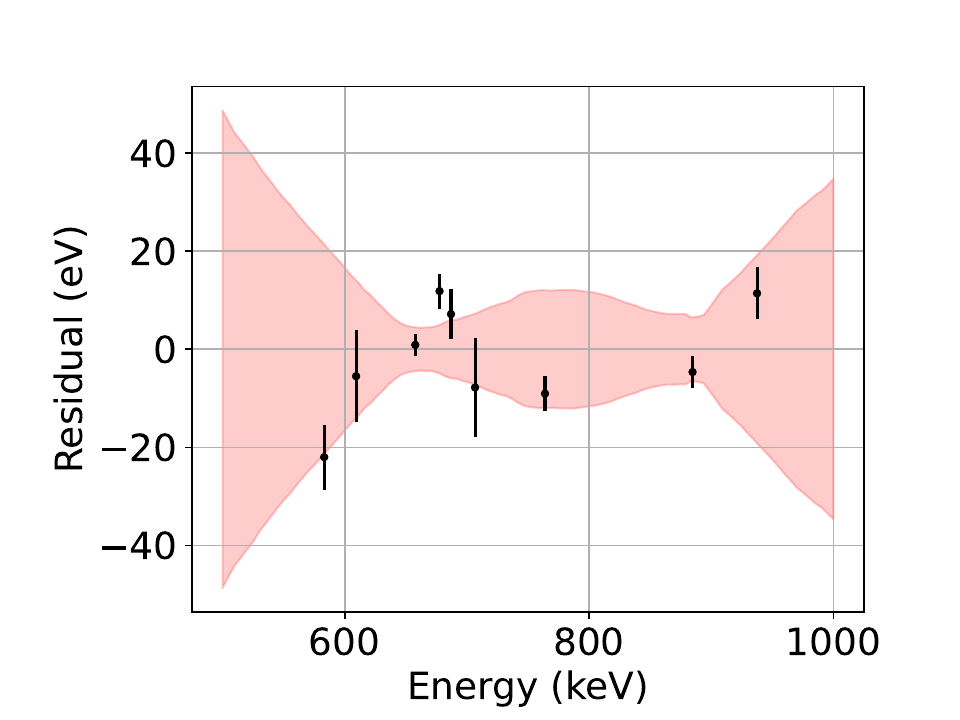}
                 \caption{Ge03}
             \end{subfigure}
             \qquad
             \begin{subfigure}[b]{0.45\textwidth}
                 \centering
                 \includegraphics[width=\textwidth]{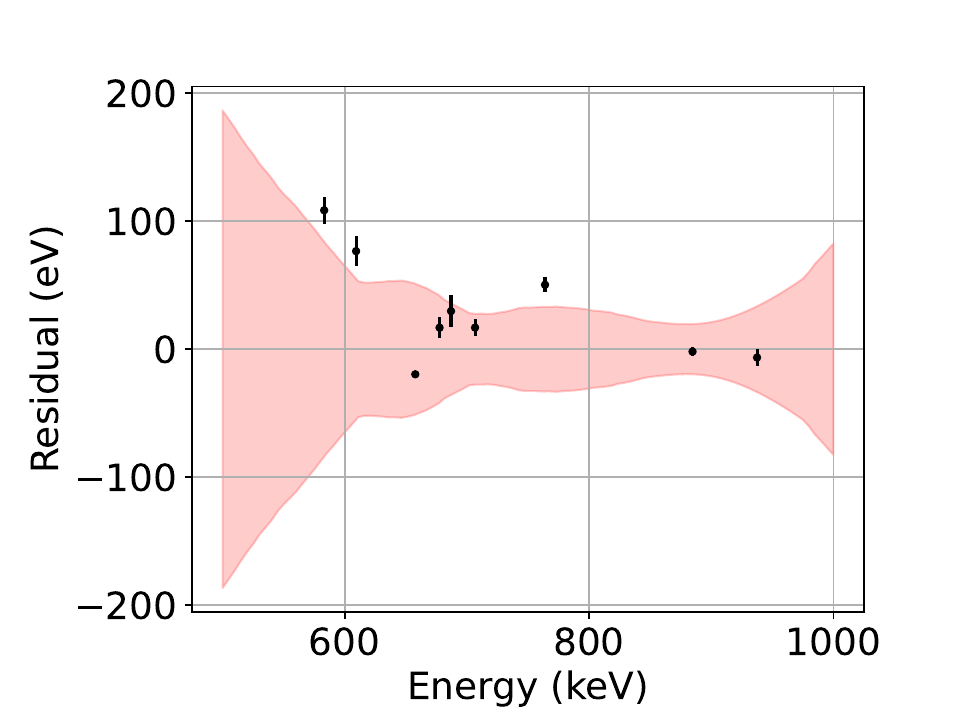}
                 \caption{Ge06A}
             \end{subfigure}
    
            \caption{Residuals (black points) and predicted error (shaded red region) for the calibration of Ge03 (BEGe) and Ge06A (TIGRESS-type clover) during the \textsuperscript{40}K run. Error bars contain the statistical uncertainty on the fitted centroid and the literature uncertainty on the calibration peaks.}
            \label{fig:exp_calibration}
        \end{figure}

        The residuals of such a calibration fit in two of the detectors are shown in Figure~\ref{fig:exp_calibration} together with the predicted uncertainty. Visually, the estimated uncertainties seem reliable for both detectors. Depending on the non-linearity of the detector and the input range for the ADC, the precision of the energy calibration changes. The two detectors shown are: Ge03 (BEGe) with good energy precision and Ge06A (TIGRESS-type clover) with a poorer energy precision. The calibration uncertainty was below $50~\si{\eV}$ in the region of interest for most detectors.
    
\subsection{Spectrum clean up}
    \subsubsection{Rejection logic}
        In order to clean up the prompt spectra, several rejections can be made. The cuts applied for this dataset are: 

        \begin{enumerate}
            \item Muon pile-up protection cut
            \item Electron veto cut
            \item Multiplicity cut
        \end{enumerate}
    
        The muon beam at PSI is approximately uniformly distributed in time ($\sim20~\si{\nano\second}$ between proton pulses, $\mathcal{O}(1~\si{\milli\second}$) average time between subsequent muons). When muons arrive close to one another, a germanium event may be assigned to the \textit{wrong} muon. This, for example, leads to prompt events being present far from $t=0$. In order to remove such events from the spectrum, muons that have another muon within the range $[-3, +3]~\si{\micro\second}$ are rejected. This window has a half-width equal to about twice the muon half-life, such that also most longer-lived time structures are removed.
        
\begin{figure}
            \centering
            \includegraphics[width=0.5\linewidth]{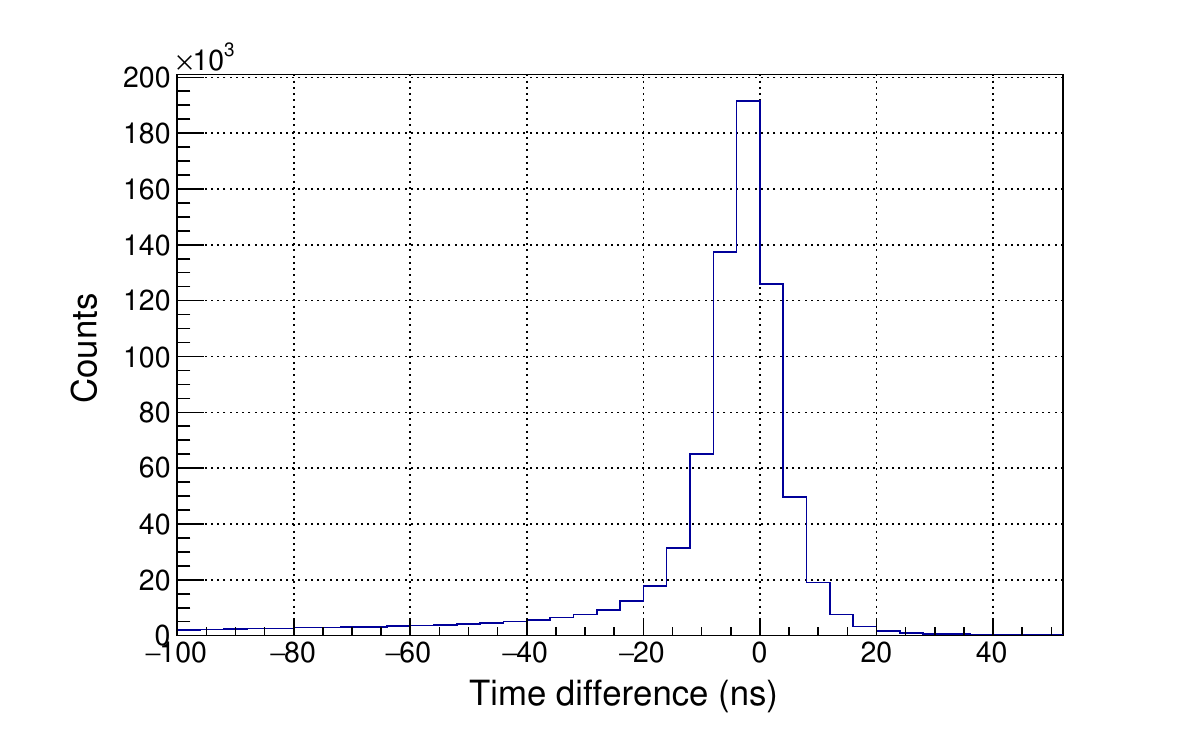}
            \caption{Time difference between germanium events and electron events where the germanium energy is in the range $[600, 1000]~\si{\kilo\eV}$.}
            \label{fig:exp_electron_timediff}
        \end{figure}

        When a muon decays to an electron, it provides the electron with up to $52.8~\si{\mega\eV}$ of kinetic energy. Such electrons induce background in the HPGe detectors that is not relevant to this work. By placing a set of scintillators around the measured sample, it is possible to tag events that are coincident with an electron, such that they can be removed from the spectrum. To find a good time window, the time difference spectrum was made for photon energies in the range $[600, 1000]~\si{\kilo\eV}$ (above the $511~\si{\kilo\eV}$ peak). Based on Figure \ref{fig:exp_electron_timediff} a veto window of $[-50, 20]~\si{\nano\second}$ was chosen.
    
        Finally, a multiplicity cut was made for multi-segmented detectors (Miniball and clover-type detectors). In these detectors, a photon may Compton scatter between the different segments, such that the detection has a multiplicity higher than one. Most commonly, the energy deposited in the different crystals is summed (add-back), but this slightly increases the width of the peaks and worsens the energy calibration. Since the energy calibration is critical for this work, multiplicity $>1$ events of multi-segmented detectors were rejected.

    \subsubsection{Coincidence time cut}\label{sec:exp_time_cut}
\begin{figure}
            \centering
            \includegraphics[width=0.5\linewidth]{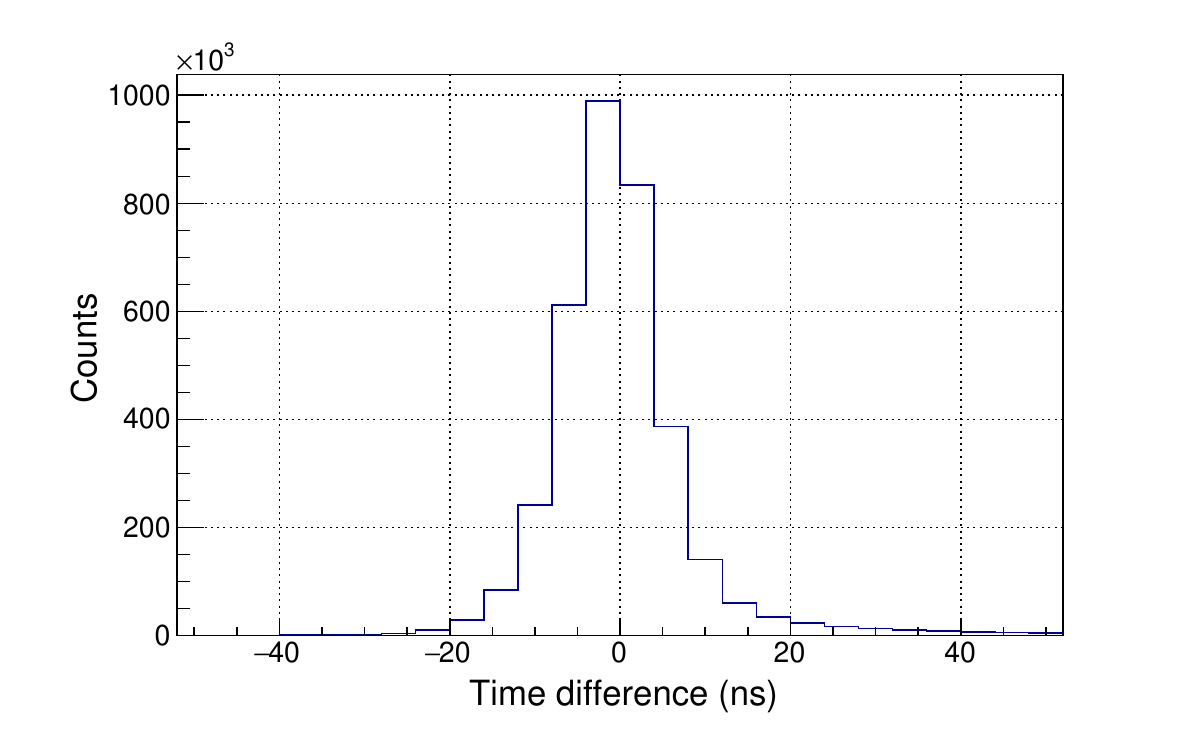}
            \caption{Time difference of germanium events with respect to the trigger muon in the range $[710, 715]~\si{\kilo\eV}$.}
            \label{fig:exp_timediff}
        \end{figure}

        Once the calibration is performed, a time cut was made which contains only few non-coincident events. A large fraction of the background at the $np1s$ peaks for these data is prompt with respect to the muon. For the \textsuperscript{35}Cl measurement, the primary background originates from Compton scattering of higher $np1s$ peaks, while for the \textsuperscript{37}Cl measurement it is mostly induced by Compton scattering of higher energy Ag muonic x rays. Since much of the background is prompt, and thus cannot be differentiated from the signal via a time cut, the exact boundaries of the time cut are not critical. However, broader windows induce a larger background from calibration sources and muon capture peaks. To find a good time cut, the time difference spectrum was evaluated around the $2p1s$ peak of \textsuperscript{39}K, which is shown in Figure~\ref{fig:exp_timediff}. Based on this spectrum, a time window of $[-25, 25]~\si{\nano\second}$ was chosen.
    
\subsection{Fitting approach}\label{sec:exp_fitting_approach}
    Once the spectra are filtered and a calibration has been made, the $np1s$ peaks can be fitted. For this work, the fitting was performed through Bayesian inference using a Poisson likelihood in the python package \texttt{pymc}. The general idea of a Bayesian approach is to insert some prior knowledge into all parameters. The goal of the fit is to find the probability distribution for the values a parameter is most likely to be (posterior distribution). In most cases, no exact solution exist, such that a Monte Carlo sampling approach is employed in dedicated packages. This section describes the fitting process and the chosen priors.

    Priors can be chosen in the form of an informative prior or in the form of an uninformative prior. If there is a good physical reason that a parameter should not deviate substantially from some value, for example the energy difference or amplitude ratio between two peaks in a doublet, one can opt for an informative prior. It is possible to overturn this prior if the fit shows a clear preference further away from the center of the prior distribution. In conventional least-squares and likelihood fitting approaches, such information is typically transformed into either fixing a parameter or setting a narrow window in which it may vary. This also intrinsically invokes prior knowledge, but in a less-controlled manner. Completely fixing such parameters may lead to an underestimation of the uncertainty of other parameters in the fit. Alternatively, it is possible to set an uninformative prior. While a completely uninformative prior is not straightforward to set, one can choose the prior distribution sufficiently wide such that the posterior distribution becomes data dominated.

    The fit model used was the same hypermet function as in Section \ref{sec:exp_hypermet}. For the calibration peaks, the background is also taken to be linear, recentered around the expected peak position $E_0$. The background in such case becomes

    \begin{equation}
        b(E) = c_0 + c_1 (E - E_0).
    \end{equation}
    
    This recentering reduces the correlation between $c_0$ and $c_1$ which improves the sampling process, without changing the fit model mathematically. For the muonic x-ray peaks, the background can be very low, such that linear background may result in the fit model becoming negative in the region of interest. When using a Poisson likelihood, this results in divergences, which is undesirable. To resolve this, these peaks were fitted with a constant background, deemed sufficient for these cases. 
    
    Many of the parameters cannot physically be negative. Accordingly, these parameters were evaluated using Gamma distribution priors, as such a distribution is only non-zero at positive values. For the line shape, certain parameters should be constrained in order to reduce potential divergences. Since experimenting with priors on $\beta/\sigma$ and $f_T/f_G$ showed slow convergence, they were taken to be fixed to the value obtained from the hypermet fit from Section \ref{sec:exp_hypermet}. Tests indicated that this assumption was sufficient to the current level of precision. To help the choice of priors, an initial least-squares fit was performed using a Gaussian + linear model. The resulting value and error of parameters in this initial fit are noted as $param'$ and $\delta param'$. Parameters that are taken from the coupled hypermet fit are noted as $param''$.
    
    The remaining parameters in the model are: The Gaussian width ($\sigma$), the step height $A$, the Gaussian mean ($\mu$), the integral of the main peak ($I$), and the constant and linear components of the background ($c_0$ and $c_1$). In the case of peak doublets, two additional parameters are needed: The difference in centroid energy between the two peaks $d\mu$ and the ratio of amplitudes $side$. The choice of prior on these last two parameters is taken to be more informative for the fitting of the $np1s$ peaks, which will be discussed in more details in Section \ref{sec:exp_np1s}. The prior choices made for this work are:

    \begin{itemize}
        \item $\sigma$: Gaussian width should not be negative. Gamma distribution with mean equal to the calculated value from the $\sigma(E)$ relation and standard deviation equal to 5\% (same as in Section \ref{sec:exp_hypermet}).
        \item $A$: Uniform distribution in the range $[A''/4,~ 4 A'']$. This parameter may slightly vary as a function of energy, but due to limited statistics, this relation could not be extracted from the coupled hypermet fit (Section \ref{sec:exp_hypermet}).
        \item $\mu$: This should act as an uninformative prior. Normal distribution with mean $\mu'$ and standard deviation $Max\{0.5~\si{\kilo\eV},~ 2\delta \mu' \}$. 
        \item $I$: Integral should not be negative. This should act as a mostly uninformative prior. Gamma distribution with mean $I'$ and standard deviation $Max\{I'/5,~ 2\delta I' \}$. 
        \item $c_0$: Background must be positive in the full range. This should act as an uninformative prior. Gamma distribution with mean $c_0'$ and standard deviation $Max\{c_0'/5,~ 2 \delta c_0'\}$. 
        \item $c_1$: This should act as an uninformative prior. Normal distribution with mean $c_1'$ and standard deviation $Max\{c_1'/2, ~ 2 \delta c_1'\}$. 
        \item $d\mu$: Generous interval allowing for potential errors in literature values. Uniform distribution in the range $[d\mu_{lit} - 50~\si{\eV},~ d\mu_{lit} + 50~\si{\eV}] $.
        \item $side$: This should fully cover the range of reasonable values. Using a uniform distribution in the range $[1\mathrm{e}{-6},~ 0.5]$ leads to significant side peaks, while remaining less than half of the height of the main peak.
    \end{itemize}

    Using the above priors, separate fits are made for each detector. Next, the extracted centroids are converted to $\si{\kilo\eV}$ units using the results from Section \ref{sec:exp_calibration}. The statistical error of the fit and the calibration error from the bootstrapping are added in quadrature for the total uncertainty $\sigma_{meas}$. Finally, a weighted average is taken over the different detectors. If $\chi_{\nu}^2$ is larger than one, the error is scaled with $\sqrt{\chi_{\nu}^2}$ to capture potential missing sources of uncertainty. Such cases are likely induced by shared ADC components, leading to a significant correlation between extracted energies in different detectors.

\subsection{Calibration reliability}\label{sec:exp_bias}
    While the bootstrapping should produce an accurate uncertainty for each detector separately, there may be a correlation between the errors of the different detectors (e.g., shared FPGA chips). Consequently, the weighted average over multiple detectors can give an inaccurate result. Using the bootstrapping methods in Section \ref{sec:exp_boostrapping} and the fitting approach in Section \ref{sec:exp_fitting_approach}, one can try to estimate an additional bias error. In order to do so, the following process was performed for each calibration peak: 1) The bootstrap calibration fit is made for all detectors using all calibration peaks except for one. 2) Bayesian fits are performed  with \texttt{pymc} following the methods described in Section \ref{sec:exp_fitting_approach}. 3) A weighted average is made over all detectors. Once this has been done for all calibration peaks, one can check how much the literature values deviate from the predicted values. This should provide an estimate to the underestimation in error made by taking the weighted average.

    We define the \textit{deviation} $\delta$ as the difference between the predicted energy of the calibration peak (from the weighted average) and its literature value. The uncertainty of the deviation is given by $\sigma_{tot} = \sqrt{\sigma_{meas}^2 + \sigma_{lit}^2}$. In order to estimate a bias uncertainty, the \textit{excess} $\epsilon$ is defined as

    \begin{equation}
        \epsilon = 
        \begin{cases}
            0 \text{, if $|\delta| < \sigma_{tot}$} \\
            |\delta - \sigma_{tot}| \text{, else}
        \end{cases}
    \end{equation}
    
    Information about the deviation of different calibration peaks is given in Table \ref{tab:exp_deviation}. Since the lowest and highest energies are substantially outside of the calibration range, the excess is expected not to be representative. Accordingly, they are not taken into account for an estimate of the bias error.
    
    \begin{table}[h]
        \centering
        \caption{Deviation, excess, and reduced chi square obtained through the procedure described in the text. Averaging over 19 detectors ($\nu=18$).}
        \begin{tabular}{c|c|c|c|c}
            Energy (keV) & Source & Deviation (eV) & Excess (eV) & $\chi_{\nu}^2$\\
            \hline
            583.187(2)   & \textsuperscript{208}Tl \rule{0pt}{2.5ex}
            & 53(11) & 41.61 & 1.43 \\
            609.316(7)   & \textsuperscript{214}Bi \rule{0pt}{2.5ex}
            & $-$26(10) & 16.40 & 1.08 \\
            657.7600(11) & \textsuperscript{110m}Ag \rule{0pt}{2.5ex}           
            & $-$5(6) & 0.0    & 2.92 \\
            677.6239(12) & \textsuperscript{110m}Ag \rule{0pt}{2.5ex}          
            & $-$10(4) & 6.58  & 0.62 \\
            687.0114(18) & \textsuperscript{110m}Ag \rule{0pt}{2.5ex}
            & $-$5(4) & 0.32   & 1.20 \\
            706.6780(15) & \textsuperscript{110m}Ag \rule{0pt}{2.5ex}
            & 17(4) & 13.12  & 1.32 \\
            763.9452(17) & \textsuperscript{110m}Ag \rule{0pt}{2.5ex}
            & $-$13(8) & 4.97  & 2.81 \\
            884.6819(13) & \textsuperscript{110m}Ag \rule{0pt}{2.5ex}
            & $-$1(20) & 0.0   & 0.085 \\
            937.485(3)   & \textsuperscript{110m}Ag \rule{0pt}{2.5ex}
            & 6(24) & 0.0    & 0.32
        \end{tabular}
        \label{tab:exp_deviation}
    \end{table}

   From the results, it seems that the uncertainties for higher energies may be overestimated given their corresponding $\chi_{\ni}^2$. To assign an error to the bias in the averaging process, two approaches were evaluated. The first method estimates the bias by taking the average of the excesses, which results in a value of $6.64~\si{\eV}$. A second method estimates the bias by taking the maximal excess divided by two, which results in a value of $8.20~\si{\eV}$. The two approaches provide very similar values. As a conservative estimate, the bigger of the two values is taken.

    A final uncertainty to take into account is one that represents the error from the literature energies. Most of the energy differences are known to a higher precision than the absolute number. Accordingly, the additional error is taken to be the uncertainty of the best-known calibration peak ($1.1~\si{\eV}$). This uncertainty, as well as the bias uncertainty should only affect the absolute energy and not small differences in energy (e.g., isotope shifts).

\subsection{np1s fitting}\label{sec:exp_np1s}
    \subsubsection{The choice of priors}
        In this section, the fitting of the $np1s$ peaks will be explained. Here, we need to use an approach which is similar to that applied for the bias investigation in Section \ref{sec:exp_bias}. Since the background does not vary significantly over energy in these spectra, the linear component of the background is omitted. Due to the fine structure splitting, all peaks that will be fitted consist of doublets that are not fully resolved. Here, $\mu$ is taken to be the $np$\textsubscript{3/2}$1s$ energy. In contrast to the calibration peaks, the difference in energy between the peaks of the doublet is not known experimentally to a high precision. However, atomic calculations can predict the fine structure splitting and the amplitude ratio between the peaks. The predicted fine structure splitting from these calculations is given in Appendix~\ref{app:exp_hyperfine}, while the $2p$\textsubscript{1/2}$1s$/$2p$\textsubscript{3/2}$1s$ amplitude ratio is equal to 0.5 following a statistical distribution. However, both of these may still vary, for example due to nuclear polarization or intensity changes in the hyperfine levels. The natural line width for these systems was estimated to be about $10~\si{\eV}$, which was deemed insignificant compared to the resolution of the detector to have a significant effect on the fits.
    
        Next, we want to check how closely the fine structure splitting and amplitude ratio reproduce experiments. Or alternatively, to what experimental precision we can evaluate the quality of the calculations. This was done by using the detector with the highest energy resolution (Ge03) to fit the $2p1s$ of \textsuperscript{39}K, which has the largest fine structure splitting and the highest statistics, with broad prior distributions for $d\mu$ and $side$. The findings from this check can then be used in order to construct more informative priors for the actual $np1s$ fits without fixing the parameters to their theoretical value. This is particularly important in the cases where the peaks become less resolved, such as for the $np1s$ peaks in Cl.
        
        During this check, the priors were taken to be Gamma distributions, similar to those described in Section~\ref{sec:exp_fitting_approach}, with standard deviations equal to $50~\si{\eV}$ and 0.05 for $d\mu$ and $side$, respectively. For these tests, no nuclear polarization was taken into account as its effect is deemed negligible compared to the experimental accuracy for the fine splitting. Additionally, the fine structure was assumed to be constant as a function of radius, and the hyperfine structure was neglected. The fit results show that both $d\mu$ and $side$ are within $1\sigma$ from the atomic theory calculations. The resulting standard deviation on $d\mu$ and $side$ are approximately equal to $5~\si{\eV}$ and 0.0065. As conservative estimates, the standard deviation of the priors was taken to be $10~\si{\eV}$ and 0.010 for the remainder of the analysis. The fit was then performed again with adjusted narrower priors, showing no significant bias.

    \subsubsection{Extraction of energies}
        During the fitting process, it was observed that $\mu$ has a significant correlation with $d\mu$ and $ratio$. This is because the model cannot identify the exact position of the individual peaks. This issue can be overcome by introducing the center-of-gravity energy $E_p$ using
    
        \begin{equation}
            E_p = \frac{\mu + ratio \times (\mu - d\mu)}{1 + ratio}.
        \end{equation}
    
\begin{figure}[h]
            \centering
            \includegraphics[width=0.9\linewidth]{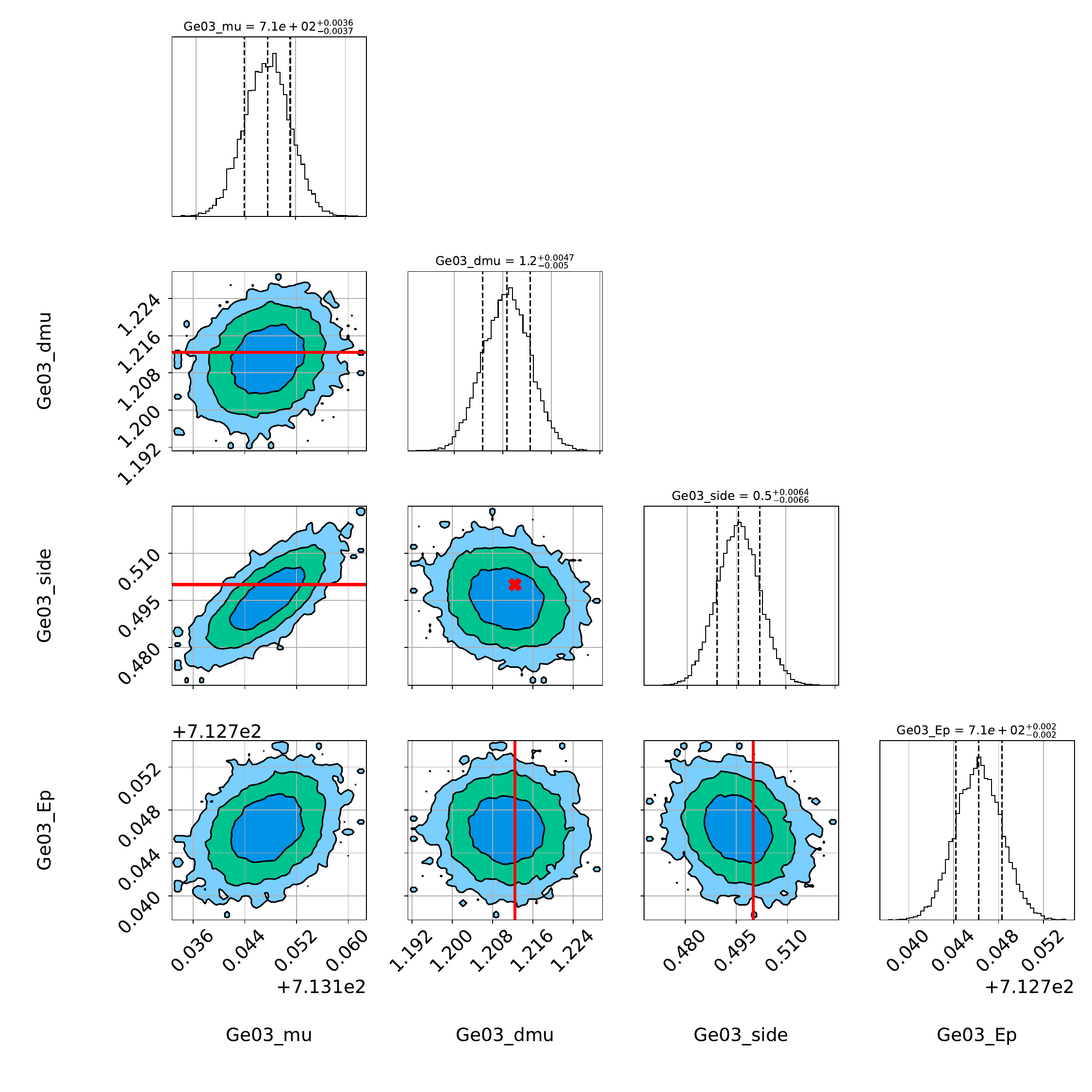}
            \caption{Cornerplot of fitresults for a $2p-1s$ fit on \textsuperscript{39}K including $E_p$ with very loose fine structure constraints. Red marker and lines indicate theory values for fine structure. Contours mark the 1, 2, and 3$\sigma$ intervals.}
            \label{fig:exp_K_loosePrior}
        \end{figure}

\begin{figure}[h]
            \centering
            \includegraphics[width=0.9\linewidth]{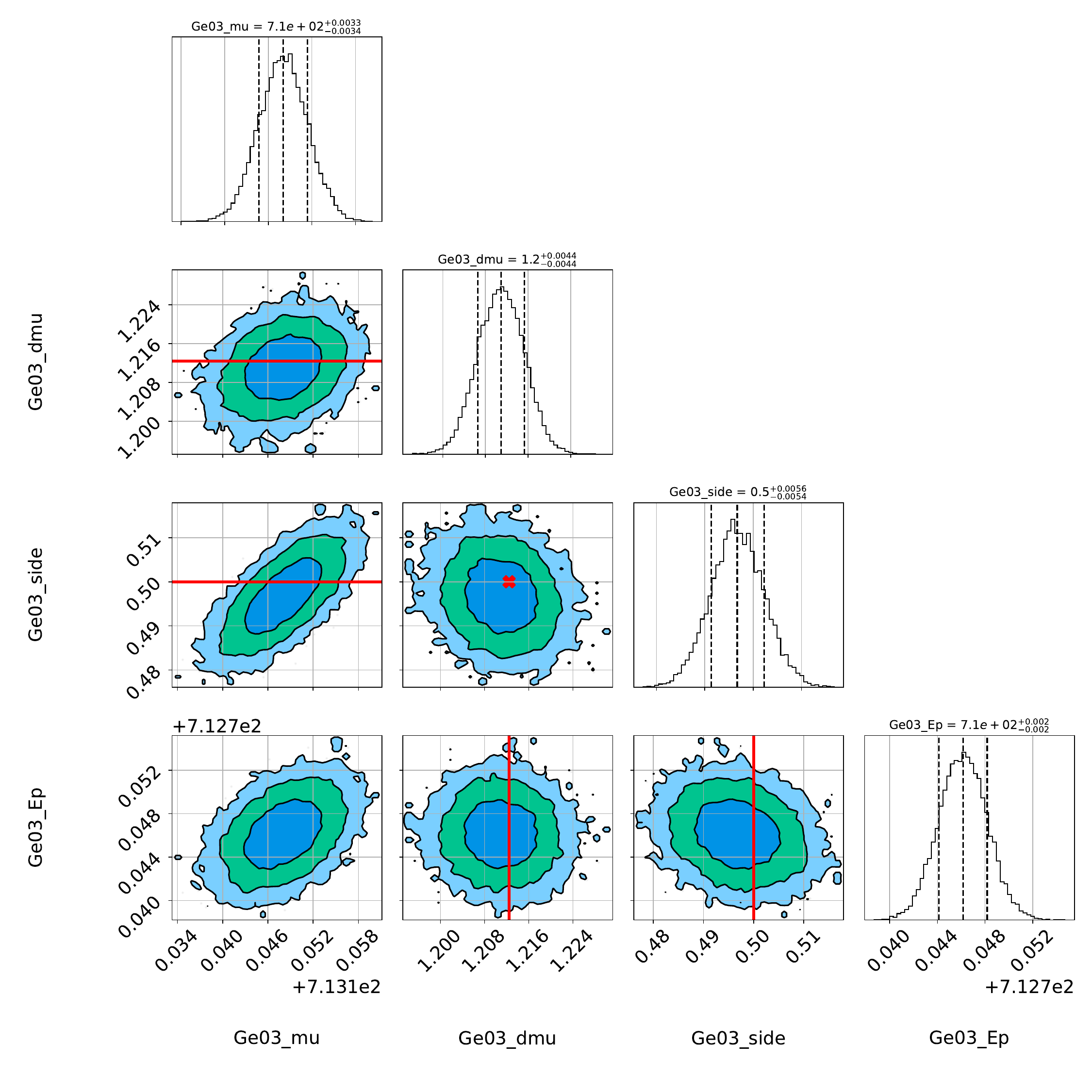}
            \caption{Cornerplot of fitresults for a $2p-1s$ fit on \textsuperscript{39}K including $E_p$ with strict fine structure constraints. Red marker and lines indicate theory values for fine structure. Contours mark the 1, 2, and 3$\sigma$ intervals.}
            \label{fig:exp_K_strictPrior}
        \end{figure}

        This quantity is a more direct representation of the experimental data. As a result, it shows less sensitivity to prior choices, less correlation to $d\mu$ and $ratio$, and provides lower uncertainties. The corner plots comparing the broad and narrow priors for \textsuperscript{39}K are shown in Figures~\ref{fig:exp_K_loosePrior}, and \ref{fig:exp_K_strictPrior}. When evaluating the broad priors, the uncertainty on $E_p$ compared to the uncertainty on $\mu$ is about a factor two lower for \textsuperscript{39}K (and a factor four when repeating the check for \textsuperscript{35}Cl). It should also be noted that this effect is expected to increase with decreasing $d\mu/\sigma$. This means that the improvement is more significant for detectors with a lower resolution, or for peaks with a smaller fine structure splitting (higher $n$). This effect, however, reduces when choosing narrower priors thus relying more on QED input.
    
        % K loose priors
        % K strict priors

    \subsubsection{Effect of time cut}
\begin{figure}
            \centering
            \begin{subfigure}[b]{0.45\textwidth}
                 \centering
                 \includegraphics[width=\textwidth]{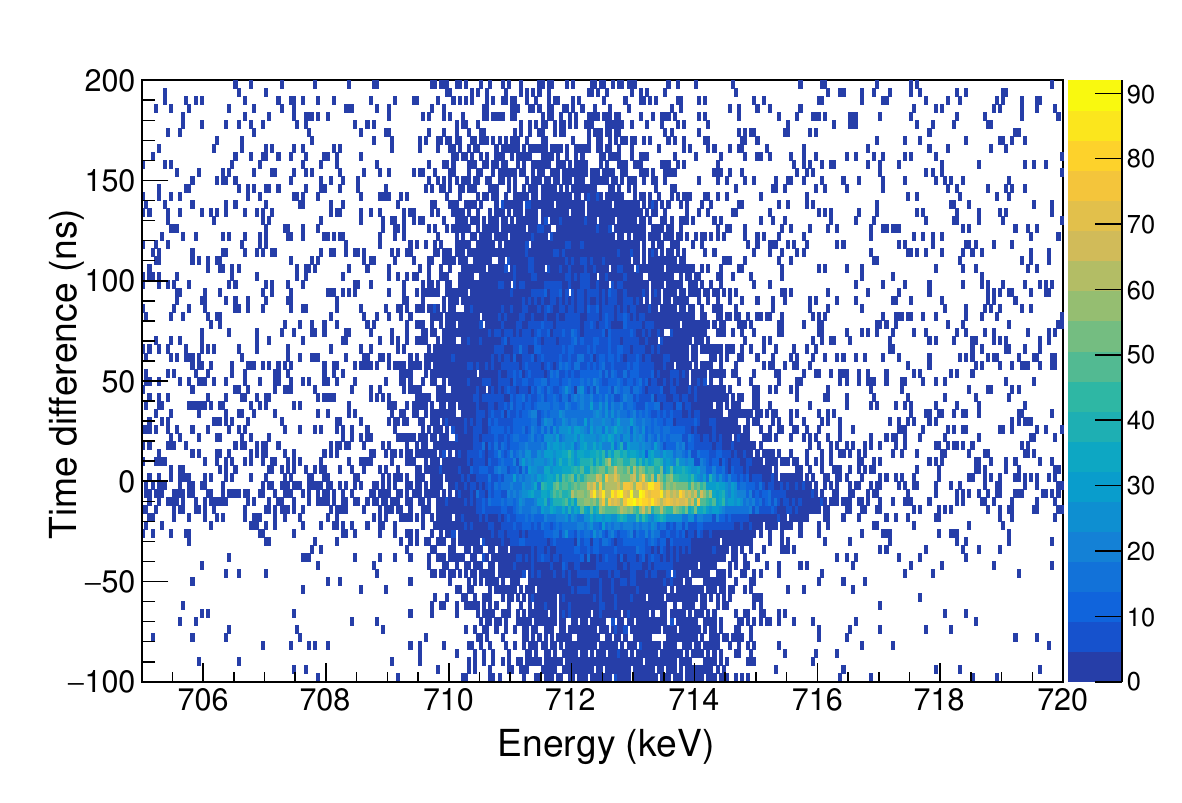}
                 \caption{}
             \end{subfigure}
             \qquad
             \begin{subfigure}[b]{0.45\textwidth}
                 \centering
                 \includegraphics[width=\textwidth]{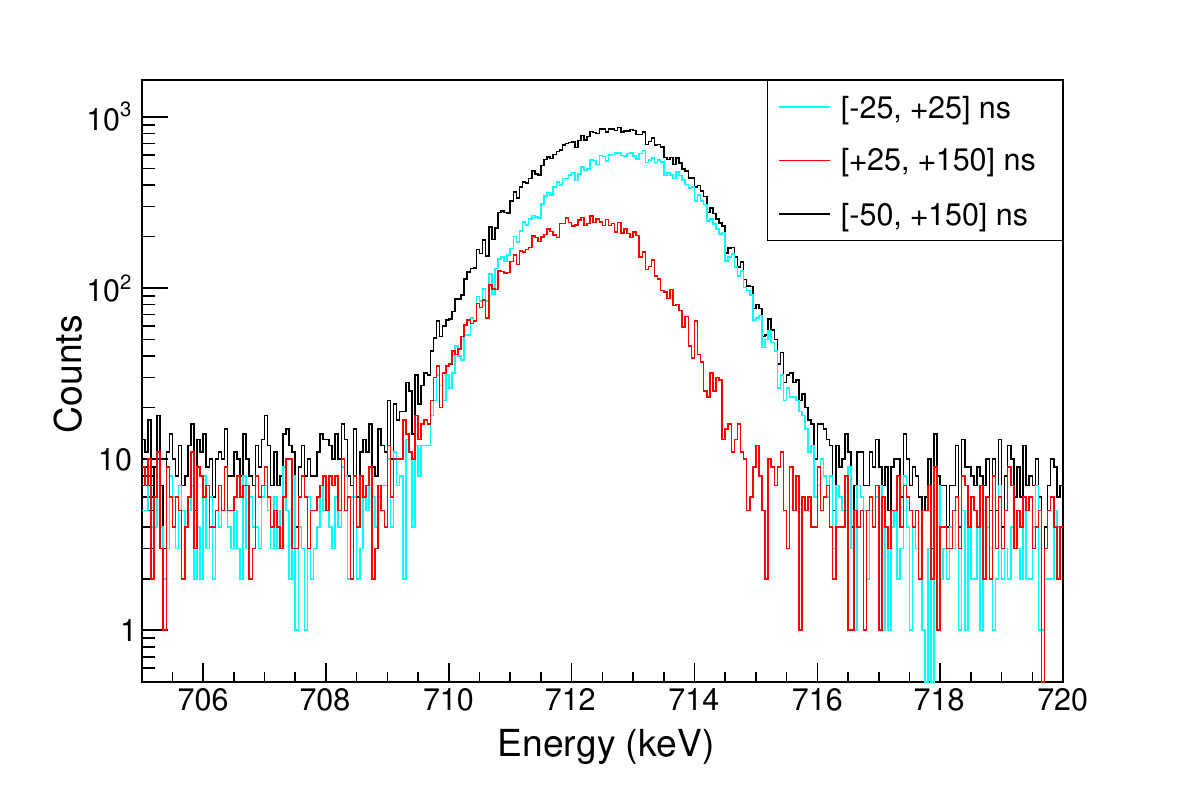}
                 \caption{}
             \end{subfigure}
    
            \caption{Spectra showing the effect of the time cut on the centroid position. The energy versus time plot (a) shows a smaller centroid at later times. Different time slices are shown in (b).}
            \label{fig:exp_timecut_systematics}
        \end{figure}

        The energy versus time spectrum shows a slight shift in centroid energy based on the time difference with the incoming muon. This is displayed in Figure~\ref{fig:exp_timecut_systematics} for a BEGe~3820 type HPGe detector. As will later be shown, this particular detector (Ge09) suffers the most from this systematic effect. 
    
        The shift is induced by slower rise time events. A pulse with a slow rise time intrinsically reaches a lower maximum amplitude, such that it is assigned a smaller energy by the ADC. Additionally, timing algorithms assign times that are delayed compared to their fast rise time equivalent pulses. As such, the time cut leads to a selection of a sub-ensemble of rise times. Since no such rise time cut is made for the calibration sources, a systematic shift $\delta_{Cut}$ will be present in the extracted energies, described by
    
        \begin{equation}
            \delta_{Cut} = E_{Cut} - E_{True},
        \end{equation}
    
        where $E_{Cut}$ is the energy extracted in the chosen time cut (Section \ref{sec:exp_time_cut}) and $E_{True}$ is the energy that would be extracted without a time cut. Since a time cut remains necessary to have clean spectra, a broad time cut ($[-50, 150]~\si{\nano\second}$) was chosen rather than a spectrum without any cut. The energy extracted from this time cut should closely resemble $E_{True}$. The effect of $\delta_{Cut}$ is expected to be most prominent for detectors with large rise time variation and bad timing resolutions (e.g., BEGe detectors). Since the time resolution improves with higher energies, the correction is expected to reduce with energy. For this work, the correction was assumed to be a linear function in energy for simplicity, such that
    
        \begin{equation}
            \delta_{Cut} = a_0 + a_1 E.
        \end{equation}
    
        It is possible to probe (and correct) for this systematic shift by comparing the extracted centroid between the broad and narrow time cuts for high-statistics data. The peaks used to extract $a_0$ and $a_1$ were the $2p1s$, $3p1s$, and $4p1s$ of \textsuperscript{35}Cl as well as the $2p1s$ and $3p1s$ of \textsuperscript{39}K. The $4p1s$ of \textsuperscript{39}K was not used, as the broad time cut features a minor muon capture peak near the peak of interest. The resulting values for $a_0$ and $a_1$ are shown in Table \ref{tab:exp_shift_timecut}. The statistical uncertainty of this fit was substantially lower than the other sources of uncertainty, so that it was omitted for this analysis.
    
        \begin{table}
            \centering
            \caption{Parameters extracted for the systematic shift based on the choice of time window.}
            \begin{tabular}{c|cc|c}
                Detector & $a_0$ & $a_1$ & shift in \textsuperscript{39}K \\
                & $(\si{\eV})$ & ($10^{-5}$)& $2p1s$ $(\si{\eV})$ \\
                \hline
                Ge01  & 145.6(111) & $-5.97(157)$ & 103.0(11) \\
                Ge02  & 56.6(111)  & $-6.35(156)$ & 11.3(6) \\
                Ge03  & 86.1(108)  & $-6.08(154)$ & 42.7(8) \\
                Ge04  & 15.6(52)   & $-1.45(75)$  & 5.3(4) \\
                Ge05  & 40.4(31)   & $-4.12(47)$  & 11.0(5) \\
                Ge06A & 14.9(118)  & $-2.02(168)$ & 0.5(9)\\
                Ge06B & -0.5(105)  & $-0.08(162)$ & -1.1(17) \\
                Ge06C & 16.3(155)  & $-2.1(22)$   & 1.1(10) \\
                Ge06D & -12.7(122) & $1.63(177)$  & -1.1(12) \\
                MB07A & 1.0(85)    & $0.13(126)$  & 1.9(10) \\
                MB07B & 14.8(67)   & $-2.09(95)$  & -0.2(4) \\
                MB07C & 21(24)     & $-2.5(34)$   & 2.6(12) \\
                Ge08  & 53.6(43)   & $-5.74(68)$  & 12.7(8) \\
                Ge09  & 236(24)    & $-8.5(32)$   & 175(3) \\
                Ge10  & 43.4(71)   & $-4.42(104)$ & 12.0(8) \\
                Ge11  & 45.6(44)   & $-3.64(66)$  & 19.7(6) \\
                Ge12  & 27.7(49)   & $-1.82(73)$  & 14.8(6) \\
                Ge13  & 52.0(40)   & $-4.18(62)$  & 22.3(7) \\
                Ge14  & 63.4(83)   & $-5.39(118)$ & 25.0(6)\\
            \end{tabular}
            \label{tab:exp_shift_timecut}
        \end{table}
    
\begin{figure}
            \centering
            \includegraphics[width=0.7\linewidth]{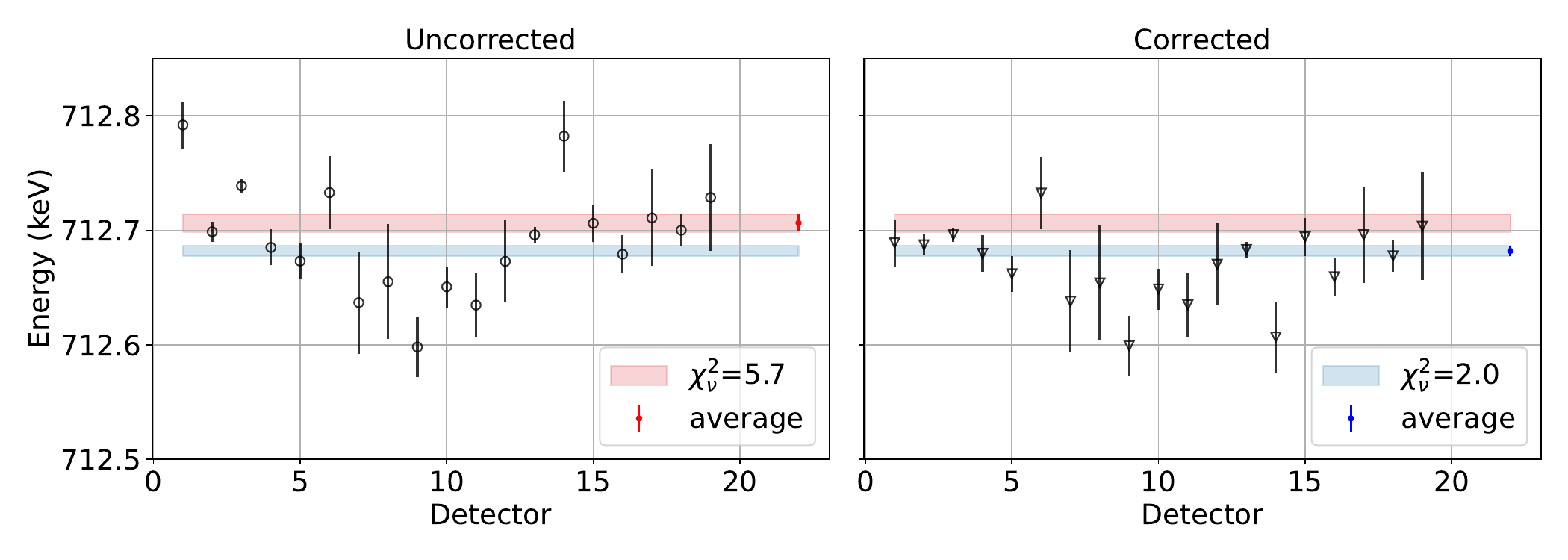}
            \caption{Weighted average of the $2p1s$ center-of-gravity energy of \textsuperscript{39}K before and after correcting for the applied time cut of $[-25, +25]~\si{\nano\second}$. BEGe detectors are the most affected by this correction (detector numbers 1, 3, and 14).}
            \label{fig:exp_av_systematics}
        \end{figure}

        The largest differences occur for the three BEGe detectors (Ge01, Ge03, and Ge09), which shift by $\mathcal{O}(100~\si{\eV})$. However, many of the other detectors shift by $\mathcal{O}(10~\si{\eV})$.  Figure~\ref{fig:exp_av_systematics} displays the weighted average for the \textsuperscript{39}K $2p1s$ energy including and excluding the correction. By correcting the energies before averaging, the $\chi_{\nu}^2$ of the average reduces from 5.7 to 2.0 indicating a much better agreement between different detectors.
    
    \subsubsection{Fit results} \label{sec:exp_energies}
\begin{figure}
            \centering
            \includegraphics[width=0.7\linewidth]{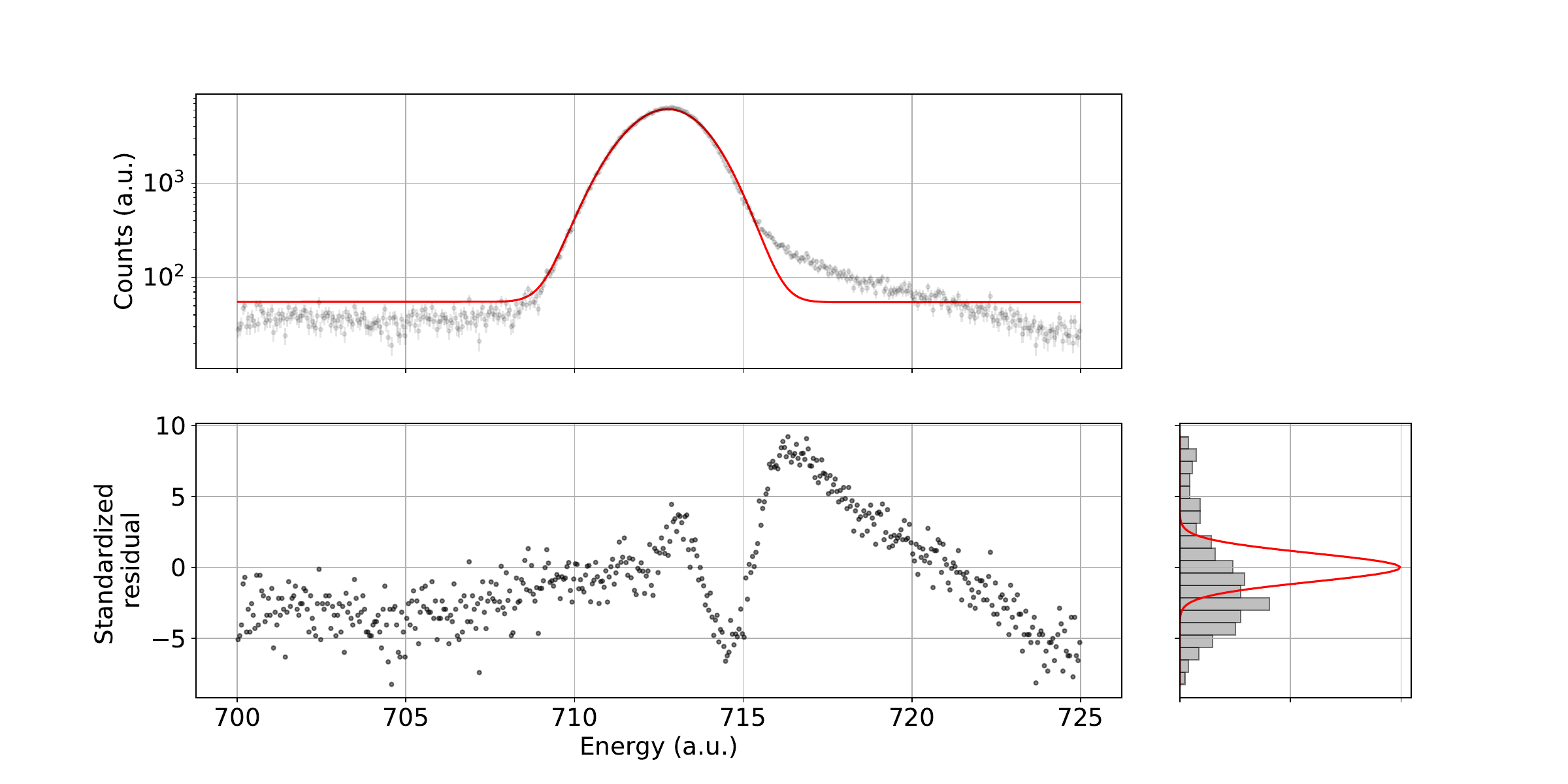}
            \caption{Energy spectrum of $2p1s$ in \textsuperscript{39}K in Ge08, showing the high energy tail.}
            \label{fig:exp_highE_tail_2p1s}
        \end{figure}

\begin{figure}
            \centering
            \includegraphics[width=0.7\linewidth]{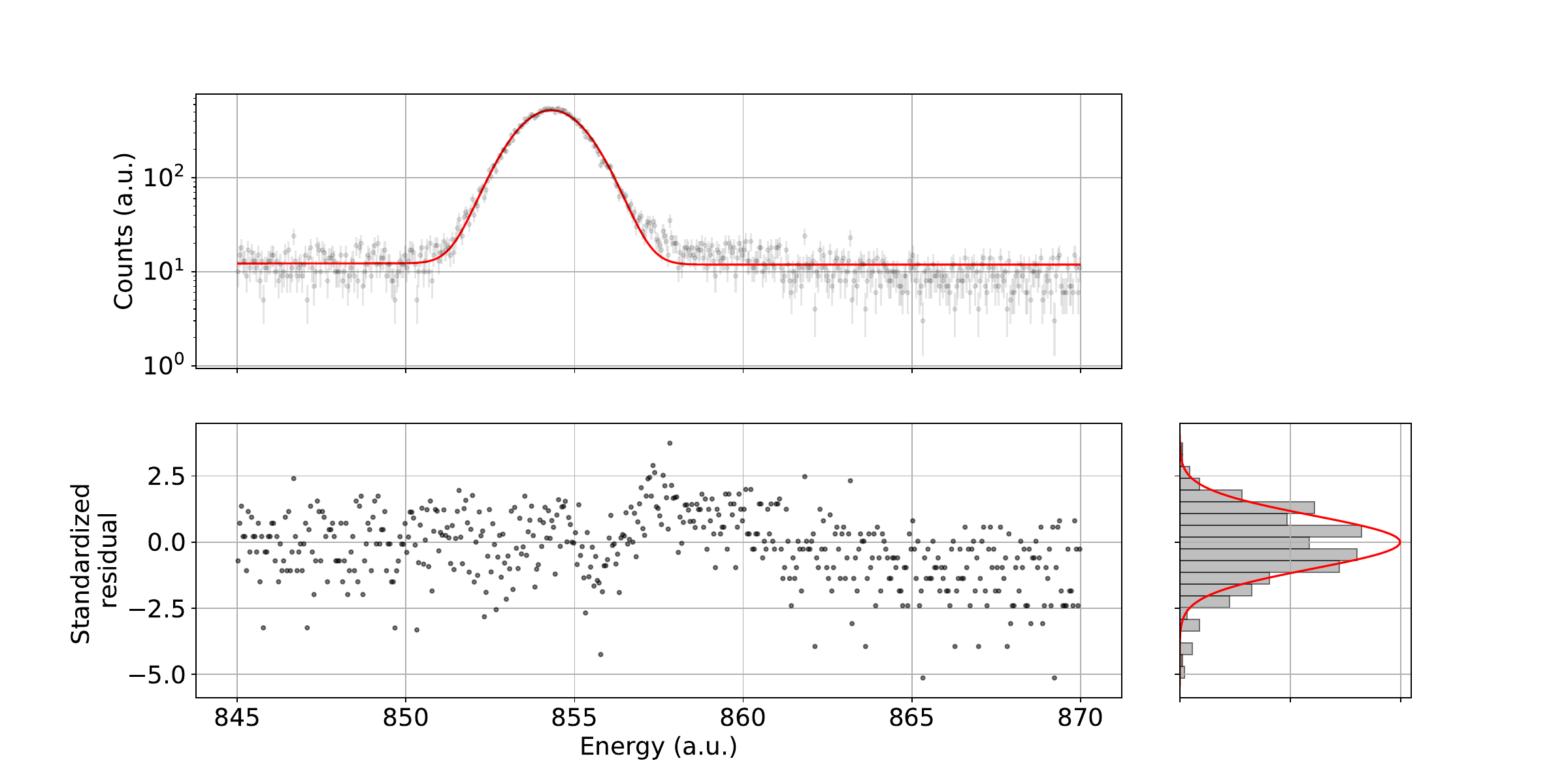}
            \caption{Energy spectrum of $3p1s$ in \textsuperscript{39}K in Ge08, showing much less significant tailing at high energy.}
            \label{fig:exp_highE_tail_3p1s}
        \end{figure}

        With these methods established, the muonic x-ray peaks of interest were fitted. The $2p1s$, $3p1s$, and $4p1s$ are visible without significant overlap for both Cl isotopes. For \textsuperscript{37}Cl, more care has to be taken, as the 7f3d peaks from silver (AgCl) are less than $10~\si{\kilo\eV}$ below the $2p1s$ of Cl. The fit described the data extremely well for all detectors except for Ge08 (100\% Coaxial HPGe detector), which displayed a high-energy tail. This is likely an electronic noise effect, as there were some issues with its preamplifier before the experiment. This tail is relatively minor, and as such, is only significant in measurements with very high statistics (such as \textsuperscript{39}K). The spectrum of Ge08 is shown for the $2p1s$ and $3p1s$ fit of \textsuperscript{39}K in Figure~\ref{fig:exp_highE_tail_2p1s} and Figure~\ref{fig:exp_highE_tail_3p1s}, respectively. Other detectors did not show such a high-energy tail, leading to the corresponding fit describing the data almost perfectly.
    
        We want to show the total sum of all of our data, which is done by summing all of the fit models and all of the data. For this visualization, Ge08 is not taken into account, in order to assess the quality of the other fits. The resulting fits for the $np1s$ peaks of \textsuperscript{35, 37}Cl are shown in Appendix~\ref{app:exp_fits}. For higher statistics peaks, slight wings can be spotted in the residuals, which are likely induced by residual gain drift, or other broadening effects. An overview of the extracted centroid energies is given in Table~\ref{tab:exp_fitresults}. The total $\chi^2$ of the fits can be calculated by summing the $\chi^2$ of all individual fits. For $\chi_{\nu}^2$, the sum of all data points in each detector and the sum of all non-floating parameters has to be taken into account. The corresponding $\chi_{\nu}^2$ of the fits are quoted both including and excluding Ge08 to probe the effect of its high energy tail.

        \begin{table}
            \centering
            \caption{Fit results of the $np1s$ energies (in $\si{\kilo\eV}$) amd isotope shifts (in $\si{\eV}$). The round brackets ( ), square brackets [ ], and curly brackets \{ \} represent the total measurement uncertainty (fit + calibration), bias uncertainty, and the uncertainty on the best known peak, respectively. Averaging over 19 detectors ($\nu=18$).}
            \begin{tabular}{cc|c|cc|c}
                Isotope & Transition & Energy & Fit $\chi_{\nu}^2$ & Fit $\chi_{\nu}^2$ & Averaging \\
                & & & All & No Ge08 & $\chi_{\nu}^2$ \\
                \hline
                
                \textsuperscript{35}Cl \rule{0pt}{2.5ex} & $2p1s$ & 578.8682(130)[82]\{11\} & 1.36 & 1.13 & 2.53 \\
                                       & $3p1s$ & 692.0950(66)[82]\{11\}  & 0.70 & 0.68 & 1.36 \\
                                       & $4p1s$ & 731.6640(94)[82]\{11\}  & 0.73 & 0.73 & 1.23 \\
                \hline
                \textsuperscript{37}Cl \rule{0pt}{2.5ex} & $2p1s$ & 578.7349(87)[82]\{11\}  & 1.15 & 1.13 & 1.70 \\
                                       & $3p1s$ & 691.9973(154)[82]\{11\} & 1.10 & 1.10 & 0.60 \\
                                       & $4p1s$ & 731.549(264)[82]\{11\}  & 0.98 & 0.96 & 0.85 \\
                \hline
                \textsuperscript{37}Cl - \textsuperscript{35}Cl \rule{0pt}{2.5ex} & $2p1s$ & $-$141.8(115) & / & / & 0.38 \\
                                                                & $3p1s$ & $-$100.4(166) & / & / & 0.58 \\
                                                                & $4p1s$ & $-$114.1(278) & / & / & 0.78 \\
            \end{tabular}
            \label{tab:exp_fitresults}
        \end{table}

        It should be noted that for the high intensity peaks, the calibration uncertainty dominates the measurement uncertainty. While the statistics of the $np1s$ peaks are lower for the \textsuperscript{37}Cl measurement, it displayed lower calibration uncertainties due to the longer measurements (and hence, increased calibration statistics). A breakdown of the uncertainty in individual detectors is given in Appendix~\ref{app:exp_uncertainty}. Finally, the muonic isotope shifts can be extracted, which are also given in Table~\ref{tab:exp_fitresults}. This is done by taking the difference in energy within each detector and subsequently performing a weighted average. The resulting $\chi_{\nu}^2$ values indicate near-statistical scatter.

\subsection{Additional systematics}
    \subsubsection{Effect of hyperfine splitting}
\begin{figure}
            \centering
            \includegraphics[width=0.6\linewidth]{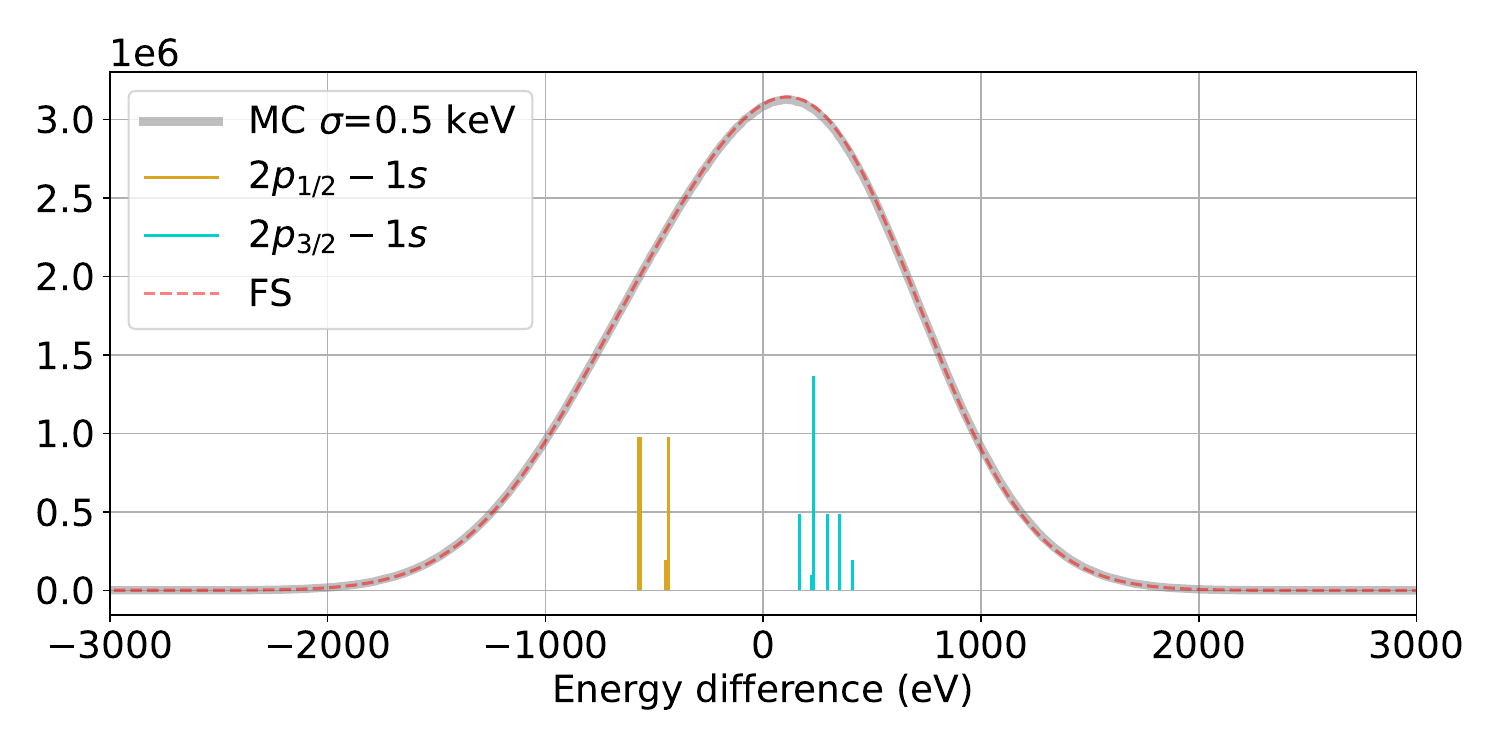}
            \caption{Monte Carlo generated hyperfine structure spectrum for \textsuperscript{35}Cl (grey). The exact line shape of the FS is plotted as a red dashed line, and the individual hyperfine peaks with their corresponding relative amplitudes are shown.}
            \label{fig:hyperfine_Cl}
        \end{figure}

        While the muonic x-ray peaks were fitted with doublets, justified by the fine structure (FS), they in principle contain more peaks due to the hyperfine structure (HFS). However, since this splitting is completely unresolved, the effects should in first order cancel out. To confirm this, a Monte Carlo simulation was made for the $2p1s$ peak of \textsuperscript{35, 37}Cl. Here, spectra were sampled including and excluding hyperfine splitting using a Gaussian line shape with width $\sigma=0.5~\si{\kilo\eV}$. This resolution is similar to the highest resolution detectors available for this experiment, which should suffer the biggest systematic shifts from hyperfine splitting. QED calculations for the hyperfine structure are given in Appendix~\ref{app:exp_hyperfine}. For these tests, the shifts due to fine structure and hyperfine structure from these calculations were used together with a statistical distribution for the corresponding hyperfine intensities. The generated energy spectra for \textsuperscript{35}Cl is shown in Figure~\ref{fig:hyperfine_Cl}.
    
        Visually, the generated hyperfine spectra do not significantly deviate from the FS spectrum. The sampled spectra of the FS and HFS were fitted, and $E_p$ was extracted. The difference $\Delta$ between these two are given in Table \ref{tab:exp_hyperfine}. This was tested down to a statistical uncertainty of $\sim 35~\si{\milli\eV}$, showing no deviation within the statistical precision. Consequently, the systematic effect corresponding to the hyperfine structure is approximated to be less than $0.1~\si{\eV}$, falling well below our experimental uncertainty.
    
        \begin{table}
            \centering
            \caption{Results from Monte Carlo investigation on the systematics induced by the HFS.}
            \begin{tabular}{c|c}
                Isotope & Deviation ($\si{\milli\eV}$)\\
                \hline
                \textsuperscript{35}Cl \rule{0pt}{2.5ex} & +4(28) \\
                \textsuperscript{37}Cl \rule{0pt}{2.5ex} & +13(27)
            \end{tabular}
            \label{tab:exp_hyperfine}
        \end{table}

        Beyond the conventional hyperfine structure, higher order effects may shift the center-of-gravity of the transition. Following the procedure from Ref.~\cite{pachucki2024second}, this effect was studied for the $1s$ state. In the article, they arrive to the following relationship for this state:

        \begin{equation}\label{eq:exp_hfs_2}
            E_{\text{hfs2, $1s$}} = m (Z \alpha)^6 (\frac{g m}{2 M})^2 I (I+1) \frac{10}{3} (0.54008 + \ln{m Z \alpha r_C}).
        \end{equation}

        Here, $m$ is the mass of the orbiting particle, $Z$ is the proton number of the nucleus, $\alpha$ is the fine structure constant, $g$ is the g-factor of the nucleus, $M$ is the mass of the nucleus, $I$ is the spin of the nucleus, and $r_C$ is the rms charge radius (which can be taken approximately for these estimates). The resulting shifts in centroid energy are given in Table~\ref{tab:exp_higher_order}. The magnitude of the effect is in the order of a few $\si{\milli\eV}$, which is negligible for our experimental precision.

        \begin{table}
            \centering
            \caption{Higher order hyperfine correction estimates for the $1s$ state.}
            \begin{tabular}{c|c}
                Isotope & $1s$ shift ($\si{\milli\eV}$) \\
                \hline
                \textsuperscript{35}Cl \rule{0pt}{2.5ex} & -3.6 \\
                \textsuperscript{37}Cl \rule{0pt}{2.5ex} & -2.2
            \end{tabular}
            \label{tab:exp_higher_order}
        \end{table}
    
        It can thus be concluded that hyperfine effects do not have a significant effect on the extracted centroid energy.

    \subsubsection{Effect of impurities}
        The enriched targets used during these measurements did not have $100\%$ isotopic purity. Consequently, the center-of-gravity energies may suffer a systematic shift. As a result, the isotope shift measured before are actually effective isotope shifts $IS'$, which are shifted compared to the true isotope shift $IS$ due to the purity fraction $f$. We can however, adjust the energies in first approximation.
    
        \begin{align}
            IS' &= \left[f_1 E_{\text{Isotope 1}} + (1-f_1) E_{\text{Isotope 2}} \right] - \left[f_2 E_{\text{Isotope 2}} + (1-f_2) E_{\text{Isotope 1}}\right] \nonumber \\
            &= (f_1 + f_2 - 1) (E_{\text{Isotope 1}} - E_{\text{Isotope 2}}) \nonumber \\
            &= (f_1 + f_2 - 1) IS \nonumber \\
            \implies IS &= \frac{IS'}{f_1 + f_2 - 1} \label{eq:exp_isotopeshift}
        \end{align}
    
        The results imply that the measured isotope shift is always smaller than the true one, as $f_1$ and $f_2$ are strictly smaller than one. In the edge cases that $f_1=f_2=1$ and $f_1=f_2=0$, the isotope shift remains the same aside from a negative sign in the latter. Furthermore, if the same spectrum would be considered twice (such that $f_2 = 1 - f_1$ and $IS'=0$), the true isotope shift is undefined. Similarly, the effective energies $E'$ measured in Section \ref{sec:exp_energies} should be adjusted to true energies $E$.
    
        \begin{align}
            E' &= f E + (1-f)(E + IS) \nonumber \\
               &= E (f + 1 - f) + (1 - f) IS \nonumber \\
            \implies E &= E' + (f - 1) IS \label{eq:exp_absoluteshift}
        \end{align}
    
        The impurity induces an offset on the isotope shift and the absolute energies, though one can correct for it assuming that the weighted average is extracted. The purity information is given in Table~\ref{tab:exp_purity} together with approximate energy shifts. These shifts are minor, but were taken into account as it is a straightforward assessment.
    
        \begin{table}
            \centering
            \caption{Information about the purity of the enriched targets and the approximate centroid shift.}
            \begin{tabular}{c|c|c}
                Isotope & Purity (\%) & Approximate \\
                & & $2p1s$ shift ($\si{\eV}$)\\
                \hline
                \textsuperscript{35}Cl \rule{0pt}{2.5ex} & 99.32(5)   & 1 \\
                \textsuperscript{37}Cl \rule{0pt}{2.5ex} & 99.28(5)   & 1
            \end{tabular}
            \label{tab:exp_purity}
        \end{table}

        Using this purity information, the transition energies and isotope shifts are recalculated. The resulting $np1s$ energies and isotope shifts are given in Table~\ref{tab:exp_energies}. A breakdown of the uncertainties is given in Table~\ref{tab:exp_uncertainties}. Here $\sigma_{\text{stat+cal}}$, $\sigma_{\text{bias}}$, $\sigma_{\text{lit}}$, $\sigma_{\text{f}}$, and $\sigma_{\text{tot}}$ respectively represent the combined statistical and calibration uncertainty, the bias uncertainty induced by averaging over detectors, the uncertainty of the best known calibration peak, the uncertainty associated with the impurity of the target, and the total uncertainty (obtained by adding individual uncertainties in quadrature).

        \begin{table}
            \centering
            \caption{Extracted absolute $np1s$ energies (in $\si{\kilo\eV}$) and isotope shifts (in $\si{\eV}$) of \textsuperscript{35, 37}Cl.}
            \begin{tabular}{cc|l}
                Isotope & Transition & Energy ($\si{\kilo\eV}$) \\
                \hline
                \textsuperscript{35}Cl \rule{0pt}{2.5ex} & $2p1s$ & 578.869(16) \\
                                       & $3p1s$ & 692.096(11) \\
                                       & $4p1s$ & 731.665(13) \\
                \hline
                \textsuperscript{37}Cl \rule{0pt}{2.5ex} & $2p1s$ & 578.734(12) \\
                                       & $3p1s$ & 691.997(18) \\
                                       & $4p1s$ & 731.55(3) \\
                \hline
                \textsuperscript{37}Cl - \textsuperscript{35}Cl \rule{0pt}{2.5ex} & $2p1s$ & $-$144(12) \\
                                                                & $3p1s$ & $-$102(17) \\
                                                                & $4p1s$ & $-$120(30)
            \end{tabular}
            \label{tab:exp_energies}
        \end{table}

        \begin{table}
            \centering
            \caption{Breakdown of the experimental uncertainties on the muonic transition energies and isotope shifts.}
            \begin{tabular}{cc|cccc|c}
                Isotope & Transition & $\sigma_{\text{stat+cal}}$ & $\sigma_{\text{bias}}$ & $\sigma_{\text{lit}}$ & $\sigma_{\text{f}}$ & $\sigma_{\text{tot}}$ \\
                & & $(\si{\eV})$ & $(\si{\eV})$ & $(\si{\eV})$ & $(\si{\eV})$ & $(\si{\eV})$ \\
                \hline
                \textsuperscript{35}Cl \rule{0pt}{2.5ex} & $2p1s$ & 13.0 & 8.2 & 1.1 & 0.11 & 15.5 \\
                & $3p1s$ & 06.6  & 8.2 & 1.1 & 0.13 & 10.6 \\
                & $4p1s$ & 09.4  & 8.2 & 1.1 & 0.20 & 12.6 \\
                \hline
                \textsuperscript{37}Cl \rule{0pt}{2.5ex} & $2p1s$ & 08.7  & 8.2 & 1.1 & 0.2 & 12.0 \\
                & $3p1s$ & 15.4 & 8.2 & 1.1 & 0.2 & 17.5 \\
                & $4p1s$ & 26.4 & 8.2 & 1.1 & 0.3 & 27.7 \\
                \hline
                \textsuperscript{37}Cl - \textsuperscript{35}Cl \rule{0pt}{2.5ex} & $2p1s$ & 11.6 & / & / & 0.11 & 11.6 \\
                & $3p1s$ & 16.8 & / & / & 0.07 & 16.8 \\
                & $4p1s$ & 28.1 & / & / & 0.09 & 28.2
            \end{tabular}
            \label{tab:exp_uncertainties}
        \end{table}

%%%%%%%%%%%%%%%%%%%%%%%
%%%%%%%%%%%%%%%%%%%%%%%
%%%%%%%% Q E D %%%%%%%%
%%%%%%%%%%%%%%%%%%%%%%%
%%%%%%%%%%%%%%%%%%%%%%%
\section{QED calculations}
    The estimation of the $np1s$ transition energies with  \( n = 2, 3, 4 \) was performed using the state-of-the-art multiconfiguration Dirac Fock and general matrix elements (mcdfgme) code~\cite{indelicato2024mcdfgme}. This code treats the bound muon as a relativistic Dirac particle interacting with the nucleus, governed by the Dirac equation:

    \begin{equation}\label{eq:Dirac}
        \left[ \boldsymbol{\alpha} \cdot \mathbf{p} + \beta m_r + V(r) \right] | \psi_{n\kappa m} \rangle = E_{n\kappa} | \psi_{n\kappa m} \rangle,
    \end{equation}

    where, \( | \psi_{n\kappa m} \rangle \) and \( E_{n\kappa} \) are the muon Dirac wavefunction and energy, respectively, for a level with quantum numbers \( n \) (principal), \( \kappa \) (relativistic angular momentum), and \( m \) (magnetic).  \( \boldsymbol{\alpha} \) and \( \beta \) are the Dirac matrices. \( \mathbf{p} \) is the muon momentum operator and $m_r$ is the reduced mass of the system. The total potential \( V(r) \) includes several contributions, detailed in the following discussion. 

    Each potential taken into account when solving the Dirac equation is taken into account non-perturbatively (\textit{to all orders}). Accordingly, it is beneficial to include potentials that show slow convergence for the system of interest. Smaller corrections can be added on top perturbatively. This section only provides an example calculation for the relevant effects. Each effect was considered for a range of input RMS radii. A spreadsheet of these calculations was added as an additional supplement.

\subsection{Contributions calculated non-perturbatively}
        The first contribution to the potential comes from the Coulomb interaction (V$_\text{Coulomb}$) of the muon with the extended nuclear charge distribution. To approximate it, we modeled the nuclear charge using a two-parameter Fermi distribution

    \begin{equation}\label{eq:theo_Fermi2}
        \rho(r) = \frac{\rho_0}{1 + \exp\left[4\ln{3} \left(\frac{r-c}{t}\right)\right]}\,.
    \end{equation}
    Here, $\rho_0$, $c$ and $t$ are the normalization factor, nuclear radius at half density, and skin thickness, respectively. In order to estimate the effect of the nuclear radius on the transition energies, we performed calculations for a range of $c$ values corresponding to a range of RMS radii $r_c$. We set $t =2.3~\si{\femto\meter}$. The model-dependency coming from the choice of this simple distribution is dealt with separately by the Barret-moment recipe.

    The strongest QED contributions, namely those related to normal one loop vacuum polarization (VP), are also taken into account non-perturbatively. Their leading contribution is given by the Uehling potential of magnitude $\alpha(Z\alpha)$, which for a spherical charge distribution can be expressed as~\cite{uehling1935polarization,klarsfeld1977analytical},

    \begin{equation}\label{eq:Uehling}
        \text{VP11}(r) = -\frac{2\alpha (Z\alpha) }{3}
        \frac{1}{r}
        \int_0^\infty dr^\prime \, r^\prime  \rho(r^\prime)
        \int_1^\infty dt \, \left(\frac{1}{t} + \frac{1}{2t^3}\right) \frac{\sqrt{t^2 - 1}}{t^2}  
        \left[ \exp \left( -\frac{2}{\lambda_i} |r - r'| t \right) - \exp \left( -\frac{2}{\lambda_i} |r + r'| t \right) \right].
    \end{equation}
    Here, $\lambda_i$ denotes the Compton wavelength of the fermion field which polarizes the vacuum. We consider either an electron (eVP11) or a muon ($\mu$VP11). The tau particle is too heavy to have a sizable effect. Both potentials are included in the Dirac equation such that the total potential is given by
    
    \begin{equation}\label{potential}
        V(r) = \text{V}_\text{Coulomb} + \text{eVP11} + \mu \text{VP11}.
    \end{equation}
    Solving equation \ref{eq:Dirac} yields the Coulomb–Uehling wavefunctions and their corresponding eigenvalues. These energies are presented in the first row of Table~\ref{tab:energy_levels_cl35} and Table~\ref{tab:energy_levels_cl37} for $\mu$\textsuperscript{35}Cl and $\mu$\textsuperscript{37}Cl, respectively.

\subsection{Contributions calculated perturbatively}
    Additional corrections were obtained using first order perturbation, with the unperturbed wave functions obtained by solving the Dirac equation. The first correction we consider is related to the Wichmann-Kroll potential (WK) of coupling $\alpha(Z\alpha)^3$~\cite{indelicato2013nonperturbative, wichmann1956vacuum, huang1976calculation}. Higher-order terms ($\alpha(Z\alpha )^5$, $\alpha(Z\alpha)^7$) were evaluated and found to be insignificant compared to the total uncertainty budget.
    Next, a correction was made to account for the K\`{a}ll\"{e}n and Sabry (KS) potential~\cite{fullerton1976accurate} created by two-loops of electron pairs. Here, the correction from combining one electron and one muon loop was neglected since it is $\mathcal{O}(0.1~\si{\eV})$.
    
    Another considerable contribution to the energy levels in muonic mid-$Z$ atoms is the Hadronic Vacuum Polarization (HVP), which arises from hadronic pair-production. The HVP correction was calculated by modifying the Uehling-type potential with a hadronic contribution in point-like nucleus approximation implemented by using the parameterization given in~\cite{breidenbach2022hadronic}. This correction is not pronounced for the $np$ orbitals as their overlap with the nucleus is small in the region where HVP is significant.
    Moreover, the contribution of the one-loop muon self-energy (SE\textsubscript{Point}) was calculated, along with corresponding finite-size correction (SE\textsubscript{FNS}). Methods for these calculations are discussed in Refs.~\cite{indelicato2013nonperturbative, karshenboim2018alpha}.

    Finally, recoil corrections beyond the reduced mass were considered. In a non-relativistic approximation, the use of the reduced mass completely accounts for the finite nuclear mass for both radiative and non-radiative terms. Beyond this approximation, the residual linear non-radiative recoil for a finite-size system (approximated by the charge distribution) was calculated, denoted as $\Delta$Rec. The calculation is done to all orders in the binding parameter using the methods described in Ref.~\cite{2023-uRec}. The result is largest for the $1s$ state, amounting to a $0.4\,$eV increase of the binding energy for both isotopes.

\subsection{Main missing contributions and uncertainty estimation}
    Although thorough QED corrections have been considered in the present work, some higher-order effects are still not accounted for. Corrections that are quadratic in $m_\mu/m_N$ are not included. Some of them are unknown, such as the nuclear self-energy beyond $Z=2$~\cite{2011-RadRec}, and some depend on the exact definition of the charge radius, which is not yet established for nuclei with $I>1$~\cite{2011-DF,2011-RadDef}. We acknowledge these gaps by taking an uncertainty of $3~\si\electronvolt$ for any transition that involves the $1s$ state. This uncertainty is also larger than the other missing terms that were considered: 1) the SE-VP diagrams (see Ref.~\cite{indelicato2019qed}) $\mathcal{O}(\alpha \times \text{SE}) \approx 1~\si{\eV}$, 2) the eVP effect on nuclear polarization $\mathcal{O}(2\alpha\times\text{NP}) \approx 1.5~\si\electronvolt$, and 3) irreducible three-loop corrections, estimated as $\mathcal{O}(\alpha \times \text{KS}) \approx 0.3~\si\electronvolt$. Adding all the corresponding uncertainties in quadrature provides a total uncertainty of about $4~\si{\eV}$ after rounding.

    \begin{table}
        \centering
        \caption{Total Binding energy (in \si{\electronvolt}) and contributions from different corrections for $\mu^{35}$Cl levels. Calculations are shown for the resulting Barrett radius, described by the 2pF with ($c=3.4933~\si{\femto\meter}$, $t=2.3~\si{\femto\meter}$).}

        \begin{ruledtabular}
        \begin{tabular}{r|rrrrrrrr|r}
            Level & Coul. + Ue. & SE\textsubscript{Point} & SE\textsubscript{FNS} & HVP & WK & KS & $\Delta$Rec & Screening & Total \\
            \hline
            $1s$       & $-$782,471.8 & 204.8  & $-$62.7 & $-$6.3 & 3.2 & $-$38.0 & $-$0.1 & 1,296.1 & $-$781,074.7 \\
            $2p_{1/2}$ & $-$204,133.1 & $-$0.7 & 0.1     & 0.0    & 0.6 & $-$4.0  & 0.0    & 1,295.7 & $-$202,841.4 \\
            $2p_{3/2}$ & $-$203,354.7 & 1.0    & 0.1     & 0.0    & 0.6 & $-$4.1  & 0.0    & 1,295.7 & $-$202.061.4 \\
            $3p_{1/2}$ & $-$90,542.5  & $-$0.2 & 0.0     & 0.0    & 0.2 & $-$1.2  & 0.0    & 1,295.0 & $-$89,248.7 \\
            $3p_{3/2}$ & $-$90,313.1  & 0.3    & 0.0     & 0.0    & 0.2 & $-$1.1  & 0.0    & 1,295.0 & $-$89,018.7 \\
            $4p_{1/2}$ & $-$50,867.6  & $-$0.1 & 0.0     & 0.0    & 0.1 & $-$0.5  & 0.0    & 1,294.5 & $-$49,573.6 \\
            $4p_{3/2}$ & $-$50,771.1  & 0.1    & 0.0     & 0.0    & 0.1 & $-$0.5  & 0.0    & 1,294.5 & $-$49,476.9
        \end{tabular}
        \end{ruledtabular}

        \label{tab:energy_levels_cl35}
    \end{table}

    \begin{table}
        \centering
        \caption{Total Binding energy (in \si{\electronvolt}) and contributions from different corrections for $\mu^{37}$Cl levels. Calculations are shown for the resulting Barrett radius, described by the 2pF with ($c=3.5127~\si{\femto\meter}$, $t=2.3~\si{\femto\meter}$).}

        \begin{ruledtabular}
        \begin{tabular}{r|rrrrrrrr|r}
            Level & Coul. + Ue. & SE\textsubscript{Point} & SE\textsubscript{FNS} & HVP & WK & KS & $\Delta$Rec & Screening & Total \\
            \hline
            $1s$       & $-$782,383.4 & 204.9  & $-$63.0 & $-$6.3 & 3.2 & $-$38.0 & $-$0.1 & 1,295.4 & $-$780,987.2 \\
            $2p_{1/2}$ & $-$204,168.6 & $-$0.7 & 0.1     & 0.0    & 0.6 & $-$4.1  & 0.0    & 1,295.0 & $-$202,877.7 \\
            $2p_{3/2}$ & $-$203,390.2 & 1.0    & 0.1     & 0.0    & 0.6 & $-$4.1  & 0.0    & 1,295.0 & $-$202,097.6 \\
            $3p_{1/2}$ & $-$90,558.2  & $-$0.2 & 0.0     & 0.0    & 0.2 & $-$1.2  & 0.0    & 1,294.3 & $-$89,265.1 \\
            $3p_{3/2}$ & $-$90,328.9  & 0.3    & 0.0     & 0.0    & 0.2 & $-$1.1  & 0.0    & 1,294.3 & $-$89,035.2 \\
            $4p_{1/2}$ & $-$50,876.5  & $-$0.1 & 0.0     & 0.0    & 0.1 & $-$0.5  & 0.0    & 1,293.9 & $-$49,583.1 \\
            $4p_{3/2}$ & $-$50,779.9  & 0.1    & 0.0     & 0.0    & 0.1 & $-$0.5  & 0.0    & 1,293.9 & $-$49,486.3
        \end{tabular}
        \end{ruledtabular}

        \label{tab:energy_levels_cl37}
    \end{table}

%%%%%%%%%%%%%%%%%%%%%%%
%%%%%%%%%%%%%%%%%%%%%%%
%%%%%%%%% N P %%%%%%%%%
%%%%%%%%%%%%%%%%%%%%%%%
%%%%%%%%%%%%%%%%%%%%%%%
\section{NP correction}
    The nuclear polarization itself can be split in a nuclear part, consisting of low-lying states (LLS) and giant resonances (GR), and the nucleon part from hadronic excitations. For the nuclear part we relied on the methodology presented in Refs.~\cite{valuev2022evidence, 2024_Valuev, 2007_Haga} with a new uncertainty evaluation, while for the nucleon part the values and uncertainties were taken from Ref.~\cite{gorchtein2025hitchhiker}.

    The nuclear parameters for LLS were taken from Nuclear Data Sheets~\cite{Nucl_data_35, Nucl_data_37}, while the contributions from the GR were estimated by means of phenomenological energy-weighted sum rules~\cite{rinker1978nuclear}. Nuclear transitions with multipolarities up to $L = 5$ were included in the calculations for muonic \textsuperscript{35, 37}Cl. For the Coulomb part of nuclear polarization, the nuclear transition charge densities were described by~\cite{2007_Haga}: 1) the monopole Tassie model for $L = 0$, 2) the hydrodynamical Jensen-Steinwedel model for $L = 1$, and 3) the Tassie-Goldhaber-Teller model for $L \geq 2$. The models were chosen based on the comparison with the microscopic Skyrme calculations~\cite{skyrme_rpa} for \textsuperscript{40}Ca, from which one can also expect that the corrections obtained for \textsuperscript{35, 37}Cl are most likely to be somewhat underestimated. The transverse contribution to nuclear polarization was estimated by means of the continuity equation for nuclear charge and current transition densities, with the $(J, J+1)$ component of the current set to zero~\cite{2007_Haga}. The reliability of such estimations was controlled by performing the calculations in both Feynman and Coulomb gauges in order to check fulfillment of gauge invariance. The resulting nuclear-polarization corrections for individual energy levels of muonic \textsuperscript{35, 37}Cl are listed in Table~\ref{tab:NP_shifts}. The nuclear polarization increases the binding energy (indicated by the negative sign), which leads to larger transition energies for $np1s$ transitions. 

    \begin{table}
        \centering
        \caption{Nuclear part of the NP (in $\si{\eV}$, Coulomb + transverse, up to $L=5$) for the energy levels of muonic \textsuperscript{35, 37}Cl.}
        \begin{ruledtabular}
        \begin{tabular}{c|ccccccc}
            Isotope & $1s$ & $2p$\textsubscript{1/2} & $2p$\textsubscript{3/2} & $3p$\textsubscript{1/2} & $3p$\textsubscript{3/2} & $4p$\textsubscript{1/2} & $4p$\textsubscript{3/2} \\
            \hline
            \textsuperscript{35}Cl & $-$103.91 & $-$0.67 & $-$0.65 & $-$0.22 & $-$0.21 & $-$0.10 & $-$0.09 \\
            \textsuperscript{37}Cl & $-$100.05 & $-$0.63 & $-$0.61 & $-$0.20 & $-$0.20 & $-$0.09 & $-$0.09
        \end{tabular}
        \label{tab:NP_shifts}
        \end{ruledtabular}
    \end{table}

    For the uncertainty evaluation, the adopted approach using simple non-microscopic charge transition densities was compared to inputs from various Skyrme models for the nearby doubly magic \textsuperscript{40}Ca. As mentioned in the main text, Skyrme models were selected in such a way that they 1) Cover significant portions of the constraints on the saturation properties, 2) Represent different groups and fitting protocols. A comparison between the approximation applied in this work and different Skyrme models is shown in Table~\ref{tab:NP_skyrmes}.

    \begin{table}[]
        \centering
        \caption{Nuclear part of the NP (in eV, Coulomb only, up to $L=3$) for the $1s$ state of muonic \textsuperscript{40}Ca calculated with different models.}
        \begin{ruledtabular}
        \begin{tabular}{c|ccccccccccc}
            Our method & LNS & SKI3 & KDE0 & SKX & SLy5 & BSk14 & SAMi & NRAPR & SkP & SkM\textsuperscript{*} & SGII \\
            \hline
            $-$170.5 & $-$197.9 & $-$187.7 & $-$188.4 & $-$194.5 & $-$195.3 & $-$194.6 & $-$198.5 & $-$192.3 & $-$207.0 & $-$209.8 & $-$207.8
        \end{tabular}
        \end{ruledtabular}
        \label{tab:NP_skyrmes}
    \end{table}

    Based on these results, the relative uncertainty was conservatively estimated by taking the maximally deviating Skyrme model (SkM*), giving an uncertainty of $(209.8 - 170.5)/170.5 \approx 23\%$. Combining the values from Table~\ref{tab:NP_shifts} with nucleon polarization provides the energy shifts on the transition energies given in Table~\ref{tab:NP_total}. The fraction of the NP that behaves in a correlated way is also given in this table, as it is crucial in determining an uncertainty on the difference in NP between different isotopes. More information about such estimates is given in the main text.

    \begin{table}
        \centering
        \caption{Calculations for the nuclear polarization (in $\si{\eV}$) for the relevant transitions in \textsuperscript{35}Cl and \textsuperscript{37}Cl. $f_{\text{Corr}}$ (\%) denotes the fraction of the nuclear polarization that behaves in a correlated way.}
        \begin{tabular}{cc|cc|cc}
            Isotope & Transition & Nuclear & Nucleon & Total & $f_{\text{Corr}}$ (\%) \\
            \hline
            \textsuperscript{35}Cl & $2p1s$ & 103(24) & 11.9(1.2) & 115(25) & 90.4 \\
                                   & $3p1s$ & 104(24) & 11.9(1.2) & 116(26) & 90.4 \\
                                   & $4p1s$ & 104(24) & 11.9(1.2) & 116(26) & 90.4 \\
            \hline
            \textsuperscript{35}Cl & $2p1s$ & 99(23)  & 12.6(1.3) & 115(25) & 97.6 \\
                                   & $3p1s$ & 100(23) & 12.6(1.3) & 116(26) & 97.6 \\
                                   & $4p1s$ & 100(23) & 12.6(1.3) & 116(26) & 97.6 \\
        \end{tabular}
        \label{tab:NP_total}
    \end{table}

%%%%%%%%%%%%%%%%%%%%%%%
%%%%%%%%%%%%%%%%%%%%%%%
%%%% K , A L P H A %%%%
%%%%%%%%%%%%%%%%%%%%%%%
%%%%%%%%%%%%%%%%%%%%%%%
\section{Determination of \texorpdfstring{$k$ and $\alpha$}{ka}}\label{sec:ka_fit}
    To determine the Barrett parameters that minimize the charge distribution model dependence, the difference in potential generated by the muon in the initial and final state is fitted using

    \begin{equation}
        V_i(r) - V_f(r) = B~r^k e^{-\alpha r}.
    \end{equation}

    To calculate these potentials, the Dirac equation is first solved assuming a charge distribution. In this case, a two-parameter Fermi distribution was used with $t=2.3~\si{\femto\meter}$ and $c$ chosen such that the RMS radius matches the one quoted in Ref.~\cite{angeli2013table}, which can be done using 

    \begin{equation}
        c = \sqrt{\frac{5}{3}R_{\text{RMS}}^2 - \frac{7}{3}\pi^2 \left(\frac{t}{4 \ln{3}}\right)^2}.
    \end{equation}
    
\begin{figure}[h]
        \centering
        \includegraphics[width=0.7\linewidth]{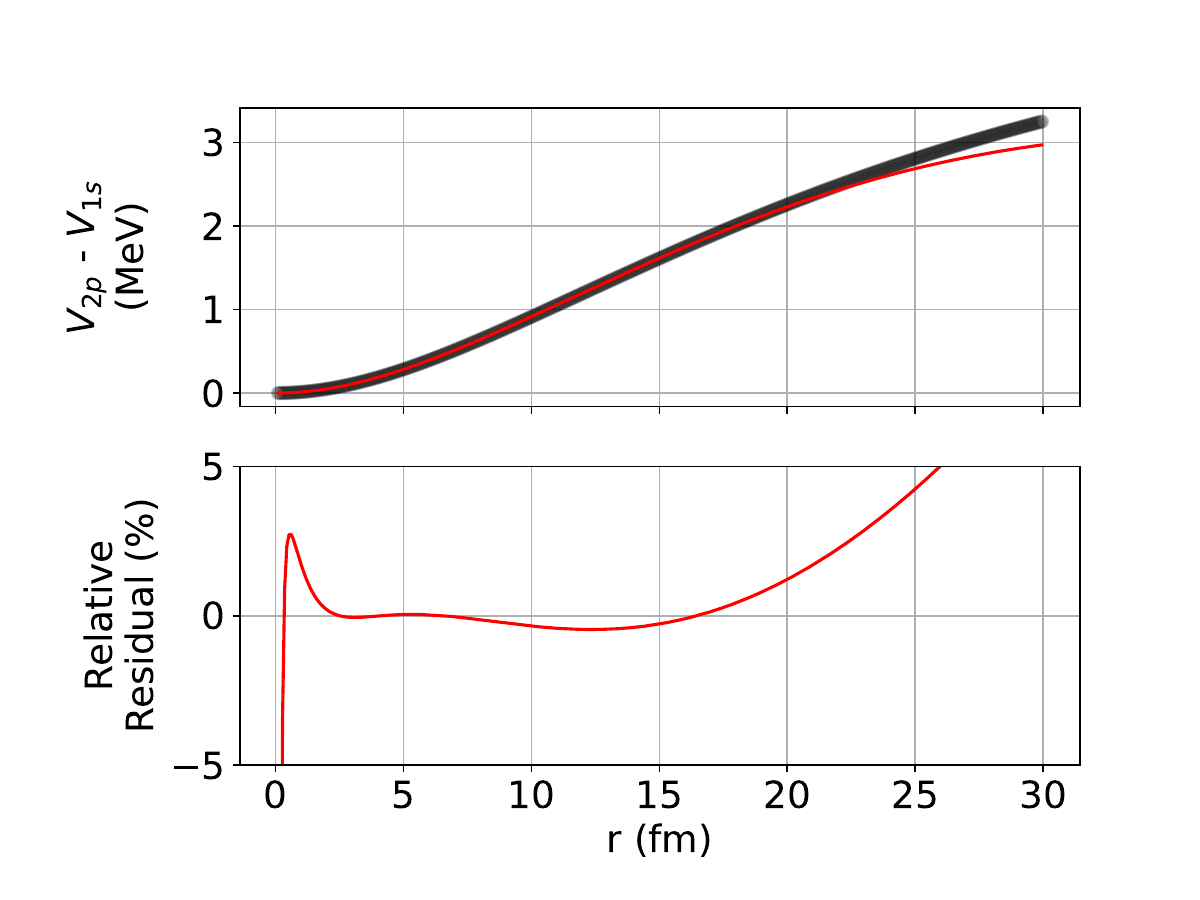}
        \caption{Fit on the difference in potentials generated by the muon in the initial ($2p$) state compared to the final ($1s$) state in \textsuperscript{35}Cl.}
        \label{fig:potentialFit}
    \end{figure}

    By solving the Dirac equation, the muonic wave functions of the different orbitals are obtained, which can be integrated to determine the potential it generates. Finally, the value at $r=0$ is subtracted. Given that the integrand that affects the energy involves the charge distribution and the Jacobian ($r^2$), the fit was made using $r^2 \rho(r)$ weights. Figure~\ref{fig:potentialFit} shows the fit for one of the potential differences and the corresponding residuals. The fit describes the calculated potentials to a precision of $\sim1\%$ within the range $r \in [2; 20]~\si{\femto\meter}$, where $r^2 \rho(r)$ is largest.

    By performing this fit with an adjusted RMS radius, it was shown that realistic variations in the charge distribution do not significantly affect the extracted Barrett parameters.
    
    \newpage

%%%%%%%%%%%%%%%%%%%%%%%
%%%%%%%%%%%%%%%%%%%%%%%
%%%%%%%%% V 2 %%%%%%%%%
%%%%%%%%%%%%%%%%%%%%%%%
%%%%%%%%%%%%%%%%%%%%%%%
\section{Nuclear shape correction}\label{sec:V2}
    For the final radius extraction, this work relies on the $V_2$ correction obtained from information about the charge distribution. Within these calculations, the numerical robustness of the integration and the reliability of the used models should be investigated.
    
    \subsection{Numerical accuracy}
\begin{figure}[h]
            \centering
            \includegraphics[width=0.45\linewidth]{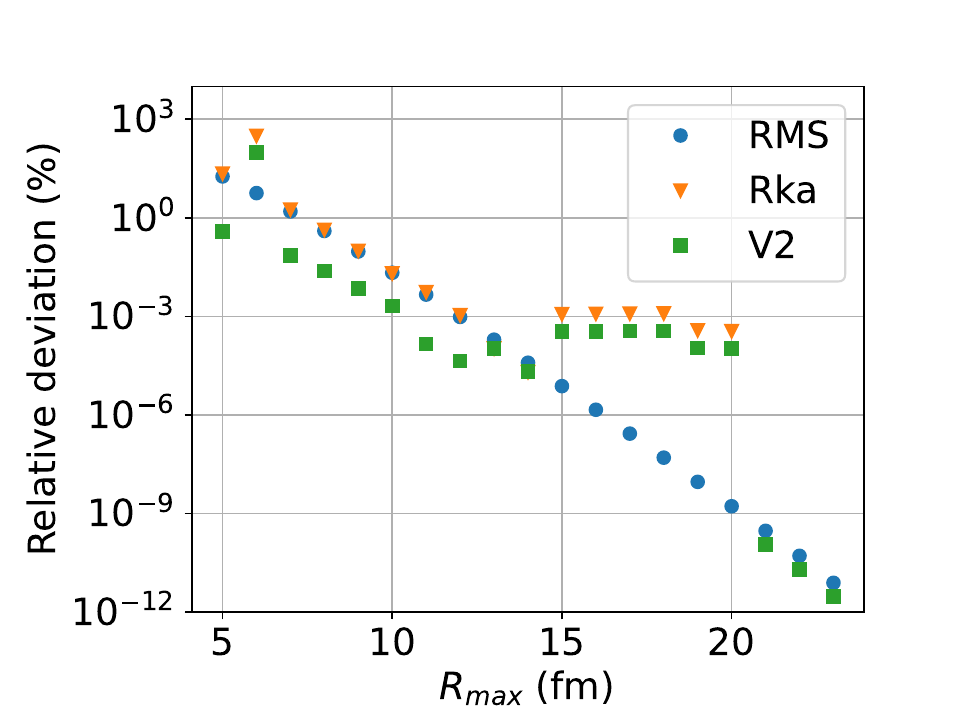}
            \qquad
            \includegraphics[width=0.45\linewidth]{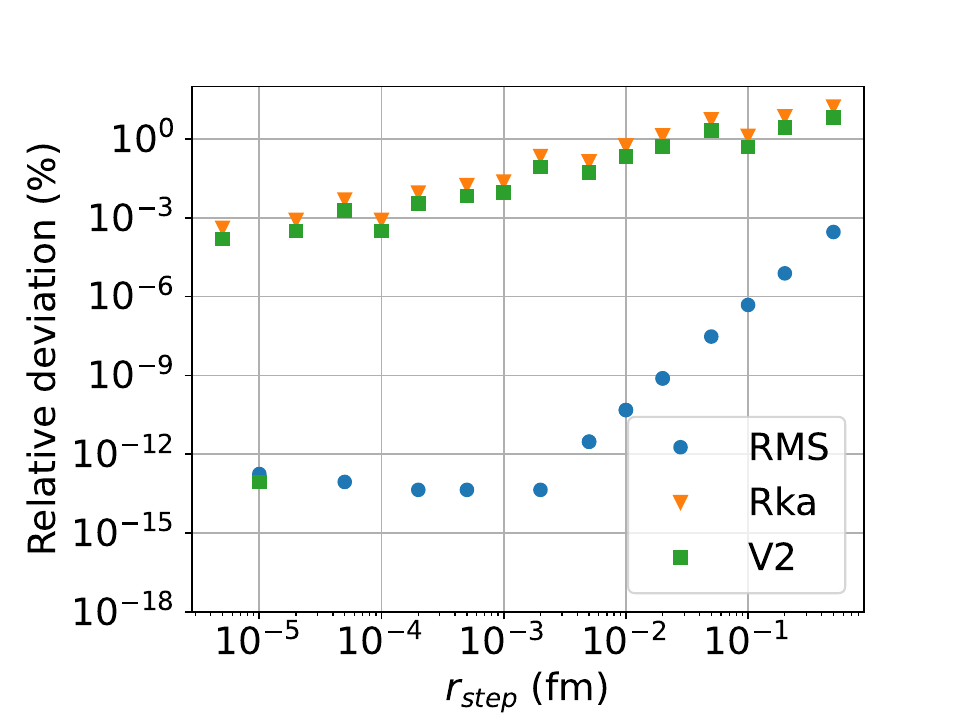}
            \caption{Evaluation of the numerical uncertainty of extracted radii and $V_2$ factors for given integration range and step size for \textsuperscript{35}Cl.}
            \label{fig:numerical_integration}
        \end{figure}

        First, a check was made on the numerical uncertainties of the extracted $V_2$ as a function of the maximal radius considered in numerical integration $R_{\text{max}}$ and the step size for Riemann summing $r_{\text{step}}$. To probe these effects, some tests were performed using a simple two-parameter Fermi model with $t=2.3~\si{\femto\meter}$ and $c$ determined from the literature RMS radius~\cite{angeli1999table}. The optimal value of $R_{\text{max}}$ was determined by varying it from $5~\si{\femto\meter}$ to $25~\si{\femto\meter}$ using a fixed step size of $10^{-5}~\si{\femto\meter}$. Similarly, the effect of the step size was investigated by varying it while keeping $R_{\text{max}}$ fixed at $25~\si{\femto\meter}$. The results of these numerical checks are shown in Figure~\ref{fig:numerical_integration}. The plateau behavior around $10^{-4}$ in the left graph is present due to a numerical quantization of the Barrett radius induced by the chosen integration step size.
    
        From these graphs, conditions were chosen such that the numerical precision is insignificant compared to the other sources of uncertainty. By performing the integration with $R_{\text{max}}=25~\si{\femto\meter}$ and an integration step size of one zeptometer ($10^{-5}~\si{\femto\meter}$), the numerical uncertainty is in the order of $10^{-3}\%$, substantially lower than other sources of uncertainty.

    \subsection{Precision of \texorpdfstring{$V_2$}{V2}}
        As indicated in the manuscript, the nuclear shape correction is important to take into account. If no high-quality electron scattering measurements (with high $q_{\text{max}}$) are available, the uncertainty on $V_2$ can become dominant in the final error budget. This is the case for the stable Cl isotopes, which was solved by relying on input from BSkG density functionals to determine the $V_2$ correction factor. In order to evaluate the reliability of BSkG models, the extracted $V_2$ was compared to some high-quality electron scattering measurements in the region. The cut-off radius for these calculations was taken to be $R_{\text{max}} = 10~\si{\femto\meter}$, which according to Figure~\ref{fig:numerical_integration} induces an uncertainty of $10^{-3}\%$ on $V_2$, well below the current uncertainty budget.
        
\begin{figure}
            \centering
            \includegraphics[width=0.8\linewidth]{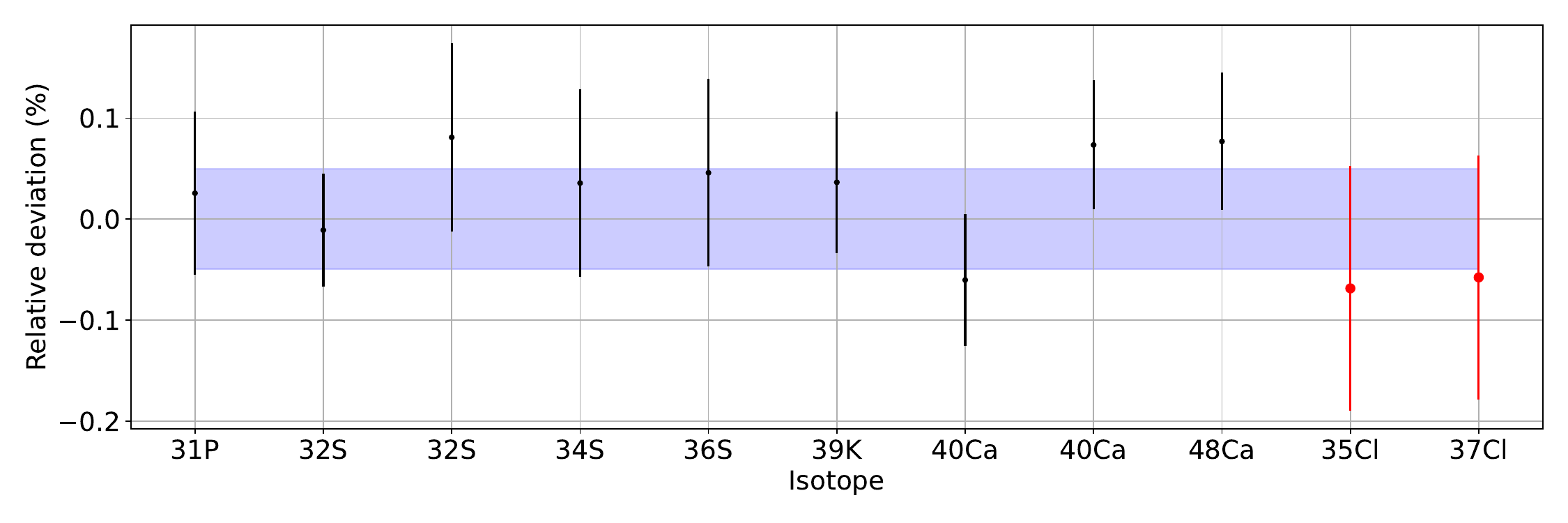}
            \caption{Relative deviation between $V_2$ extracted from BSkG4 and electron scattering. The blue band represents the average deviation. Measurements on \textsuperscript{35, 37}Cl were not used for this average, but their values are in agreement with the EDF estimates.}
            \label{fig:V2_deviation}
        \end{figure}

        While we only report on the $V_2$ from BSkG4, the results do not change significantly when when comparing to previous generations of this model, i.e. BSkG2 and BSkG3. For these comparisons, the charge distributions from electron scattering were taken from high $q_{\text{max}}$ measurements reported in Ref.~\cite{de1987nuclear}. This compilation quotes two separate high $q_{\text{max}}$ measurements for \textsuperscript{32}S and \textsuperscript{40}Ca, which also provides an indication for the scatter between different scattering experiments. Figure~\ref{fig:V2_deviation} shows the typical variation between $V_2$ calculated from BSkG4 and electron scattering, showing agreement within approximated uncertainties. As such, the uncertainty on $V_2$ was taken to be the average of the relative deviation (0.05\%). Moreover, in the case where multiple high $q_{\text{max}}$ measurements were available (\textsuperscript{40}Ca and \textsuperscript{32}S), the different experiments scatter more with one another than with BSkG4.

%%%%%%%%%%%%%%%%%%%%%%%
%%%%%%%%%%%%%%%%%%%%%%%
%%%%% E, dE(Rka) %%%%%%
%%%%%%%%%%%%%%%%%%%%%%%
%%%%%%%%%%%%%%%%%%%%%%%
\section{Fit on theory values}
    \subsection{Extracting \texorpdfstring{$E$, $\Delta E$, and $R_{k\alpha}$}{E dE, Rka}}
        In order to extract the Barrett radius and Barrett radius difference, one must construct a formulation that is bijective for $E(R_{k\alpha})$ and $\Delta E (\Delta R_{k \alpha})$. A first step is to calculate the total transition energy by combing the QED calculations, the NP, and the electron screening. These energies, however, describe the energy difference between the initial and final atomic states, which is not what is experimentally investigated. Given the conservation of momentum, the nucleus receives a small fraction of the total transition energy in the form of recoil energy. A derivation of this correction is provided in Appendix~\ref{app:exp_recoil} and the estimated effect is shown in Table~\ref{tab:recoil}. The shift induced by photon recoil is in the same order as the experimental uncertainty, such that it should be considered. Including this correction, the total emission energies and differences in emission energies for different isotopes are extracted.
    
        \begin{table}
            \centering
            \caption{Approximate recoil correction for $np1s$ peaks of Cl.}
            \begin{tabular}{cc|c}
                 Isotope & Transition & Recoil energy ($\si{\eV}$)\\
                 \hline
                 \textsuperscript{35}Cl & $2p1s$ & 5.14 \\
                                        & $3p1s$ & 7.34 \\
                                        & $4p1s$ & 8.22 \\
                \hline
                 \textsuperscript{37}Cl & $2p1s$ & 4.86\\
                                        & $3p1s$ & 6.95\\
                                        & $4p1s$ & 7.77
            \end{tabular}
            \label{tab:recoil}
        \end{table}
    
        Next, the assumed Fermi distributions used as input for the QED calculations should be transformed into Barrett radii. Using the findings on numerical limitations from Section~\ref{sec:V2}, the cutoff radius was chosen to be $25~\si{\femto\meter}$ with a step size of a zeptometer. This step size was determined in order to reduce the equivalent numerical energy uncertainty to be below $0.1~\si{\eV}$.

        \begin{figure}
            \centering
            \includegraphics[width=0.5\linewidth]{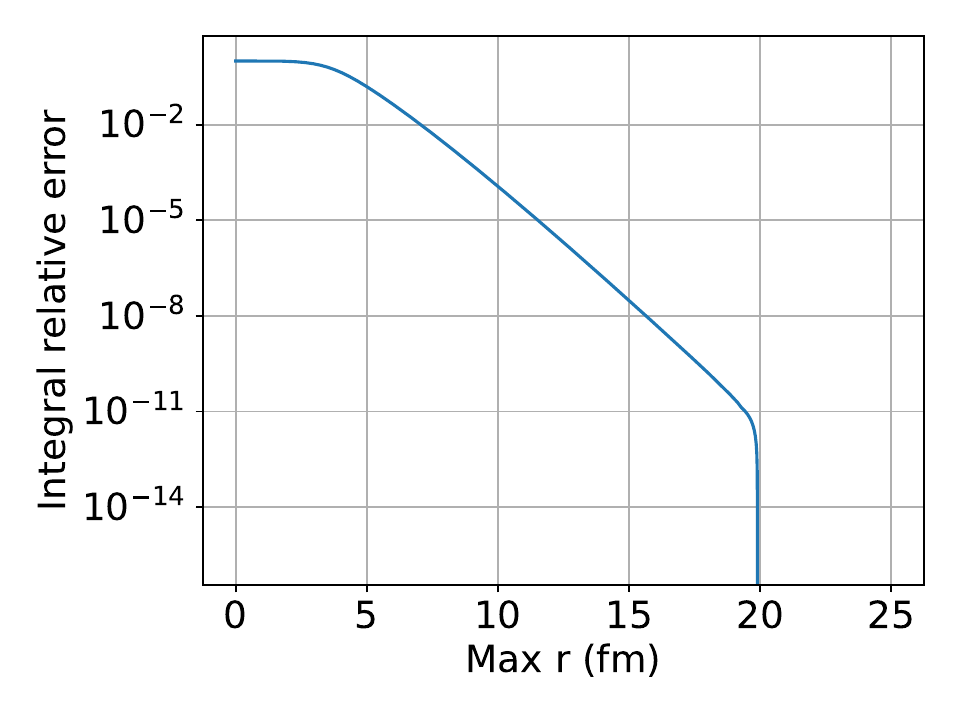}
            \caption{Integration error on the Barrett moment induced by the choice cut-off radius.}
            \label{fig:int_error}
        \end{figure}

    \subsection{Fitting \texorpdfstring{$E(R_{k\alpha})$}{E Rka}}
\begin{figure}
            \centering
            \includegraphics[width=0.7\linewidth]{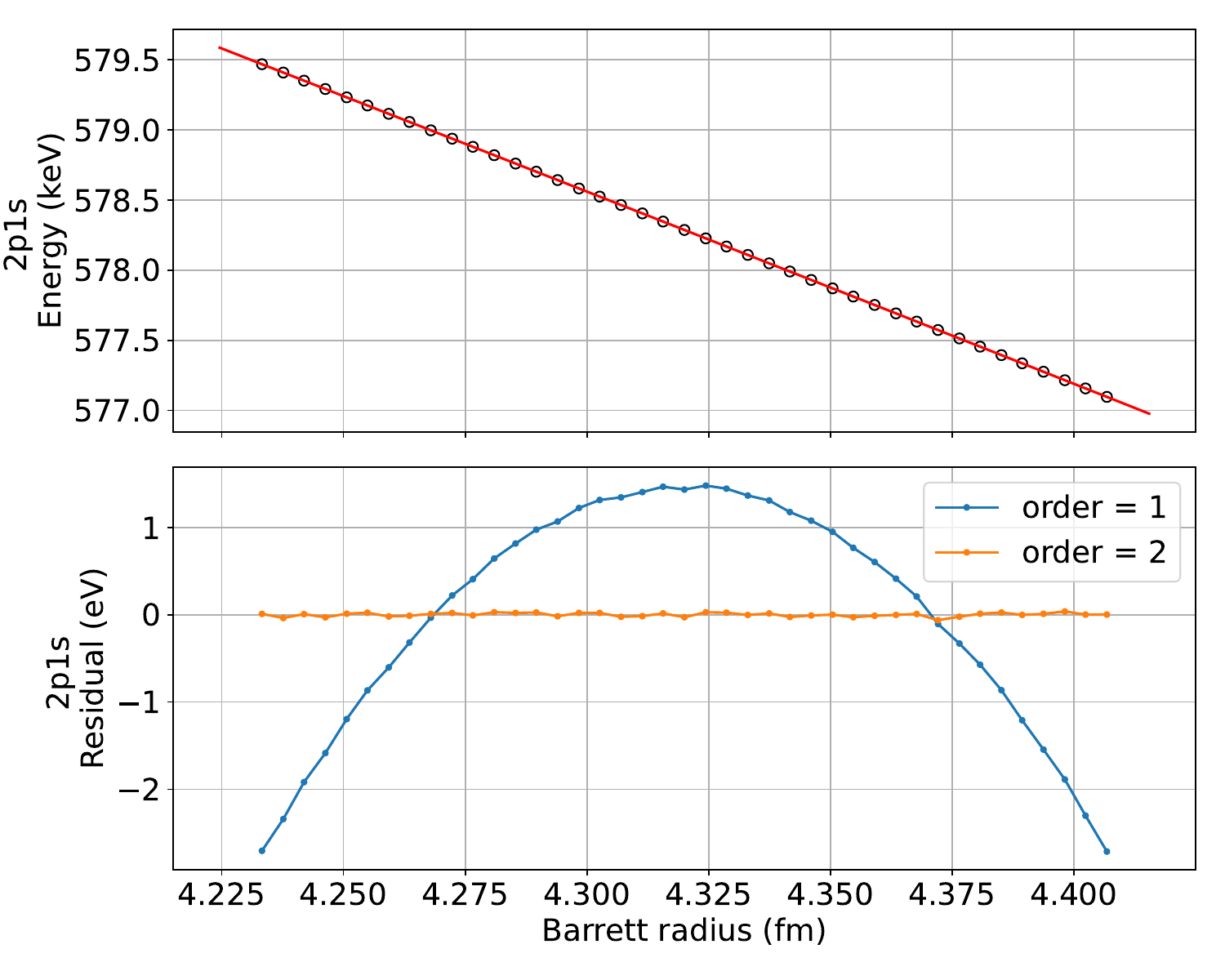}
            \caption{Emission energy as a function of Barrett radius for the $2p1s$ transition in \textsuperscript{35}Cl.}
            \label{fig:fit_E_Rka}
        \end{figure}

         To extract the Barrett radius corresponding to the measured emission energy, a fit was made on the theoretical values for $E(R_{k\alpha})$. The calculated points showed a significant (though only a few $\si{\eV}$) quadratic trend as a function of $R_{k\alpha}$. To reduce the multicollinearity of the fit, the Barrett radii were recentered around $R_{\text{cen}} = 4.3~\si{\femto\meter}$. The residuals for the fit on the $2p1s$ of \textsuperscript{35}Cl are shown in Figure~\ref{fig:fit_E_Rka}, showing minimal residuals with respect to the experimental uncertainty.

    \subsection{Fitting \texorpdfstring{$\Delta E$}{dE}}
\begin{figure}
            \centering
            \includegraphics[width=0.7\linewidth]{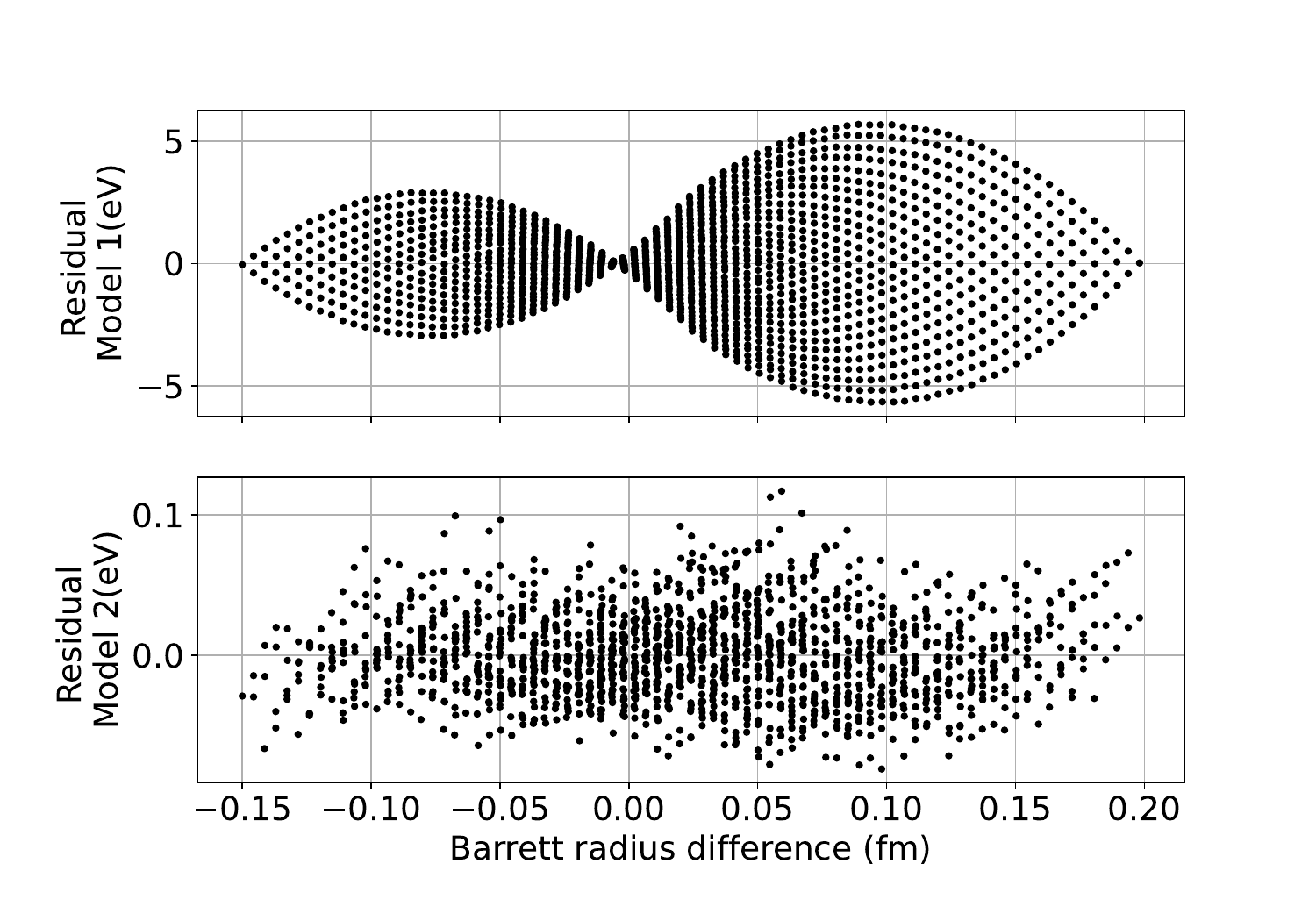}
            \caption{Description of the isotope shift with a linear model in $\Delta R_{k \alpha}$ (model 1) and a model involving $R_{k \alpha, 1}$ dependencies (model 2).}
            \label{fig:fit_dE_attempt1}
        \end{figure}

        When working with isotope shifts, the theoretical calculations can be combined to produce many more fitting points than for the absolute energies. The description of these isotope shifts, however, becomes a bit more complex. If a linear or quadratic fit is made on $\Delta E$ as function of $\Delta R_{k \alpha}$ (model 1), a strange residual trend remains, as shown in Figure~\ref{fig:fit_dE_attempt1}.

        This odd trend stems from the dependence of the absolute scale of the Barrett radii. As such, different models were evaluated that contained dependencies of the Barrett radius of a reference isotope $R_{k \alpha, 1}$. The model found to work best (model 2) is given by

        \begin{equation}
            \Delta E = b_0 + b_1 \Delta R_{k\alpha} + b_2 (R_{k \alpha, 1} - R_{\text{cen}}) + b_3 (\Delta R_{k\alpha})^2 + b_4 (R_{k \alpha, 1} - R_{\text{cen}}) \Delta R_{k \alpha},
        \end{equation}

        where the reference Barrett radius is again recentered around $R_{\text{cen}}=4.3~\si{\femto\meter}$. Using this more advanced model, the isotope shifts are described well to a precision of $0.1~\si{\eV}$. The extracted value of $\Delta R_{k \alpha}$ will have an error associated to the value of $R_{k \alpha, 1}$, but it does not significantly contribute to the total uncertainty.

%%%%%%%%%%%%%%%%%%%%%%%
%%%%%%%%%%%%%%%%%%%%%%%
%%%% C O M P A R E %%%%
%%%%%%%%%%%%%%%%%%%%%%%
%%%%%%%%%%%%%%%%%%%%%%%
\section{Self-consistency check}
    In order to check the self-consistency of the calculations, it is beneficial to evaluate quantities without substantial radius sensitivity. If theoretical predictions and experiments agree on these quantities, it provides more confidence in the extracted transition energies. Given that $p$ states carry only a very minor radius sensitivity in this mass range, one such observable is the difference in $np1s$ energy from two different initial principal quantum numbers. These quantities can be evaluated in each detector before averaging over detectors to slightly reduce systematic uncertainties. However, the bias uncertainty should not fully cancel out, as the different $np1s$ peaks are not sufficiently close in energy for this.

    The resulting energy differences are given in Table~\ref{tab:exp_np1s_diff}. The experimental uncertainty is calculated by adding the uncertainty of a weighted average over detections and the bias uncertainty (for each transition) in quadrature. No uncertainty is assumed for the QED predictions and the nuclear polarization on the difference in $p$ states is not considered for these comparisons. The QED numbers provided here include a correction for the photon recoil, as it varies by several $\si{\eV}$ between different transitions.
    
     \begin{table}
        \centering
        \caption{Energy difference between $np1s$ peaks of the same isotope. Averaging over 19 detectors ($\nu=18$). QED values were corrected for the photon recoil effect.}
        \begin{tabular}{cc|cc|cc}
            Isotope & Transitions & Difference ($\si{\kilo\eV}$) & Averaging $\chi_{\nu}^2$ & QED ($\si{\kilo\eV}$) & QED - Exp ($\si{\eV}$) \\
            \hline
            \textsuperscript{35}Cl & $3p1s$ - $2p1s$ & 113.220(22) & 3.04 & 113.2249 & 5(22) \\
                                   & $4p1s$ - $2p1s$ & 152.794(21) & 1.87 & 152.8108 & 17(21) \\
                                   & $4p1s$ - $3p1s$ & 39.571(16)  & 0.64 & 39.5859  & 15(16) \\
            \hline
            \textsuperscript{37}Cl & $3p1s$ - $2p1s$ & 113.271(21) & 0.52 & 113.2446 & $-$26(21) \\
                                   & $4p1s$ - $2p1s$ & 152.823(15) & 0.78 & 152.8375 & 14(31) \\
                                   & $4p1s$ - $3p1s$ & 39.551(33)  & 0.84 & 39.5928  & 42(33)
        \end{tabular}
        \label{tab:exp_np1s_diff}
    \end{table}

    The theoretical and experimental numbers seem to be in good agreement with each other. Note that the systematic error from the bias uncertainty is expected to induce a similar shift in the $np1s-n'p1s$ for different isotopes. Furthermore, the differences within each isotope are intrinsically correlated to one another as they always have one transition in common.

%%%%%%%%%%%%%%%%%%%%%%%
%%%%%%%%%%%%%%%%%%%%%%%
%%% A P P E N D I X %%%
%%%%%%%%%%%%%%%%%%%%%%%
%%%%%%%%%%%%%%%%%%%%%%%
\clearpage
\appendix
\section{Plots for np1s fitting} \label{app:exp_fits}
    % 35Cl
    \begin{figure}[ht!]
        \centering
        \includegraphics[width=0.8\linewidth]{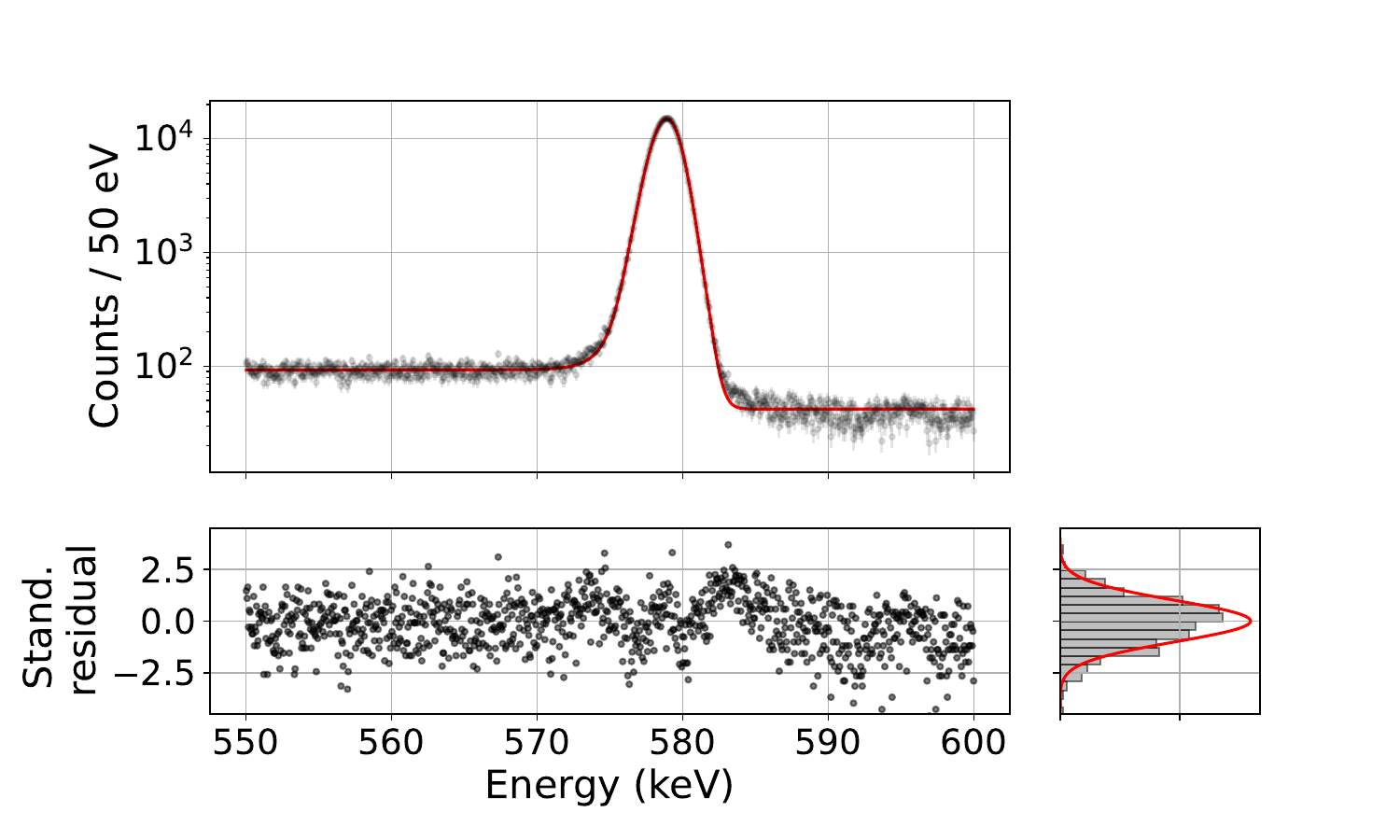}
        \caption{Energy spectrum of $2p1s$ in \textsuperscript{35}Cl summed over detectors, excluding Ge08.}
        \label{fig:exp_35Cl_2p1s}
    \end{figure}
    \begin{figure}[ht!]
        \centering
        \includegraphics[width=0.8\linewidth]{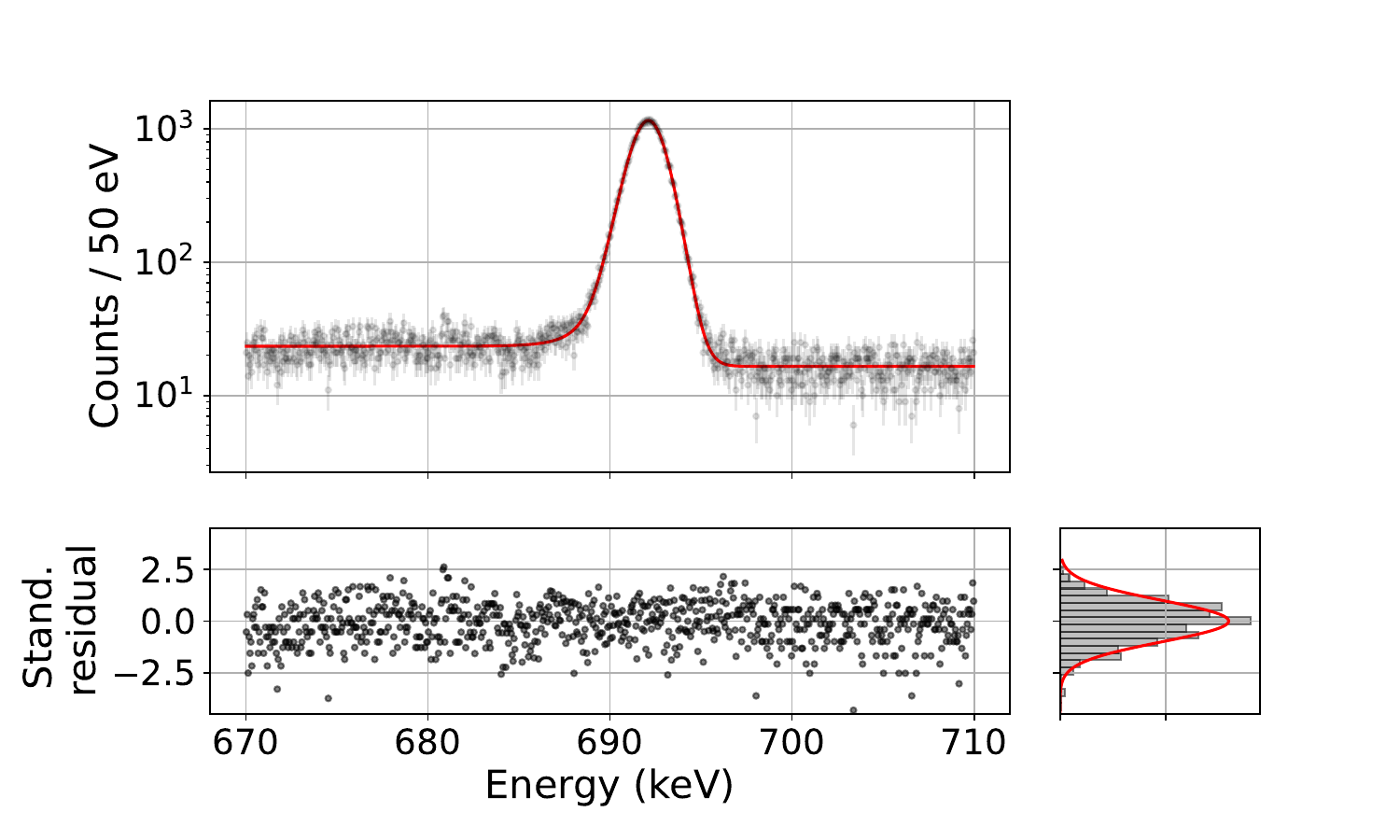}
        \caption{Energy spectrum of $3p1s$ in \textsuperscript{35}Cl summed over detectors, excluding Ge08.}
        \label{fig:exp_35Cl_3p1s}
    \end{figure}
    \begin{figure}[ht!]
        \centering
        \includegraphics[width=0.8\linewidth]{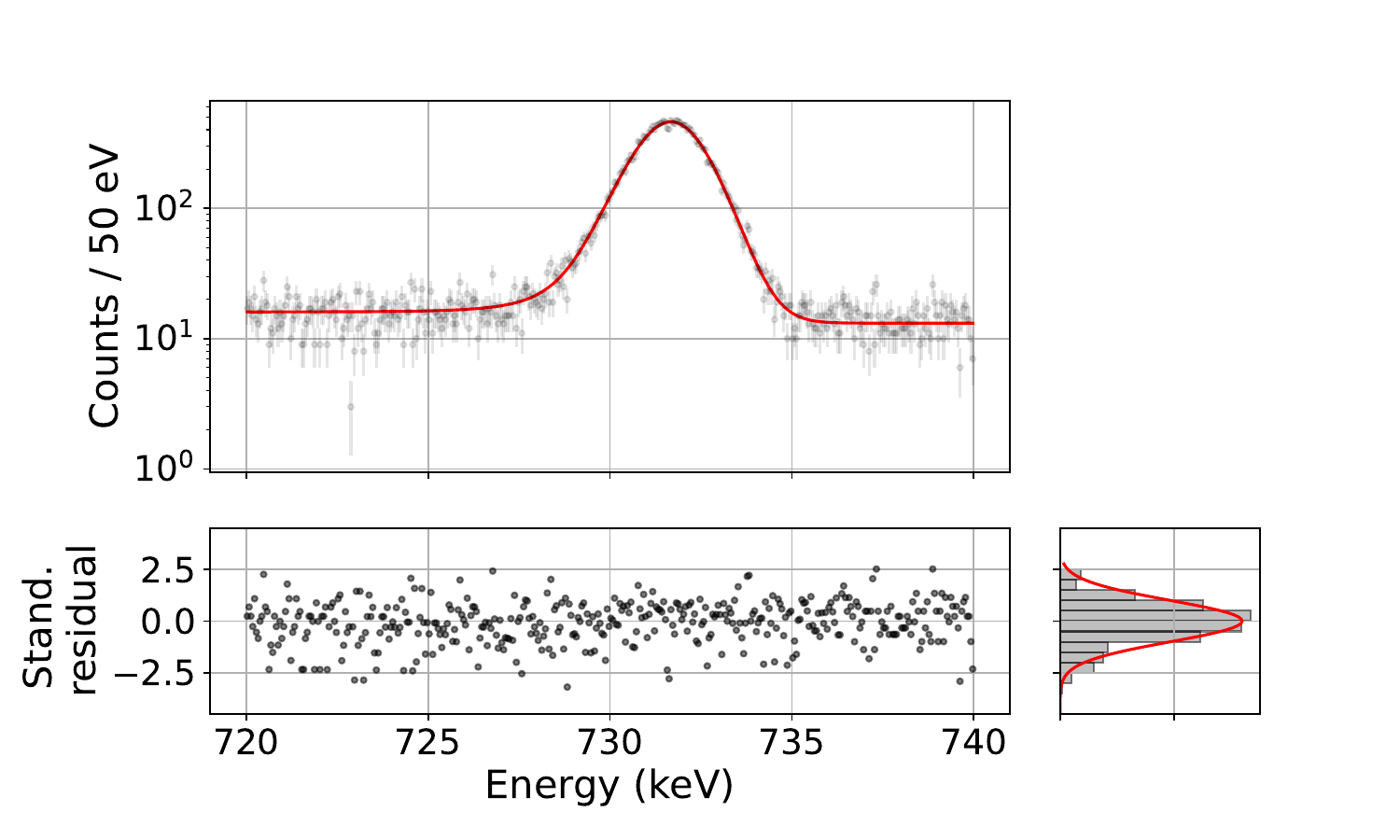}
        \caption{Energy spectrum of $4p1s$ in \textsuperscript{35}Cl summed over detectors, excluding Ge08.}
        \label{fig:exp_35Cl_4p1s}
    \end{figure}

    % 37Cl
    \begin{figure}[ht!]
        \centering
        \includegraphics[width=0.8\linewidth]{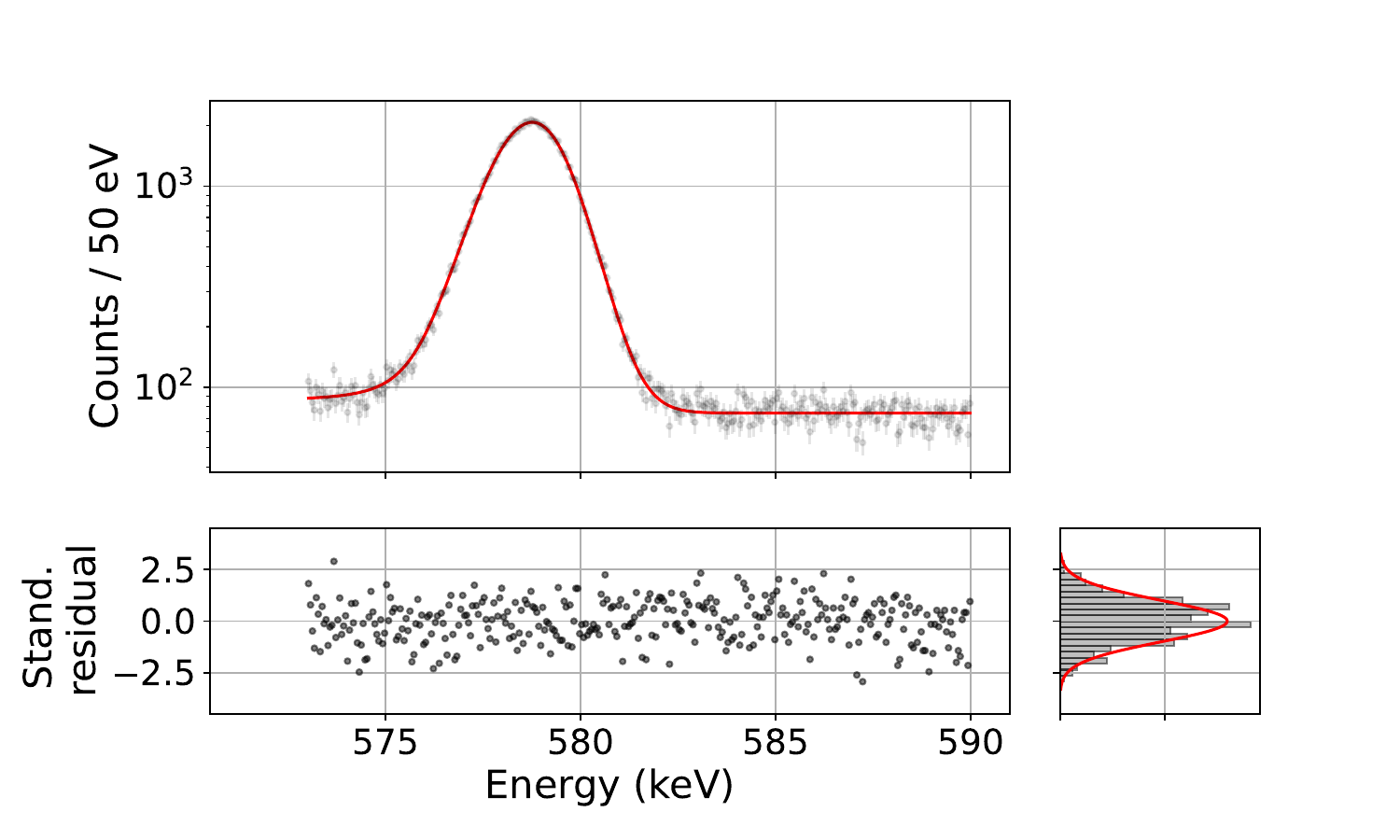}
        \caption{Energy spectrum of $2p1s$ in \textsuperscript{37}Cl summed over detectors, excluding Ge08.}
        \label{fig:exp_37Cl_2p1s}
    \end{figure}
    \begin{figure}[ht!]
        \centering
        \includegraphics[width=0.8\linewidth]{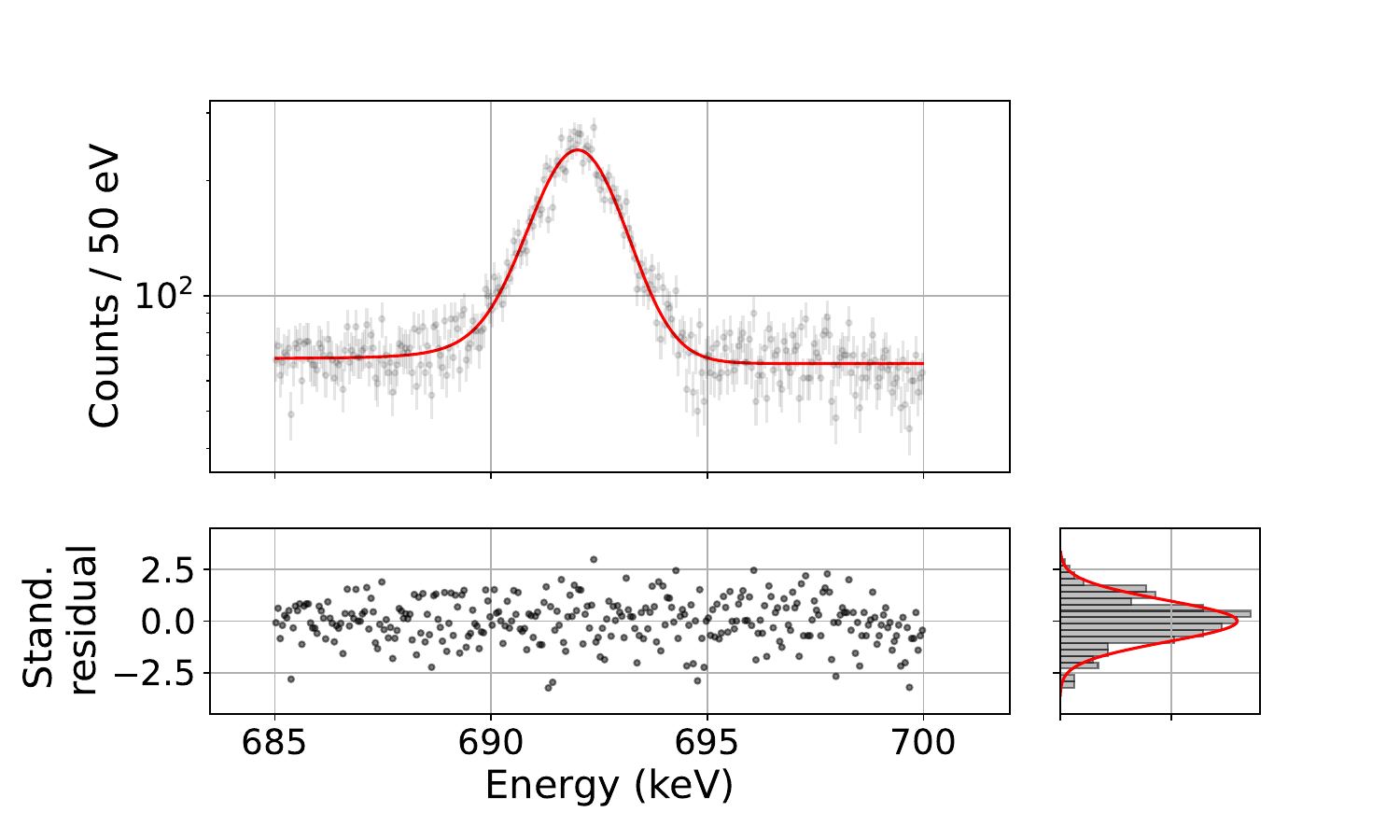}
        \caption{Energy spectrum of $3p1s$ in \textsuperscript{37}Cl summed over detectors, excluding Ge08.}
        \label{fig:exp_37Cl_3p1s}
    \end{figure}
    \begin{figure}[ht!]
        \centering
        \includegraphics[width=0.8\linewidth]{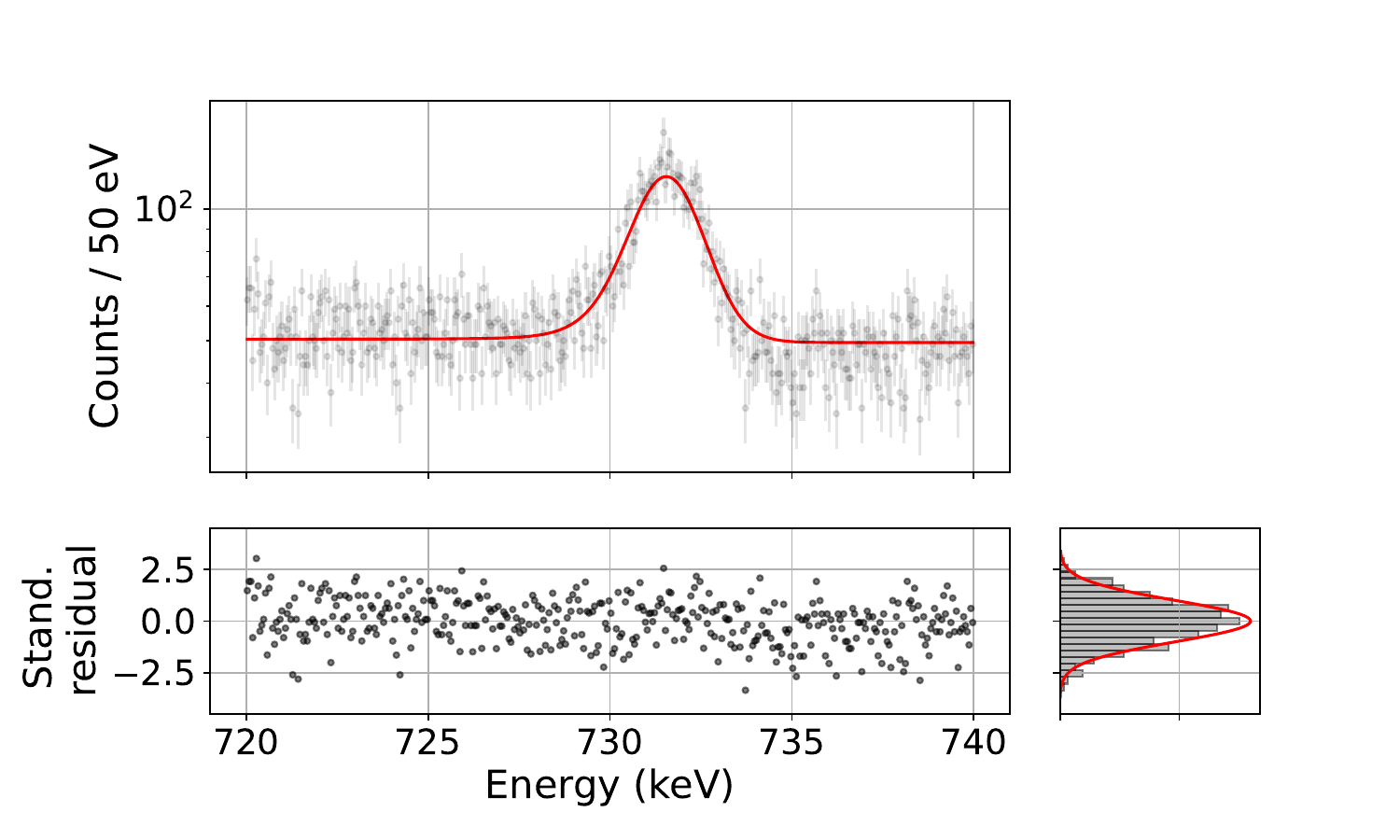}
        \caption{Energy spectrum of $4p1s$ in \textsuperscript{37}Cl summed over detectors, excluding Ge08.}
        \label{fig:exp_37Cl_4p1s}
    \end{figure}

\clearpage
\section{Uncertainty per detector} \label{app:exp_uncertainty}
    \begin{figure}[ht!]
        \centering
        \includegraphics[width=0.9\linewidth]{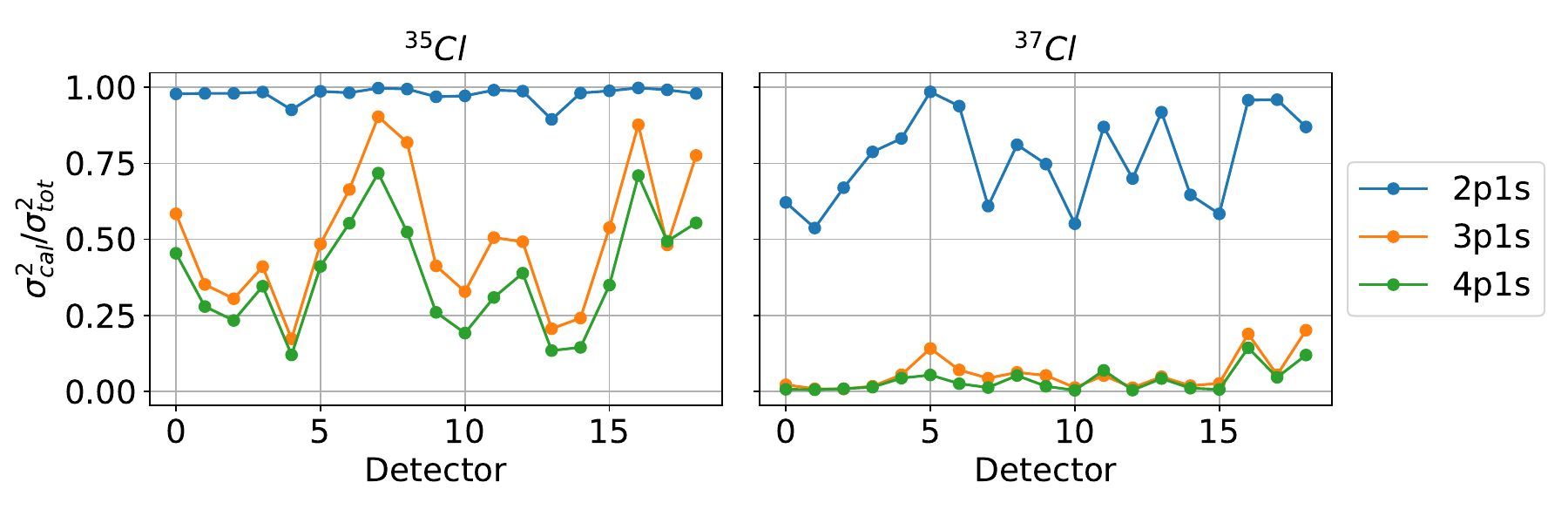}
        \caption{Ratio of calibration uncertainty in the total uncertainty per detector for the chlorine measurements}
        \label{fig:exp_unc_ratio_Cl}
    \end{figure}

\section{QED for fine and hyperfine structure} \label{app:exp_hyperfine}
    \begin{table}[ht!]
        \centering
        \caption{Hyperfine and fine structure shifts for \textsuperscript{35}Cl (I=3/2). Shifts are given compared to the center of gravity.}
        \begin{tabular}{cc|ccc|c}
            State & F & M1 ($\si{\eV}$) & E2 ($\si{\eV}$) & FS ($\si{\eV}$) & Total ($\si{eV}$) \\
            \hline
            $1s$\textsubscript{1/2} & 1 & $-$79.557 & 0 & 0 & $-$79.557\\
                                  & 2 & +47.734 & 0 & 0 & +47.734\\
            \hline
            $2p$\textsubscript{1/2} & 1 & $-$6.278 & 0 & $-$519.864 & $-$526.142 \\
                                  & 2 & +3.767 & 0 & $-$519.864 & $-$516.097 \\
            \hline
            $2p$\textsubscript{3/2} & 0 & $-$3.794 & $-$75.081 & +259.932 & +181.057 \\
                                  & 1 & $-$2.782 & $-$15.016 & +259.932 & +242.134 \\
                                  & 2 & $-$0.759 & +45.049 & +259.932 & +304.222 \\
                                  & 3 & +2.276 & $-$15.016  & +259.932 & +247.192
        \end{tabular}
        \label{tab:hfs_35Cl}
    \end{table}

    \begin{table}[ht!]
        \centering
        \caption{Hyperfine and fine structure shifts for \textsuperscript{37}Cl (I=3/2). Shifts are given compared to the center of gravity.}
        \begin{tabular}{cc|ccc|c}
            State & F & M1 ($\si{\eV}$) & E2 ($\si{\eV}$) & FS ($\si{\eV}$) & Total ($\si{eV}$) \\
            \hline
            $1s$\textsubscript{1/2} & 1 & $-$66.000 & 0 & 0 & $-$66.000\\
                                  & 2 & +39.600 & 0 & 0 & +39.600\\
            \hline
            $2p$\textsubscript{1/2} & 1 & $-$5.226 & 0 & $-$519.824 & $-$525.050 \\
                                  & 2 & +3.136 & 0 & $-$519.824 & $-$516.688 \\
            \hline
            $2p$\textsubscript{3/2} & 0 & $-$3.160 & $-$56.913 & +259.912 & +199.839 \\
                                  & 1 & $-$2.317 & $-$11.383 & +259.912 & +246.212 \\
                                  & 2 & $-$0.632 & +34.148 & +259.912 & +293.428 \\
                                  & 3 & +1.896 & $-$11.383 & +259.912 & +250.425
        \end{tabular}
        \label{tab:hfs_37Cl}
    \end{table}

\newpage

\section{Photon recoil effects}\label{app:exp_recoil}
    The measured energy and the atomic transition energy are not quite the same. The nucleus of mass $m_N$ carries part of the transition energy into recoil energy $E_R$ in order to retain momentum conservation. The relation between the measured energy $E_{\gamma}$ and the total transition energy $E$ can be derived as follows. Starting from momentum conservation, we have

    \begin{align*}
        p_{\gamma} &= p_N \\
        \implies \frac{E_\gamma}{c} &= \sqrt{2 m_N E_R} \\
        \implies \frac{E_\gamma}{c} &= \sqrt{2 m_N (E - E_\gamma)} \\
        \implies E_\gamma^2 &= 2 m_N c^2 (E - E_\gamma)
    \end{align*}

    Next, one can solve for $E$ or $E_\gamma$. Starting with the former

    \begin{align*}
        E_\gamma^2 &= 2 m_N c^2 (E - E_\gamma) \\
        \implies E &= E_\gamma + \frac{E_\gamma^2}{2 m_N c^2}
    \end{align*}

    The recoil correction is given by 

    \begin{align*}
        E_R & = E - E_\gamma \\
        & = \frac{E_\gamma ^2}{2 m_N c^2}
    \end{align*}

    Alternatively, one can solve for $E_\gamma$

    \begin{align*}
        & E_\gamma^2 = 2 m_N c^2 (E - E_\gamma) \\
        \implies& E_\gamma^2 + 2 m_N c^2 E_\gamma - 2 m_N c^2 E = 0 \\
        \implies& E_\gamma = \frac{-2 m_N c^2 \pm \sqrt{(2 m_N c^2)^2 - 4\times (-2 m_N c^2 E))}}{2}
    \end{align*}

    Since $E_\gamma > 0$, we have

    \begin{align*}
        E_\gamma &= \frac{-2 m_N c^2 + \sqrt{(2 m_N c^2)^2 - 4\times (-2m_N c^2 E))}}{2} \\
        &= -m_N c^2 + \sqrt{m_N c^2 (2E + m_N c^2)} \\
        &= m_N c^2 \left(\sqrt{1 + \frac{2E}{m_N c^2}} - 1\right)
    \end{align*}

\bibliography{biblio.bib}